\documentclass[a4paper,fleqn,usenatbib]{mnras}

\usepackage[T1]{fontenc}
\usepackage{ae,aecompl}

\usepackage{graphicx}	
\usepackage{amsmath}	
\usepackage{amssymb}	
\usepackage{subfigure}

\newcommand{\msun}{M$_\odot$}
\newcommand{\ha}{H$\alpha$}

\newcommand{\sigmasfr}{$\Sigma_{\textrm{SFR}}$}

\graphicspath{{./pictures/}}

\title[Cluster formation and evolution in M51 with LEGUS]{The Young Star Cluster population of M51 with LEGUS: I. A comprehensive study of cluster formation and evolution}

\author[M. Messa et al.]{M. Messa,$^{1}$\thanks{E-mail: matteo.messa@astro.su.se}
A. Adamo,$^{1}$
G. \"{O}stlin,$^{1}$
D. Calzetti,$^{2}$
K. Grasha,$^{2}$
E.K. Grebel,$^{3}$
\newauthor
F. Shabani,$^{3}$
R. Chandar,$^{4}$
D.A. Dale,$^{5}$
C.L. Dobbs,$^{6}$
B.G. Elmegreen,$^{7}$
\newauthor
M. Fumagalli,$^{8}$
D.A. Gouliermis,$^{9,10}$
H. Kim,$^{11}$
L.J. Smith,$^{12}$
D.A. Thilker$^{13}$
M. Tosi,$^{14}$
\newauthor
L. Ubeda,$^{15}$
R. Walterbos,$^{16}$
B.C. Whitmore,$^{15}$
K. Fedorenko,$^{17}$
S. Mahadevan,$^{17}$
\newauthor
J.E. Andrews,$^{2,18}$
S.N. Bright,$^{15}$
D.O. Cook,$^{19}$
L. Kahre,$^{16}$
P. Nair,$^{20}$
A. Pellerin,$^{21}$
\newauthor
J. Ryon,$^{22}$
S.D. Ahmad,$^{21}$
L.P. Beale,$^{21}$
K. Brown,$^{21}$
D.A. Clarkson,$^{21}$
\newauthor
G.C. Guidarelli,$^{21}$
R. Parziale,$^{5}$
J. Turner,$^{5}$
M. Weber$^{21}$
\\
$^{1}$Dept. of Astronomy, The Oskar Klein Centre, Stockholm University, Stockholm, Sweden\\
$^{2}$Dept. of Astronomy, University of Massachusetts -- Amherst, Amherst, MA 01003\\
$^{3}$Astronomisches Rechen-Institut, Zentrum f\"ur Astronomie der Universit\"at Heidelberg, M\"onchhofstr.\ 12--14, 69120 Heidelberg, Germany\\
$^{4}$Dept. of Physics and Astronomy, University of Toledo, Toledo, OH\\
$^{5}$Dept. of Physics and Astronomy, University of Wyoming, Laramie, WY\\
$^{6}$School of Physics and Astronomy, University of Exeter, Exeter, United Kingdom\\
$^{7}$IBM Research Division, T.J. Watson Research Center, Yorktown Hts., NY\\
$^{8}$Institute for Computational Cosmology and Centre for Extragalactic Astronomy, Durham University, Durham, United Kingdom\\
$^{9}$Zentrum f\"ur Astronomie der Universit\"at Heidelberg, Institut f\"ur Theoretische Astrophysik, Albert-Ueberle-Str.\,2, 69120 Heidelberg, Germany\\
$^{10}$Max Planck Institute for Astronomy,  K\"{o}nigstuhl\,17, 69117 Heidelberg, Germany\\
$^{11}$Gemini Observatory, Casilla 603, La Serena, Chile\\
$^{12}$European Space Agency/Space Telescope Science Institute, Baltimore, MD\\
$^{13}$Dept. of Physics and Astronomy, The Johns Hopkins University, Baltimore, MD\\
$^{14}$INAF -- Osservatorio Astronomico di Bologna, Bologna, Italy\\
$^{15}$Space Telescope Science Institute, Baltimore, MD\\
$^{16}$Dept. of Astronomy, New Mexico State University, Las Cruces, NM\\
$^{17}$College of Information and Computer Sciences, University of Massachusetts, Amherst, MA \\
$^{18}$Dept. of Astronomy, University of Arizona, Tucson, AZ\\
$^{19}$California Institute of Technology, Pasadena, CA\\
$^{20}$Dept. of Physics and Astronomy, University of Alabama, Tuscaloosa, AL\\
$^{21}$Dept. of Physics and Astronomy, State University of New York at Geneseo, Geneseo, NY\\
$^{22}$Dept. of Astronomy, University of Wisconsin--Madison, Madison, WI\\
}

\date{Accepted XXX. Received YYY; in original form ZZZ}

\pubyear{2017}

\begin{document}
\label{firstpage}
\pagerange{\pageref{firstpage}--\pageref{lastpage}}
\maketitle

\begin{abstract}
Recently acquired WFC3 UV (F275W and F336W) imaging mosaics under the Legacy Extragalactic UV Survey (LEGUS) combined with archival ACS data of M51 are used to study the young star cluster (YSC) population of this interacting system. Our newly extracted source catalogue contains 2834 cluster candidates, morphologically classified to be compact and uniform in colour, for which ages, masses and extinction are derived. 
In this first work we study the main properties of the YSC population of the whole galaxy, considering a mass-limited sample. Both luminosity and mass functions follow a power law shape with slope -2, but at high luminosities and masses a dearth of sources is observed. The analysis of the mass function suggests that it is best fitted by a Schechter function with slope -2 and a truncation mass at $1.00\pm0.12\times10^5$ \msun. Through Monte Carlo simulations we confirm this result and link the shape of the luminosity function to the presence of a truncation in the mass function.
A mass limited age function analysis, between 10 and 200 Myr, suggests that the cluster population is undergoing only moderate disruption. We observe little variation in the shape of the mass function at masses above $1\times10^4$ \msun\, over this age range. 
The fraction of star formation happening in the form of bound clusters in M51 is $\sim20\%$ in the age range 10 to 100 Myr and little variation is observed over the whole range from 1 to 200 Myr. 

\end{abstract}



\section{Introduction}	         
The majority of stars do not form in isolation but in areas of clustered star formation \citep[e.g.][]{lada2003}. In some cases, the densest areas of these large regions result in gravitationally bound stellar systems, commonly referred to as star clusters. These bound systems can survive for hundreds of Myr. To distinguish them from ancient stellar objects like the globular clusters (GCs), we refer to them as young stellar clusters (YSCs). They usually populate star-forming galaxies in the local universe (e.g. \citealp{larsen2006b}) and their physical properties (ages, masses) can in principle be used to determined star formation histories (SFHs) of the hosting galaxies \citep[e.g.][]{miller1997,goudfrooij2004,konstantopoulos2009,glatt2010}.

Over the past 20 years, studies of the distributions of YSC luminosities and masses in local galaxies have shown that they are well-described by a power law function of the form $Ndm \propto M^\alpha dm$, with a slope $\alpha \sim-2$, observed both for low-mass clusters in the Milky Way \citep{piskunov2006}, in the Magellanic Clouds \citep{baumgardt2013,degrijs2006} and in M31 \citep{fouesneau2014}, and for sources up to masses of $\sim10^5-10^6$ \msun\ in nearby spirals and starburst galaxies \citep{chandar2010,konstantopoulos2013,whitmore2010}. This result is expected if star formation happens in a hierarchical manner, dominated by interstellar medium (ISM) turbulence, and the clusters occupy the densest regions (e.g. \citealp{elmegreen2006}; see also \citealp{elmegreen2010b} for a review).

Despite observational and theoretical progress over the past few decades, many questions concerning the properties of YSC populations remain open. Among these: is cluster formation only driven in space and time by size--of--sample effects \citep[e.g.][]{hunter03}, with an increasing number of clusters found in galaxies with higher star formation rate (SFR)? Will the galactic environment (ISM conditions, gas fraction, galaxy type) where clusters form leave an imprint on the final properties of the YSC populations? When we look at YSC populations in local spirals \citep[e.g.][]{larsen2004}, merger systems \citep[e.g.][]{whitmore1999} and dwarf galaxies \citep[e.g.][]{billett2002} it is challenging to discern the role played by statistical sampling \citep[e.g.][]{fumagalli2011} and environment.

Even the exact shape of the mass function is still debated, in particular concerning its high mass end. Some early studies \citep[e.g.][]{larsen2006} have pointed out the dearth of massive YSCs if a single power law fit of slope $-2$ describes the upper-part of the YSC mass function. \cite{gieles2006} have proposed a Schechter function as a better description of the YSC mass function in local galaxies, due to a mass truncation at a characteristic mass above which the likelihood of forming massive clusters goes rapidly to zero.

In order to be able to characterise how star clusters form and evolve, 
it is important to study a statistically meaningful sample. The Legacy Extragalactic UV Survey (LEGUS) is a Cycle 21 HST Treasury program which observed 50 nearby galaxies from the UV to NIR bands, with the goal of deriving high quality star cluster catalogues, and, more in general, of studying star formation at intermediate scales, linking the smallest (stellar) scales to the larger (galactic) ones \citep[see][]{legus1}. 
In general, the large number of galaxies and galaxy properties available in LEGUS will enable us to statistically study YSC populations over a wide range of galactic environments \citep{legus2}. 

Among the most interesting galaxies in the LEGUS catalogue is NGC 5194 (also known as M51a or the Whirlpool Galaxy), because of its proximity and the number of star clusters that it hosts.
It is a spiral galaxy, catalogued as SAbc\footnote{According to the Nasa Extra-galactic Database (NED)}, almost face-on (inclination angle $i\approx22^\circ$, \citealp{colombo2014b}) at a distance of 7.66 Mpc \citep{distancem51}. M51a is interacting with the (smaller) companion galaxy NGC 5195 and it is probably this interaction that is the cause of a marked spiral geometry and a high star formation process (a SFR value of 2.9 \msun/yr is derived from published total fluxes in the far-$UV$ and 24$\mu$m, combined using the recipe by \citealp{hao2011}) sustained over time \citep[e.g.][]{dobbs2010_m51}. The two galaxies together form the M51 system. In the remainder of this paper we will use the name M51 mainly referring to the main spiral galaxy M51a.
This galaxy hosts numerous star formation complexes \citep{bastian2005b}, HII regions \citep{thilker2000,lee2011} and YSCs and it has been a benchmark in the study of extragalactic star and cluster formation. 

High-brightness blue sources in M51 have been studied already by \citet{georgiev1990}.
In more recent years, broadband and narrowband imaging with the Hubble Space Telescope (HST) Wide-Field Planetary Camera 2 (WFPC2) in various bands from $UV$ to $NIR$ were used for initial studies of the cluster population in small parts of the galaxy \citep{bik2003,bastian2005,gieles2005,lee2005}. Later optical observations with the higher resolution and more sensitive ACS camera were obtained in the $BVI$ bands and covered uniformly the entire galaxy allowing to extend the investigation of the YSC population to the whole galaxy \citep{scheepmaker2007,hwang_lee2008, chandar2011}. More recently, the coverage by the WFPC2 F336W filter ($U$-band) has been expanded, with 5 more pointings, along with \ha\, data, allowing improved age determination for a significant fraction of the cluster population\footnote{The $U$-band filter (or bluer filters) is fundamental to break the age-extinction degeneracy when SEDs are compared to stellar population synthesis models, see \citet{anders2004}} \citep{scheepmaker2009,hwang_lee2010,chandar2011,chandar16}.

All this effort led to the consensus that the star cluster population in M51 can be described by a standard mass distribution, i.e. a simple power law with slope -2. However, whether the single-power law function is also a good representation of the upper-mass end of the cluster mass function, in terms of the eventual presence of a truncation at high masses, is still under debate (compare e.g. \citealp{gieles2006} and \citealp{chandar2011}). 
The analyses of the cluster mass function evolving in time, and, more in general, of the cluster number densities evolving with time, reach  different conclusions on the disruption properties of the clusters in M51. Some studies observe a mass function evolution consistent with a disruption time dependent on the mass of the clusters \citep[e.g. the mass-dependent disruption -MDD- model by][]{gieles2009}, while in others the study of the mass function (MF) evolution seems to exclude this model, and to favour a constant disruption time of clusters \citep[e.g. mass independent disruption -MID- model by][]{chandar16}. 

The interaction of M51 has been studied using simulations in order to describe the current geometrical and dynamical properties of the star formation  \citep{salo2000,dobbs2010_m51}. 
Cluster properties have then been compared with the expectations based on simulations in order to test the models for the formation of the spiral structure (e.g. \citealp{chandar2011} ruled out the possibility of self-gravity as the cause of the generation of the spiral structure). 

Star formation in M51 has also been studied from the point of view of molecular gas via radio observations (\citealp{schuster2007,koda2009,koda2011} and \citealp{schinnerer2010,schinnerer13} among the most recent). High resolution interferometric data have been used to study in detail the properties of giant molecular clouds (GMCs) \citep{koda2012,colombo2014a}.
The possibility of studying the galaxy at high resolution at different wavelengths allows studying star formation at different ages, in particular to compare the properties of the progenitors (GMCs) and the final products (stars and star clusters). 

One of the goals of the present work is to conduct a statistically driven study of the YSC population of M51 using the new data and cluster catalogue produced by the LEGUS team. The new LEGUS dataset of M51 provides 5 new pointings in the $NUV$ (F275W and F336W) with the Wide Field Camera 3 (WFC3). The improved spatial resolution of the WFC3 and sensitivity in the $NUV$ give a better leverage on the physical determinations of the YSC properties \citep{legus1}.
In order to compare our new catalogue with previously published works we investigate, in this paper, YSC mass and luminosity functions for the whole galaxy. With the help of simulated Monte Carlo cluster populations we build a comprehensive picture of the cluster formation and evolution in the galaxy as a whole. In a forthcoming paper (Messa et al., in prep, hereafter Paper II) we test whether YSC properties change across the galaxy as a function of star formation rate (SFR) density (\sigmasfr) and gas surface density.  These results can shed light on a possible environmental dependences in the properties of the cluster population and whether studies of YSC populations can be used to characterise the galactic environment.

The paper is divided as follows: a short description of the data is given in Section~\ref{sec:data} and the steps necessary to produce the final cluster catalogue are described in Section~\ref{sec:ccp}. In Section~\ref{sec:global} the global properties of the sample (luminosity, mass and age functions) are studied, while in Section~\ref{sec:montecarlo} the same properties are analysed using simulated Monte Carlo populations. The fraction of star formation happening in a clustered fashion is studied in Section~\ref{sec:cfe}.
Finally, the conclusions are summarised in Section~\ref{sec:conclusions}.

\section {Data} 
\label{sec:data}
A detailed description of the LEGUS general dataset and the standard data reduction used for LEGUS imaging is given in \citet{legus1} and we refer the reader to that paper for details on the data reduction steps.

Here we summarize the properties of the data used in this study. The M51 system (NGC 5194 and NGC 5195) spans $\sim7 \times 10$ arcmin on the sky at optical wavelengths (at an assumed distance of 7.66 Mpc, from \citealp{distancem51}) and several pointings are therefore necessary to cover their entire angular size. 
The LEGUS dataset includes multi-band data spanning the wavelength range from near-$UV$ to near-$IR$; data for M51 cover the $UV$ (F275W), $U$ (F336W), $B$ (F435W), $V$ (F555W) and $I$ (F814W) bands. Even though no conversion is applied to the Cousins-Johnson filter system, we keep the same nomenclature, due to the similarity of the central wavelength between that system and our data. Concerning the $B$, $V$ and $I$ filters, ACS WFC archival data
available from the Mikulski Archive for Space Telescopes (MAST) have been re-processed. The data in these bands include 6 pointings that cover the entire galaxy and the companion galaxy NGC 5195 (GO-10452, PI: S. Beckwith).

\begin{figure}
\centering
\includegraphics[width=0.4\textwidth]{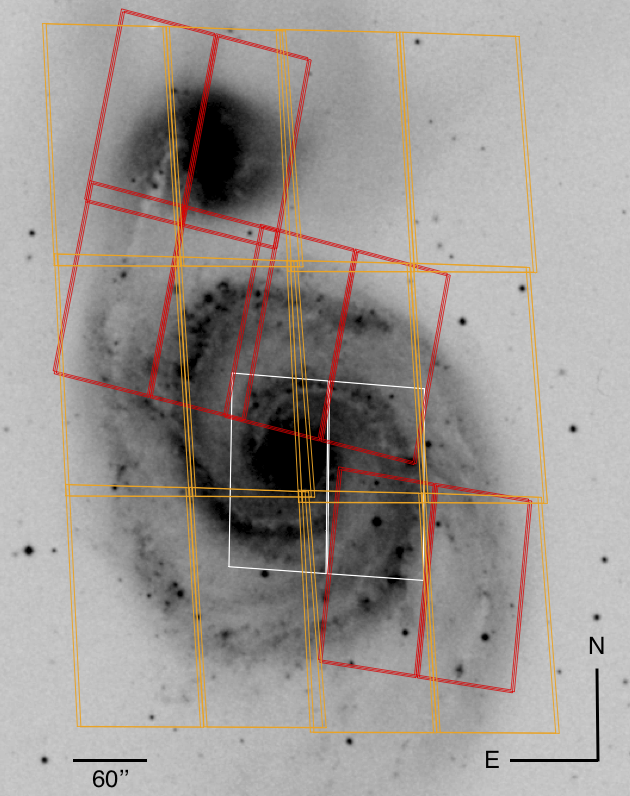}
\caption{UVIS (red) and ACS (orange) footprints on a DSS image of the NGC5194 and NGC5195 system. The UVIS (white) footprint corresponds to proposal 13340 (PI: S. Van Dyk). See also Tab. \ref{tab:data} for more info on the observations.}
\label{fig:footprints}
\end{figure}

Within the LEGUS project, the coverage has been extended to the $U$ and $UV$ bands. The new $UV$ data consist of 4 pointings covering the arms and outskirts of the galaxy combined with a deep central exposure (GO-13340, PI: S. Van Dyk) covering the nuclear region of the galaxy. Exposure times for all filters are summarized in Tab. \ref{tab:data} while the footprints of the pointings are illustrated in Figure \ref{fig:footprints}.

\begin{table}
\centering
\caption{Exposure times for the different filters and number of pointings (the exposure times refer to each single pointing). As can be noted also in Fig.~\ref{fig:footprints} the ACS data cover the entire galaxy with 6 pointings, while for the $UV/U$ band the observations consist of 5 pointings only.}
\begin{tabular}{clccl}
\hline
Instr. & Filter & Expt. & \# & Project Nr. \& PI \\
\hline
\hline
WFC3	& F275W($UV$)& 2500 s	& 4	& GO-13364\ \ 	 D. Calzetti \\
\ 		 & \			& 7147 s	& 1	& GO-13340\ \ 	 S. Van Dyk \\
WFC3	& F336W($U$)	& 2400 s	& 4	& GO-13364\ \	 D. Calzetti \\
\		 & \			& 4360 s	& 1	& GO-13340\ \ 	 S. Van Dyk \\
ACS 		& F435W($B$)	& 2720 s	& 6	& GO-10452\ \   S. Beckwith \\
ACS		& F555W($V$)	& 1360 s	& 6	& GO-10452\ \ 	 S. Beckwith \\
ACS		& F814W($I$)	& 1360 s	& 6	& GO-10452\ \ 	 S. Beckwith \\
\hline
\end{tabular}
\label{tab:data}
\end{table}

\section{Cluster Catalogue production} 
\label{sec:ccp}
\subsection{Cluster Extraction}
\label{sec:cluster_extraction}
In order to produce a cluster catalogue of the M51 galaxy we follow the procedures described in \cite{legus2}, where a detailed description of the standard reduction steps can be found. Hereafter we describe these steps along with the specific parameters used for the M51 dataset. The catalog
production is divided into 2 main parts, the cluster extraction and the cluster classification.

The cluster extraction is executed through a semi-automatic custom pipeline available inside the LEGUS collaboration. As the first step we extracted the source position of the cluster candidates in the $V$ band (used as reference frame in our analysis) with \texttt{SExtractor} \citep{sextractor}. The parameters of \texttt{SExtractor} were chosen to extract sources with at least a 10$\sigma$ detection in a minimum of 10 contiguous pixels.
In the same band, we measured the concentration index (CI) on each of the extracted sources.
We use the definition for the CI as the magnitude difference between the fluxes in circular regions of 1 and 3 pixels radius, centred on the source position. It measures how much the light is concentrated in the centre of the source and can also be used as also a tracer of the cluster size \citep[see][]{ryon2017}. 

\begin{figure*}
\centering
\subfigure{\includegraphics[width=0.9\columnwidth]{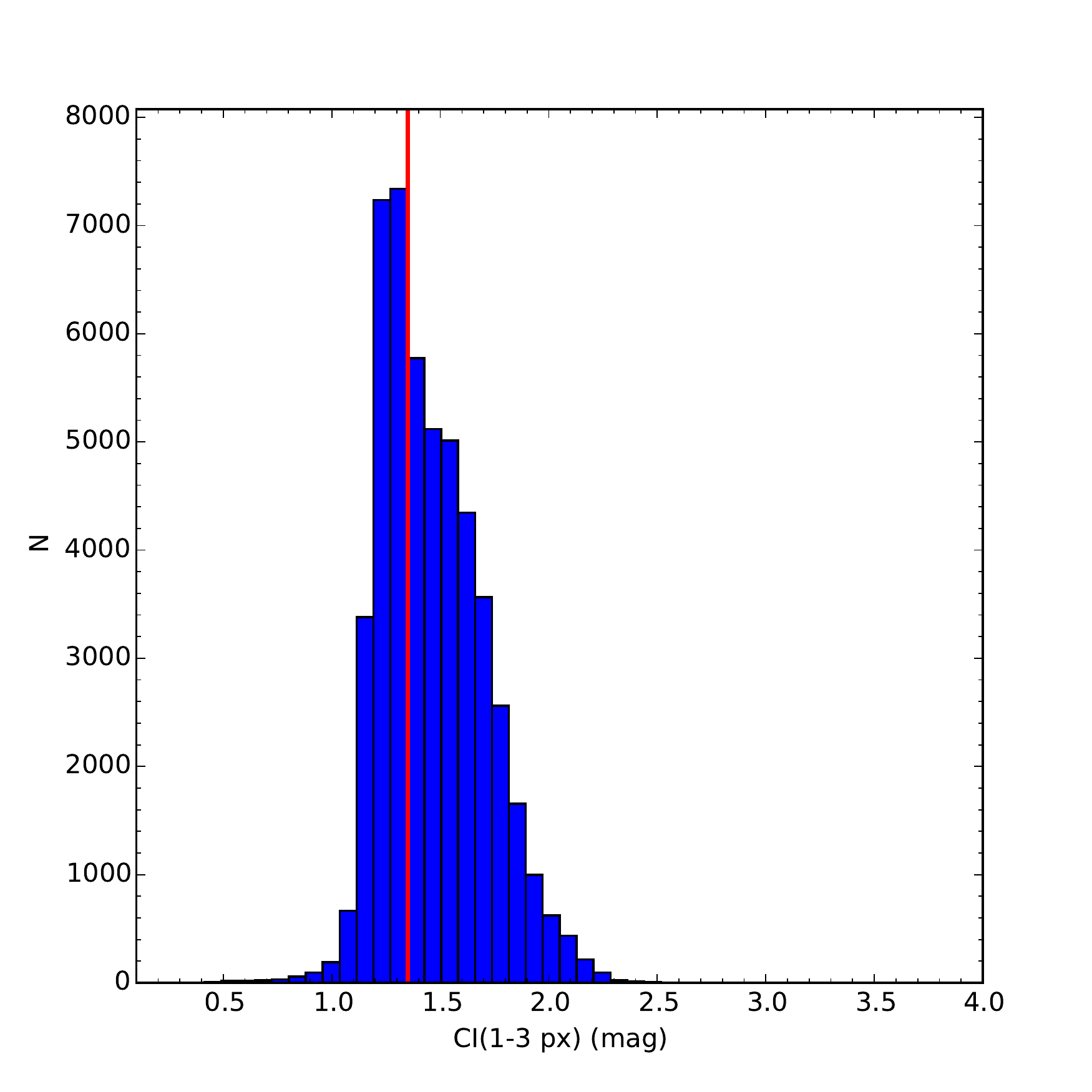}}
\hspace{1mm}
\subfigure{\includegraphics[width=0.9\columnwidth]{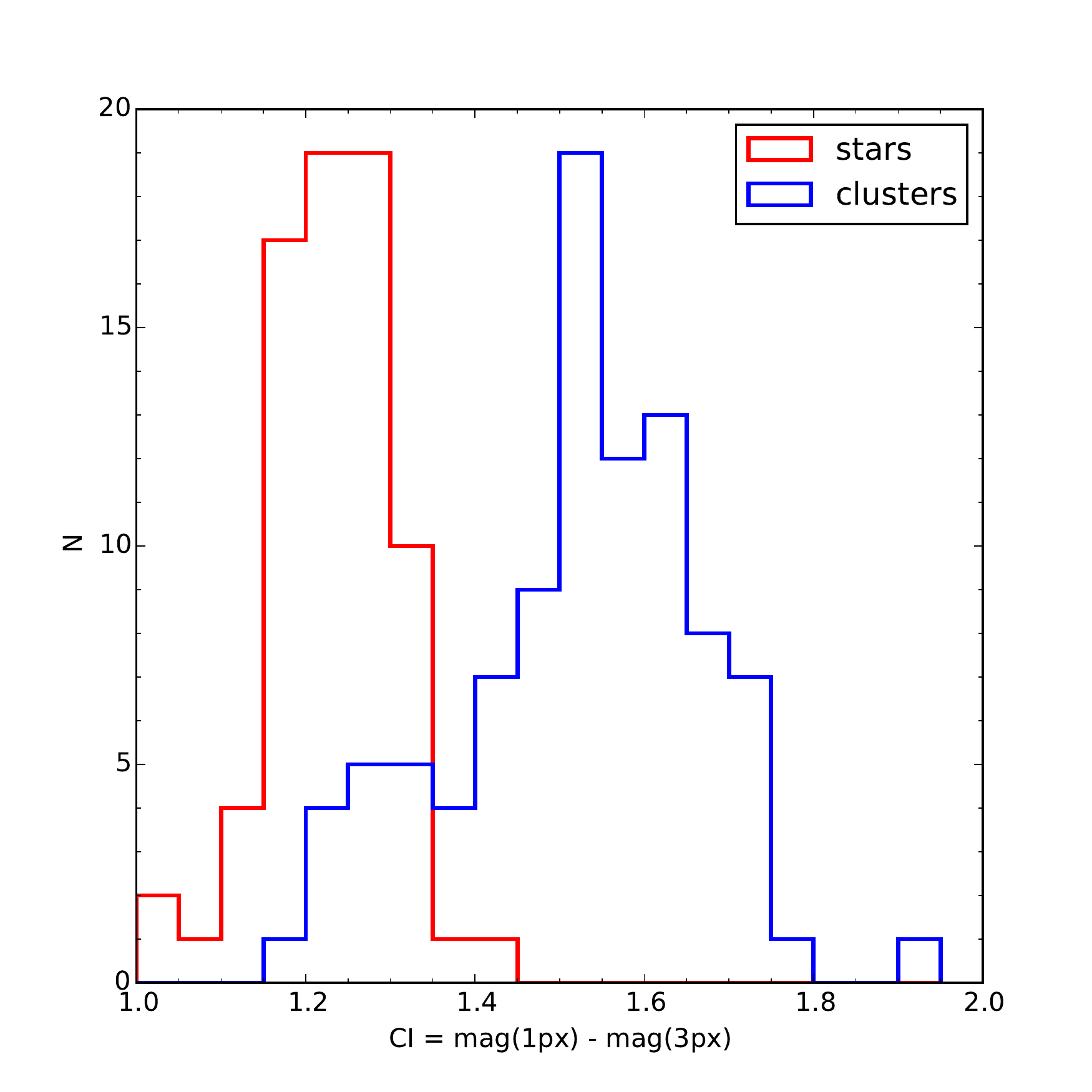}}
\subfigure{ \includegraphics[width=0.18\textwidth]{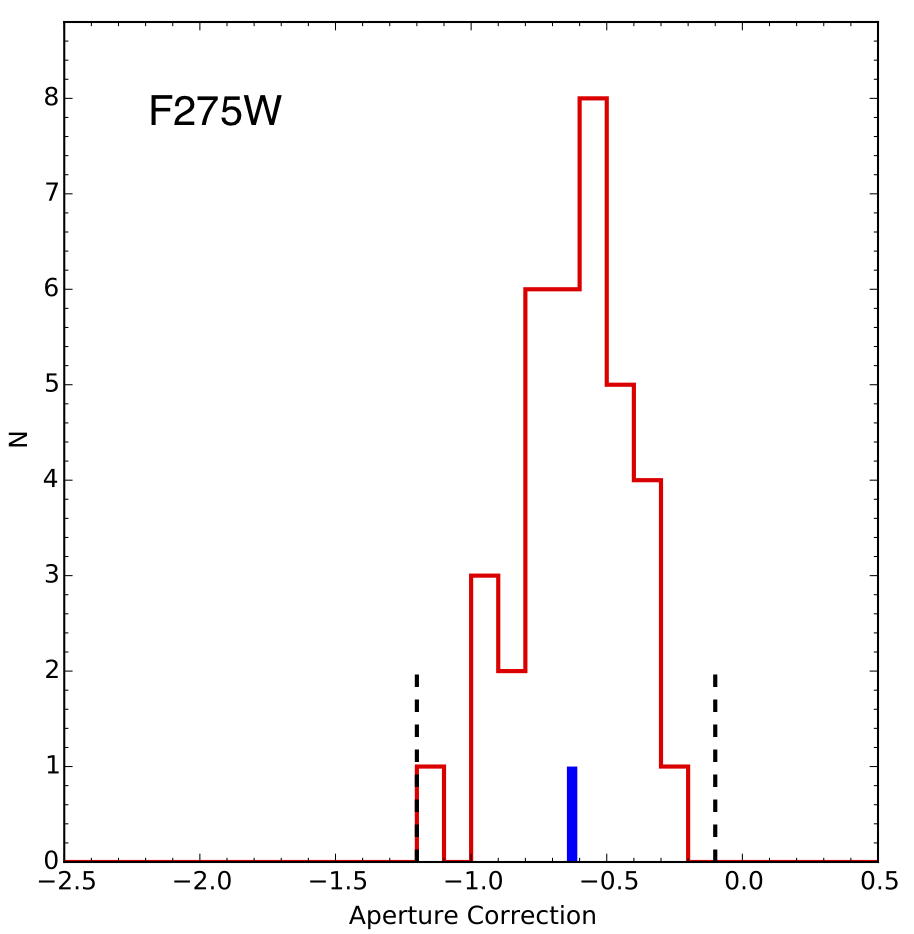}}
\subfigure{ \includegraphics[width=0.18\textwidth]{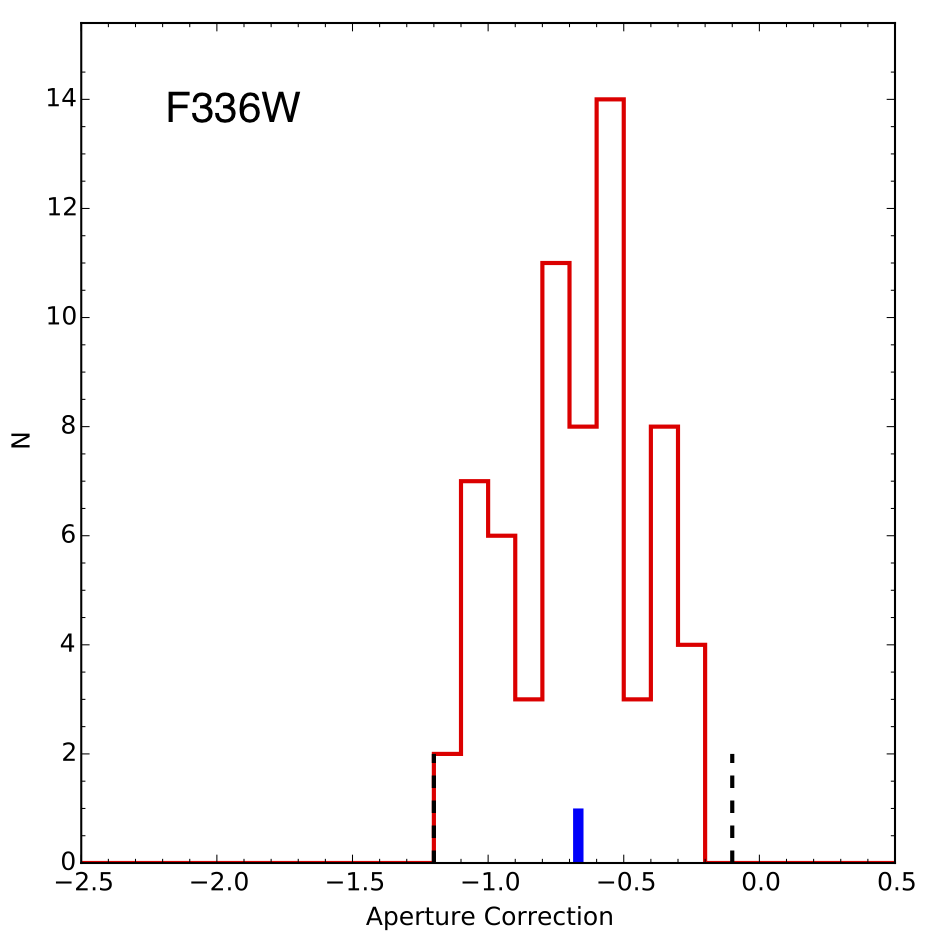}}
\subfigure{ \includegraphics[width=0.18\textwidth]{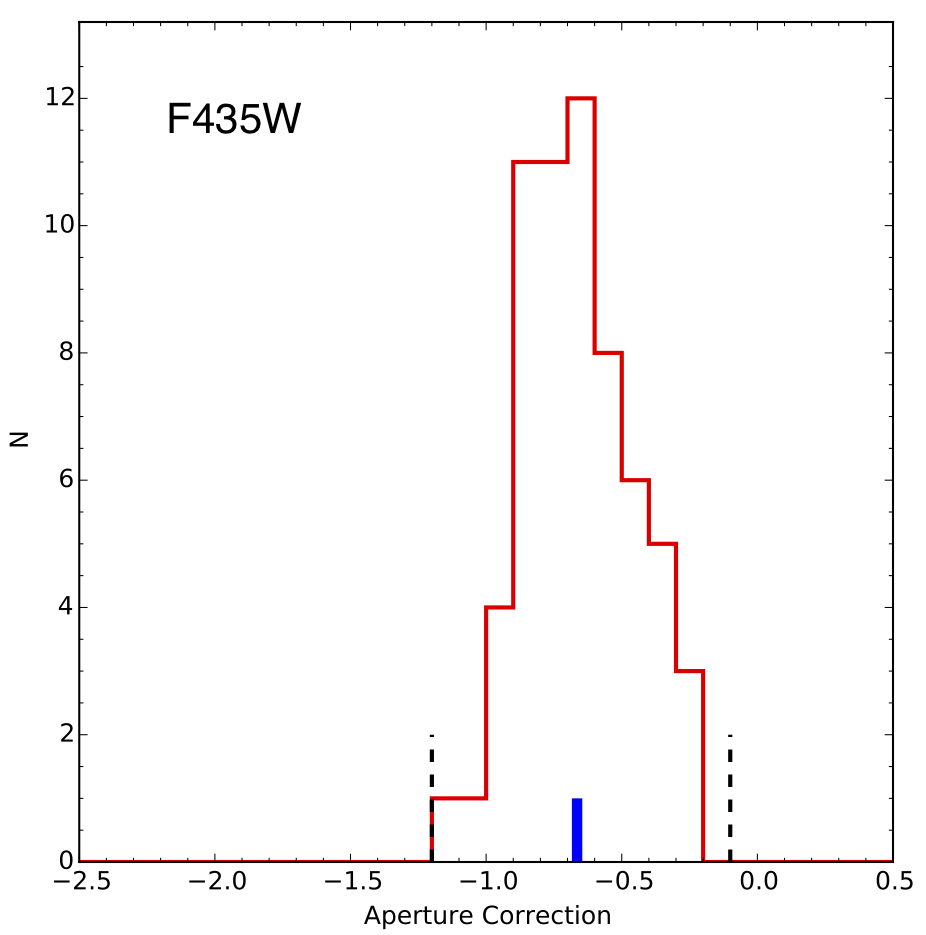}}
\subfigure{ \includegraphics[width=0.18\textwidth]{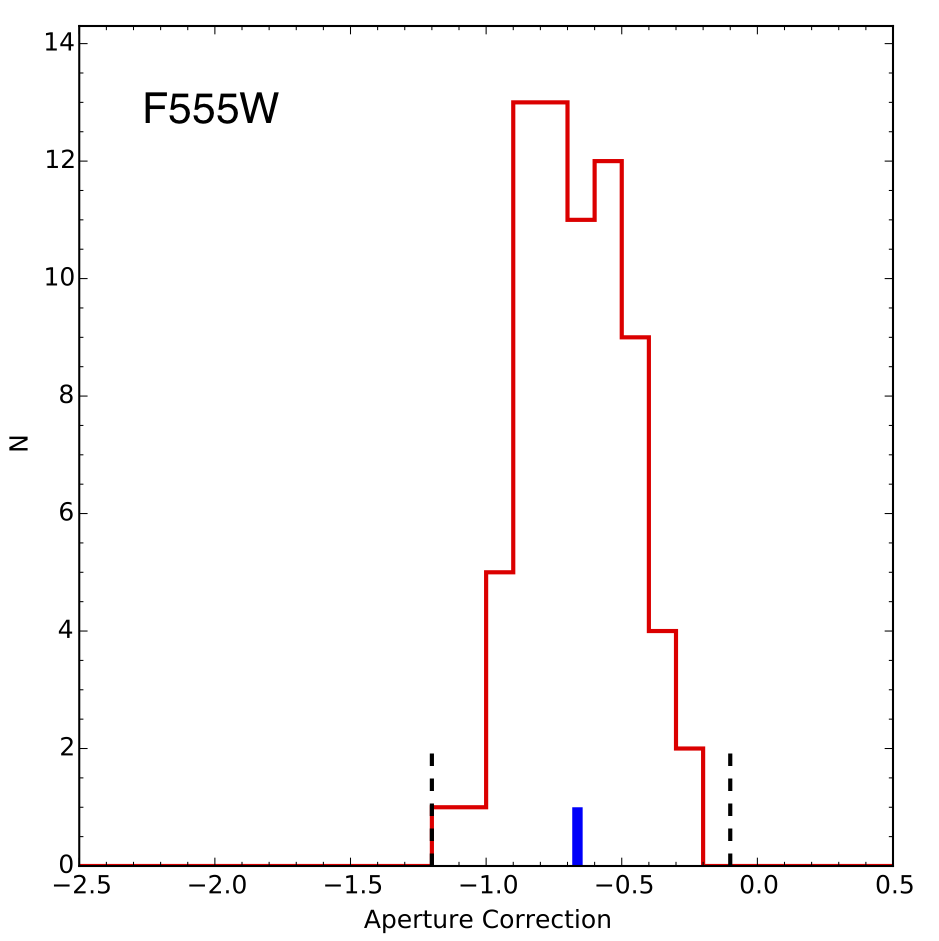}}
\subfigure{ \includegraphics[width=0.18\textwidth]{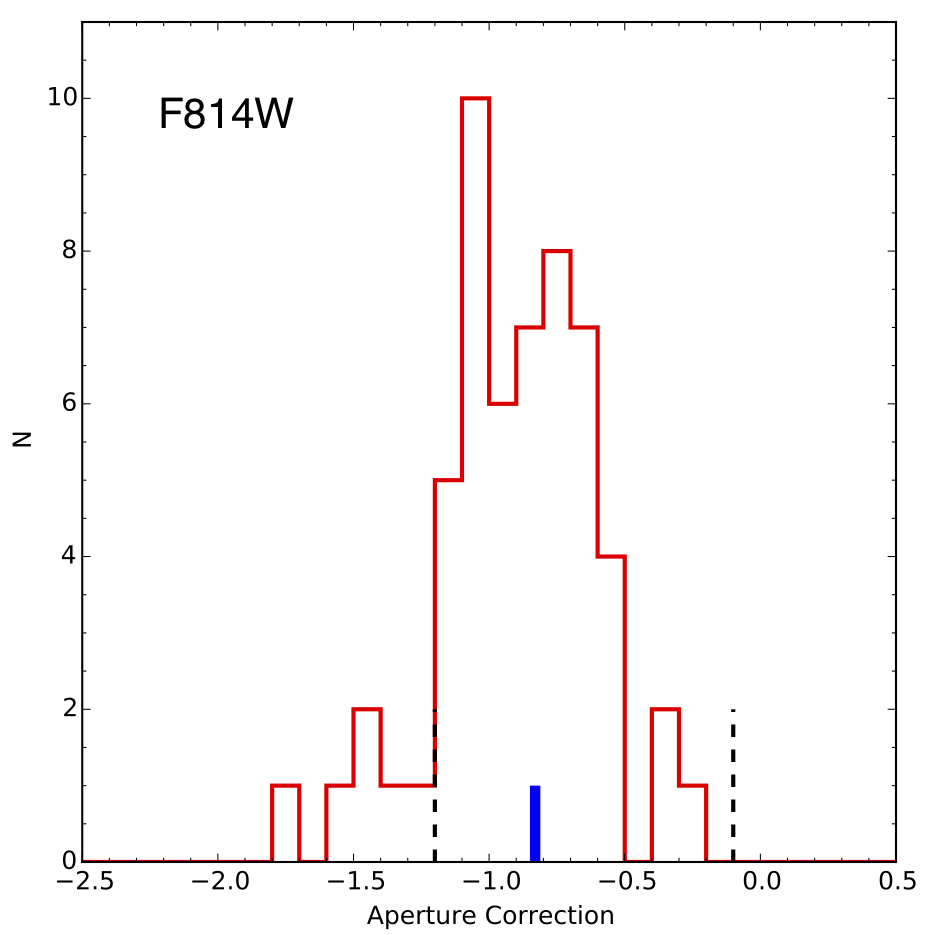}}
\caption{\textit{Top left:} Distribution of the CI values for all the sources extracted with \texttt{SExtractor}. The red solid line indicates the value of 1.35 used to select the cluster catalogue. \textit{Top right:} CI distribution for visually selected stars (in red) and clusters (in blue). \textit{Bottom:} Distributions of the aperture corrections given by visually selected clusters. Black dashed lines set the interval within which values are considered to calculate the average value (blue bin). 
}
\label{fig:histo_CI}
\end{figure*}
The distribution of the CI values for the extracted sources looks like a continuous distribution, peaked around a value of 1.3, as Fig.~\ref{fig:histo_CI} (top left) shows, but is in fact the sum of two sub-distributions, one for stars and one for clusters. To understand how the distributions of CI values changes between stars and clusters we select via visual inspection a sample of stars and clusters that are used as training samples for our analysis. In Fig.~\ref{fig:histo_CI}, top right, we show the CI distributions of stars and clusters. It becomes clear that the distributions of CIs in the two cases are different. Stars, being point sources, have CI values that do not exceed values of $\sim$1.4, while clusters have on average higher CI values and a broader distribution. The distributions also suggest that considering sources with a CI bigger than 1.35 strongly decreases the chances of erroneously including stars in the catalogue, thus facilitating the selection of most of the clusters.  Following the CI versus effective radius relation showed in Fig. 4 of \citet{legus2}, we estimate that a CI cut at 1.35 mag corresponds to a cluster effective radius of 0.5 pc. Because the distribution star cluster radii peak at $\sim$3 pc \citep{ryon2017}, placing a CI cut at 1.35 mag will not negatively impact the recovery of clusters in this system.

Aperture photometry was performed on the CI-selected sample, using fixed apertures of 4 pixels radius, and local sky is subtracted from an annulus with 7 pixels (px) interior radius and 1 px width. 
A fixed aperture correction was estimated using the photometric data of the visually selected sample of clusters. The sample was adjusted in each filter in order to consider only isolated bright clusters with well defined PSF wings. The number of visually selected sources used in each filter are listed in Tab~\ref{tab:corrections}. During the visual selection, sources were chosen to span different cluster sizes and to also include compact clusters. In this way the resulting aperture correction is not biased towards the very large clusters which are more easily detectable. For each source the single aperture correction was calculated subtracting the standard photometry (aperture: 4 px and sky at 7 px) with the total photometry within a 20 px radius (with a 1 px wide sky annulus at a radius of 21 px). The final correction in each filter was calculated taking the average of the values within an allowed range of values. The single aperture correction values of the selected sample along with the final mean value in each filter are plotted in Fig.~\ref{fig:histo_CI} (bottom) and the values are also reported in Tab.~\ref{tab:corrections}. The standard deviation ($\sigma_{\lambda}$) is added in quadrature to the photometric error of each source. 
\begin{table}
\centering
\caption{Corrections applied to the photometry of all the sources in different filters. With reddening we refer to the Galactic extinction in magnitudes (in each filter). The uncertainty on the average aperture correction has been used to estimate the final error on the magnitude.}\begin{tabular}{ccccc}
\hline
Filter		& Reddening	& \# Clusters 	& Avg ap.corr. 	& $\sigma_{\textrm{ap.corr.}}$	\\
\		& [mag]		&  			& [mag]	 	& [mag]	\\
\hline
\hline
F275W	& 0.192		& 36				& -0.628  		& 0.034 \\
F336W	& 0.156		& 66				& -0.668  		& 0.030 \\
F435W	& 0.127		& 62				& -0.665  		& 0.026 \\
F555W	& 0.098		& 71				& -0.663		& 0.023 \\
F814W	& 0.054		& 56				& -0.830		& 0.031 \\
\hline
\end{tabular}
\label{tab:corrections}
\end{table}

A final cut was made excluding sources which are not detected in at least 2 contiguous bands (the reference $V$ band and either the $B$ or $I$ band) with a photometric error smaller than 0.3 mag. The positions of the 30176 sources satisfying the CI cut of 1.35 mag and this last selection criterion are collected, along with their photometric data, in a catalogue named \textit{``automatic\_catalog\_avgapcor\_ngc5194.tab''}, following the LEGUS naming convention. Note that, being automatically-selected, this catalogue probably includes contaminant sources (e.g. foreground stars, background galaxies, stars in the field of M51).

To remove the contamination of non-clusters in the automatic catalog, we created a high-fidelity sub-catalog including all sources detected in at least 4 bands with a photometric error below 0.30 mag and having an absolute $V$ band magnitude brighter than -6 mag. Selecting only bright sources reduces considerably the number of stars in the catalogue, while the constraint on the number of detected bands allows a reliable process for the SED fitting analysis (see Section~\ref{sec:sed}). Note that, differently from the standard LEGUS procedure, we applied the -6 mag cut to the 
average--aperture--corrected magnitudes and not to the CI--based--corrected ones (see \citealp{legus2} for a description of the CI--based correction).
This choice is motivated mainly by the use of the average--aperture--corrected catalogue as the reference one: when testing the completeness level of our catalogue (see Section~\ref{sec:completeness}) we noticed that applying the cut on the average--aperture--corrected magnitudes improves the completeness, being more conservative (i.e. allowing the inclusion of more sources). This high-fidelity catalogue counts 10925 sources, which have been all morphologically classified (see Section~\ref{sec:classification}).

\subsection{Morphological classification of the cluster candidates, human versus machine learning inspection} 
\label{sec:classification}

Sources in the high fidelity catalogue were visually inspected, in order to morphologically classify the cluster candidates and exclude stars and interlopers that passed the automatic selection.
Like for the other galaxies of the LEGUS sample, visually inspected sources were divided into 4 classes, described and illustrated in \citet{legus2}. Briefly, class 1 contains compact (but more extended than stars) and spherically symmetric sources while class 2 contains similarly compact sources but with a less symmetric light distribution. Both these classes include sources with a uniform colour. Class 3 sources show multi-peaked profiles with underlying diffuse wings, which can trace the presence of (small and compact) associations of stars. Sources in class 3 can have colour gradients. Contaminants like single stars, or multiple stars that lie on nearby pixels even if not part of a single structure, and background galaxies are all stored in class 4 and excluded from the study of the cluster population of the galaxy.

\begin{table}
\centering
\caption{Number of sources in each class for the human-classified (2nd column) and machine-learning classified (3rd column) sources. In parentheses are the percentage of sources over the number of the visually inspected sources. The fourth column lists the percentage of the sources in each class for which the ML has assigned the same class as the humans. In the fifth column the number of sources per class classified by the machine learning algorithm is given. In brackets we include the recovered fraction with respect to the total number of sources in the catalogue (i.e. 10925 sources).}
\begin{tabular}{cllll}
\hline
Class 	& Human 			& ML				&HvsML		& ML	 (tot cat.) \\
\hline
\hline
tot		& 2487			& 2487			&\ 			& 10925		\\
\hline
1		& 360 ($14.5\%$)  	& 377 ($15.2\%$)	& $95.3\%	$	& 1324 ($12.1\%$)\\
2		& 500 ($20.1\%$) 	& 516 ($20.7\%$)	& $93.4\%$	& 1665 ($15.2\%$)\\
3		& 365 ($14.7\%$)  	& 338 ($13.6\%$)	& $92.1\%$	& 385 ($3.5\%$)\\
4		& 1262 ($50.7\%$)	& 1256 ($50.5\%$)	& $95.6\%$	& 7551 ($69.1\%$)\\
\hline
\end{tabular}
\label{tab:classification}
\end{table}
Due to the large number of sources entering the automatic catalogue we have implemented the use of a machine learning (ML) optimised classification. We have visually inspected only $\sim$1/4 of the 10925 sources, located in different regions of the galaxy. This visually inspected subsample has been used as a training set for the ML algorithm to classify the entire catalogue (details of the algorithm are discussed in a forthcoming paper by Grasha et al., in prep). 
The ML code is run on the entire sample of 10925 sources, including the already humanly--classified ones. In this way we can use the sources having a classification with both methods to estimate the goodness of the ML classification for M51. Table~\ref{tab:classification} gives the number of sources classified in each class by human and ML, as well as the comparison between the two classification (forth column). We recover a 95\% of agreement between the two different classifiers, within the areas used as training sets. To assess the goodness of the ML classification on the entire sample, we list in Table~\ref{tab:classification} between brackets the relative fraction of each class with respect to the total number of sources classified with different methods. We observe that the relative number of class 1 and 2 sources  with respect to the total number of sources (10925) classified by the ML approach is only a few per cent smaller than the fraction obtained with the control sample (2487 sources).  However, there is a striking difference in the recovery fractions of  class 3 and 4 sources. When considering the entire catalogue, the relative number of class 3 objects is much smaller (and on the contrary the one of class 4 is significantly more numerous). We consider very unlikely that there are so few class 3 objects in the total sample. So far the ML algorithm fails in recognising the most variate class of our classification scheme that contains sources with irregular morphology, multi-peaked, and some degree of colour gradient. From the absolute numbers of sources per class it is easy to conclude that the ML code is able to reclassify correctly almost all the class 3 objects given as a training sample, but is unable to recognize new class 3 sources, considering many of them as class 4 objects.
Future improvements for the classification will be produced by the use of different ML recognition algorithms. For our current analysis we will focus on the properties of class 1 and 2 cluster candidates and exclude class 3 sources, as explained in Sec \ref{sec:finalcat}.

\begin{figure}
\centering
\includegraphics[width=\columnwidth]{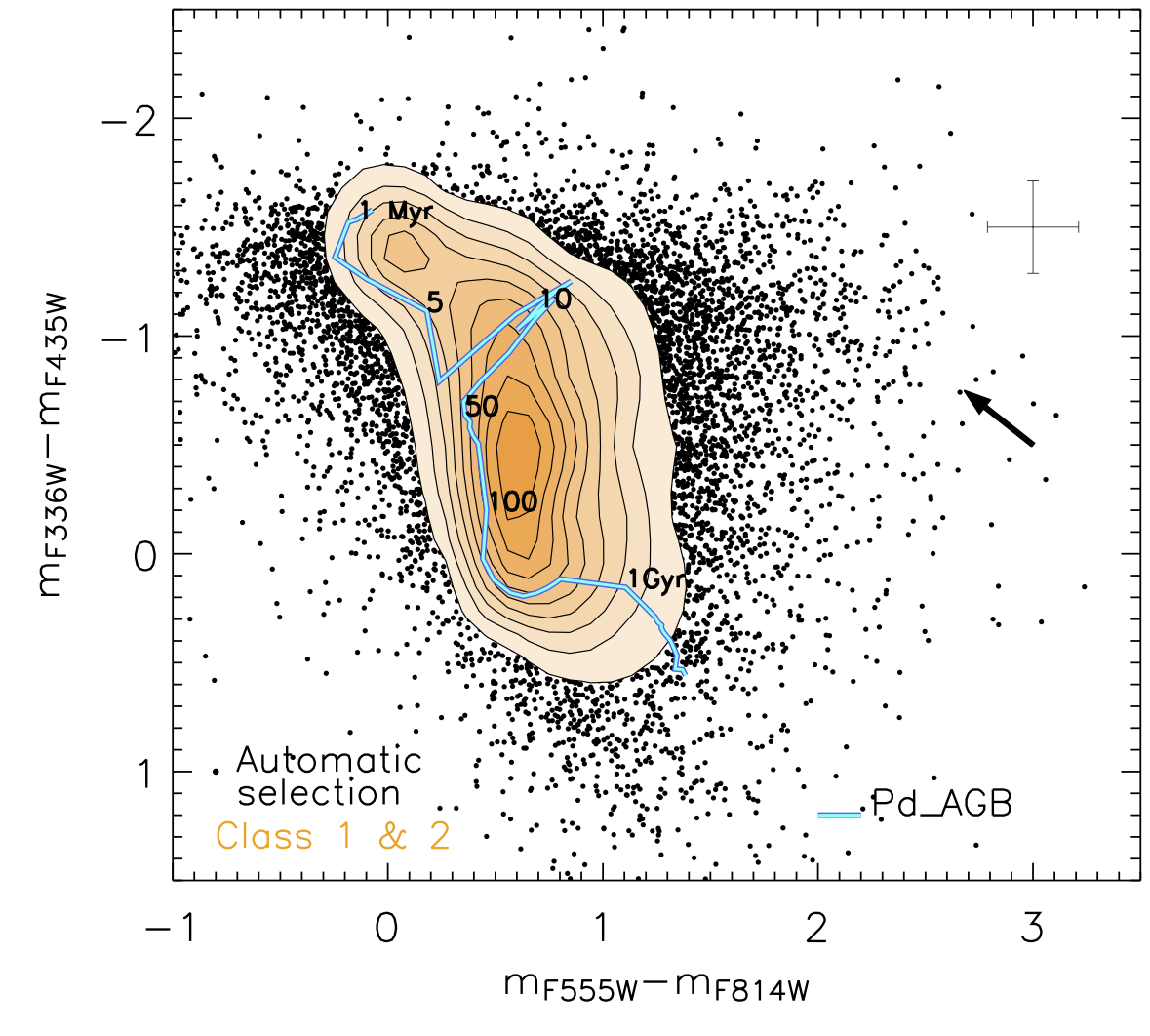}
\caption{Colour--colour diagram with $V-I$ on the x axis and $U-B$ on the y axis. Black points are all the sources in the catalogue that passed the automatic selection, while the orange shaded area gives the density of only the class 1 ad 2 sources. The SSP evolutionary track from Padova-AGB models is overplotted. The ages covered by the track goes from 1 Myr to 14 Gyr. The arrow indicates how an object would move if corrected for a reddening E($B-V$)=0.2. The cross at the top-right shows the average error in colours. }
\label{fig:ccd}
\end{figure}

We can summarise the photometric properties of the M51 cluster population using a two colour diagram (Fig~\ref{fig:ccd}).
Contours based on number densities of clusters show the regions occupied by the class 1 and class 2 M51 cluster population with respect to the location of the 10925 sources included by the automatic selection. A simple stellar population (SSP) track showing the cluster colour evolution as a function of age is also included. Sources are mainly situated along the tracks, implying the high quality of our morphological classification. Extinction spreads the observed colours towards the right side of the evolutionary tracks. Correcting for extinction, in the direction indicated by the black arrow, would move the sources back on the track, at the position corresponding to the best--fitted age.
The diagram shows a broad peak between 50 and 100 Myr. The contours are quite shallow towards younger ages and show a pronounced decline around $\sim1$ Gyr suggesting that most of the sources detected are younger than 1 Gyr. We use SED fitting analysis to derive cluster physical properties (see Section~\ref{sec:sed}), including the age distribution (Fig~\ref{fig:agemass}b).

\subsection{Completeness} 
\label{sec:completeness}

To investigate the completeness limit of the final catalogue, we use the custom pipeline available within the LEGUS collaboration as described in \citet{legus2}.
The pipeline follows closely the selection criteria adopted to produce the final automatic catalogues.
For each filter we produce frames containing simulated clusters of different luminosities and sizes which are added to the original scientific frames. Effective radii ($R_\textrm{eff}$) between 1 and 5 pc have been used, as studies of cluster sizes in similar galaxies suggest that most of the sources fall in this range \citep{ryon2015,ryon2017}. The synthetic clusters span an apparent magnitude range between 19 and 26 mag and are created using the software \texttt{BAOlab}, freely available online\footnote{\hyperref[http://baolab.astroduo.org/]{http://baolab.astroduo.org/}}. All clusters are simulated as symmetric sources with a MOFFAT15 profile \citep[see][]{baolab} considered a standard assumption for the YSC light profiles \citep{eff1987}.
Cluster extraction (via \texttt{SExtractor}) and  photometry are repeated on the resulting coadded scientific and synthetic frames. 
A signal of 10$\sigma$ in at least 10 contiguous pixel is requested to extract sources in the $B$, $V$, and $I$ bands, while a value of 5$\sigma$ over at least 10 contiguous pixels is chosen for the $UV$ and $U$ bands.
The recovery rate of sources as a function of luminosity yields the completeness.
The magnitude limits above which the relative number of the recovered sources falls below 90\% level is summarized in Tab.~\ref{tab:completeness} for each filter. 
\begin{table}
\centering
\caption{Completeness limits in each filter. The second and third column give the 90\% completeness limit calculated with the completeness test code (described in the text). The completeness was computed in the disk (area outside a 35'' radius, second column) and in the central region (area inside 35'' radius, third column). The last column gives the peak magnitude of the luminosity distribution, as plotted in Fig~\ref{fig:checkmaglimits}.}
\begin{tabular}{lccc}
\hline
Filter		& compl. (disk) & compl. (centre)	& Lum$_\textrm{peak}$	\\
\		& [mag] 		& [mag]			& [mag]	\\
\hline
\hline
F275W	& 22.17 	& 21.75 	& 21.75	\\
F336W	& 22.75	& 21.79	& 22.00	\\
F435W	& 24.17 	& 22.65 	& 23.25	\\
F555W	& 23.70	& 22.31	& 23.25	\\
F814W	& 22.70	& 21.61	& 22.25	\\
\hline
\end{tabular}
\label{tab:completeness}
\end{table}

\begin{figure*}
 \centering
 \includegraphics[width=\textwidth]{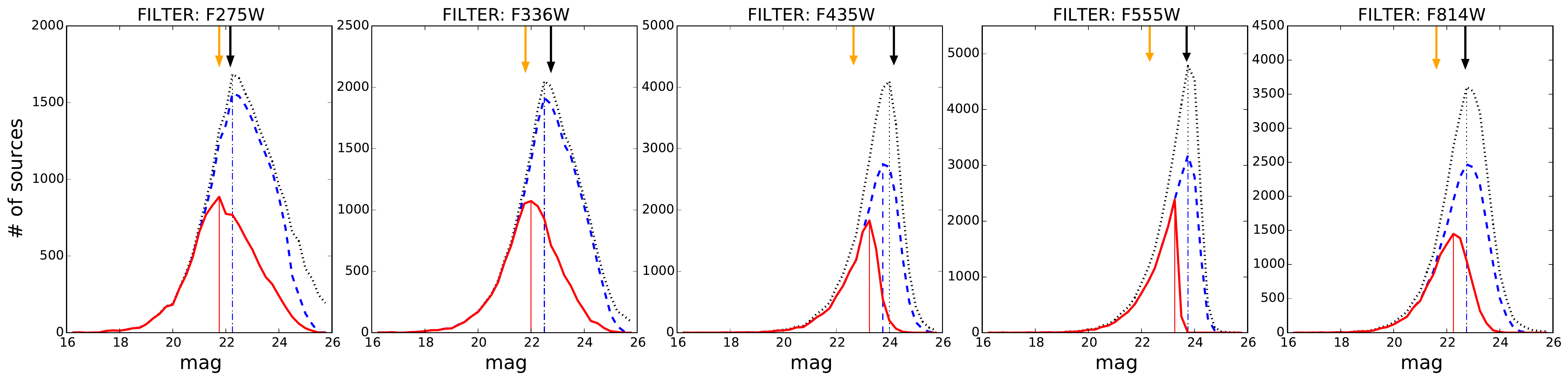}
 \caption{Luminosity distributions showing how the completeness changes when different cuts are applied in the selection process. Each distribution represents the catalogue at a different stage:  the black dotted line shows the distribution of all the sources with photometric errors smaller than 0.3 mag, the blue dashed line represents the sources left after requiring that they have a detection in at least 4 filters and the red solid line is the distribution of the sources after the -6 mag cut in the $V$ band. Vertical lines are plotted corresponding to the peak of the distribution. The peak values after the magnitude cut is applied in $V$ band are listed in Table~\ref{tab:completeness}. The sources are grouped in bins of 0.25 mag width. The black (orange) arrows show the value of the 90\% completeness limit evaluated with the completeness test code in the disk (centre) of the galaxy (see Tab.~\ref{tab:completeness}).
 }
\label{fig:checkmaglimits}
\end{figure*}

The completeness test code only gives a completeness limit estimated for each filter independently. These values should be considered as completeness limits resulting from the depth of the data. However, our cluster candidate catalogue is the result of several selection criteria that cross correlate the detection of sources among the 5 LEGUS bands. This effect can be visualized in Fig~\ref{fig:checkmaglimits}, where we show the recovered luminosity distributions of sources at different stages of the data reduction. The requirement of detecting the sources in 4 filters with a photometric error smaller than 0.30 mag diminishes the number of recovered objects, mainly due to the smaller area covered by UVIS compared to ACS.
The cut at $M_V = -6$ mag not only excludes the sources which are faint in $V$ band, but also modifies the luminosity distributions in the other filters (see differences between red solid line distributions and blue dashed ones). Both these conditions affect the resulting catalogue and modify the shape of the luminosity distributions in each filter in a complicated way at the faint limits, therefore modifying also the completeness limits with respect to our approach of treating each filter independently. 
We use the observed luminosity distributions to understand how the completeness limits change as a function of waveband and adopted selection criteria. 
We observe in each filter an increase going from bright to low luminosities and we know that incompleteness starts to affect the catalogue where we see the luminosity distribution turning over (see e.g. \citealp{larsen2002}). 

We draw the following conclusions from the analysis conducted in Fig~\ref{fig:checkmaglimits}. First, the $-6$ mag cut in the $V$ band strongly reduces the number of selected sources in all the bands. At the distance of M51, this brightness corresponds to an apparent magnitude in $V$ of 23.4 mag, thus, it is brighter than the 90\% completeness limit recovered in the $V$ band (23.84 mag, see Tab.~\ref{tab:completeness}). Secondly, the 90\% completeness limits fall rightwards of the peak in the luminosity distributions with the only exception for the F275W filter. For this reason we prefer to apply a more conservative approach and use the peak of the luminosity distribution as a limit for the cluster luminosity function analysis. Only in the case of F275W the 90\% completeness limit is brighter than the peak magnitude, therefore the latter is adopted as the completeness limit value.
We stress that the part of the luminosity distribution leftwards of the peak remains almost identical after the selection cut (check Fig~\ref{fig:checkmaglimits}) suggesting that the distributions are not affected by our selection criteria and completeness recovery at magnitudes brighter than the peak of the distributions.

We also tested completeness variations on sub-galactic scales. Our analysis shows that the completeness is worse towards the centre of the galaxy. Outside the inner region (radius larger than 35''), the $V$ band 90\% completeness value is fainter than the cut at 23.4 mag (see Tab.~\ref{tab:completeness} and Fig.~\ref{fig:checkmaglimits}). Similar results are observed in the other filters. Because of the completeness drop at radii smaller than 35'' (1.3 kpc), we have excluded from our analysis this inner region.
 
\subsection{SED fitting} 
\label{sec:sed}
Sources with detection in at least four filters were analyzed via SED fitting algorithms in order to derive physical properties of the clusters. We use uniformly sampled IMF to derive SSP models that include a treatment for nebular continuum and emission lines as described in \citet{legus2}.  
Putting together the different aperture correction methods, different stellar libraries and different extinction curves, 12 final catalogues are produced with the deterministic fitting method (and will be available online on the LEGUS website\footnote{\hyperref[https://archive.stsci.edu/prepds/legus/]{https://archive.stsci.edu/prepds/legus/}}). All catalogues use models with solar metallicity for both stars and gas and an average gas covering factor of $50\%$.

The analyses and results presented hereafter are obtained using the final catalogue with:
\begin{itemize}
\item photometry corrected by average aperture corrections;
\item Padova evolutionary tracks produced with solar metallicity stellar libraries. 
\item Milky Way extinction curve \citep{cardelli1989}.
\item  \citet{kroupa2001} stellar initial mass function.
\end{itemize}

\section{Global Properties of the cluster sample}
\label{sec:global}
\subsection{Final Catalogue} 
\label{sec:finalcat}
\begin{figure*}
 \centering
 \subfigure[]
   {\includegraphics[width=0.49\textwidth]{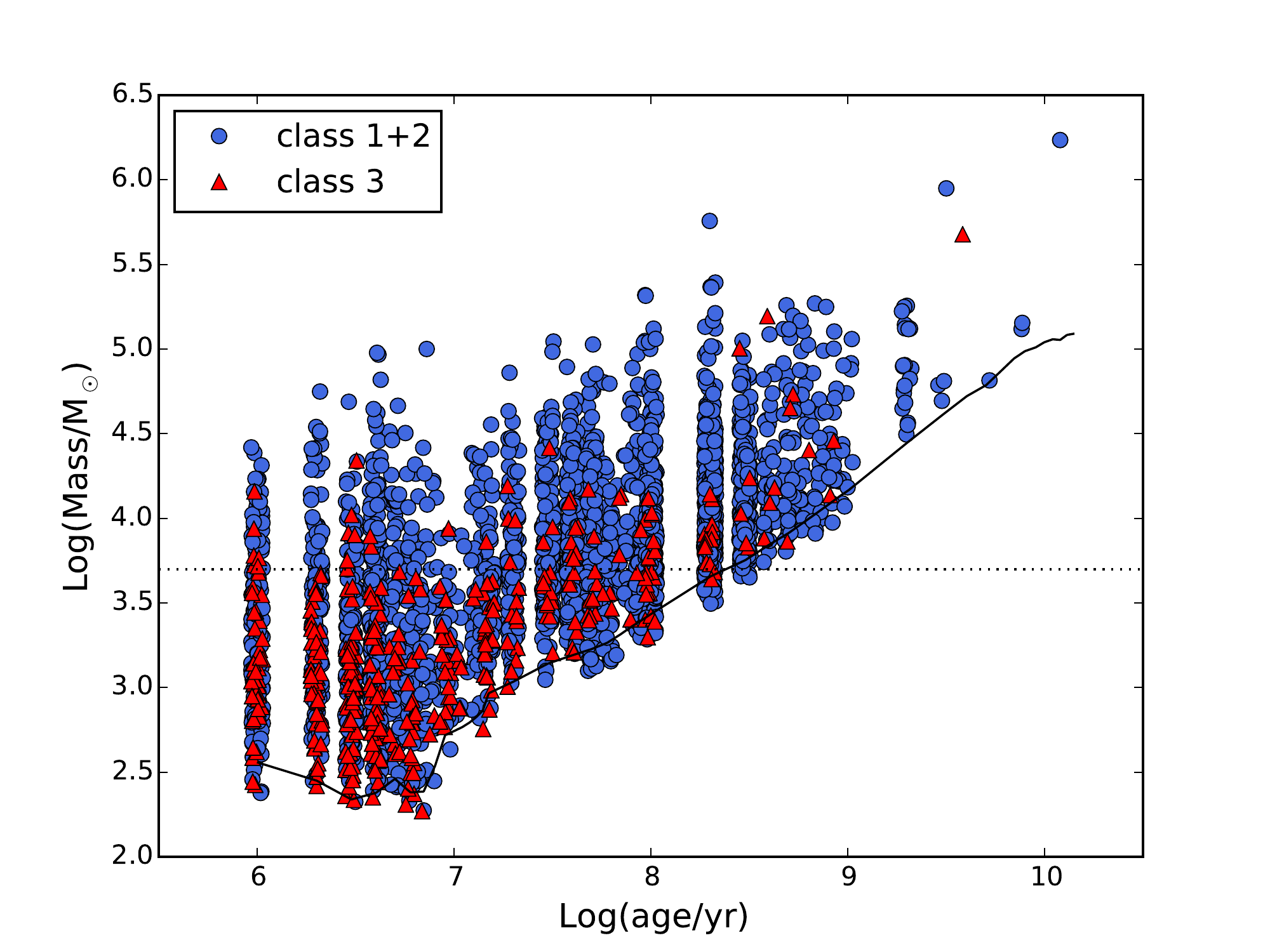}}
 \hspace{0mm}
 \subfigure[]
   {\includegraphics[width=0.49\textwidth]{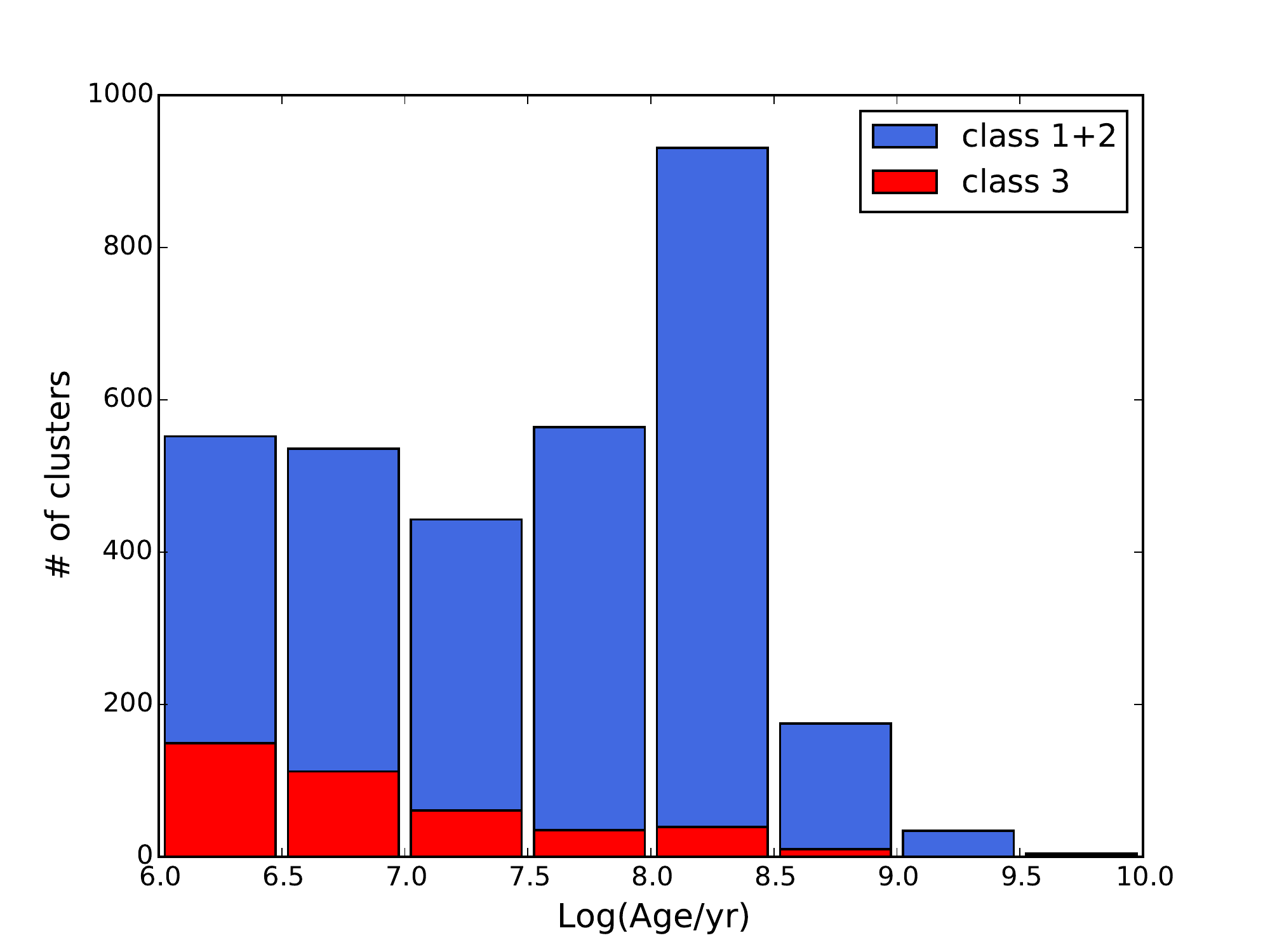}}
 \hspace{0mm}
 \caption{(a) Age-mass diagram for the sources in class 1 and 2 (blue circles) and class 3 (red triangles). The solid black line represents the mass limit as function of ages estimated from the evolutionary tracks assuming a completeness limit of 23.4 mag in V band. The dotted horizontal line shows a mass of 5000 \msun. The sources in each age bin were furnished with a small amount of artificial scatter around their respective bin to make the plot easier to read. (b) Histogram of the age distribution of the class 1+2 (blue) and class 3 (red) sources: the total height of each bin gives the total number of sources of the classes 1, 2 and 3. The number of class 3 sources drops fast in the first 10 Myr, such that in the range Log(age) = 7 - 8.5 their number is very small even if the age spanned is more that 100 Myr.}
\label{fig:agemass}
\end{figure*}

In Fig.~\ref{fig:agemass} we show the ages and masses of the sources classified as class 1, 2 and 3. In the same plot, the completeness limit of 23.4 mag in the $V$ band discussed in the previous section is converted into an estimate of the minimum mass detectable for each age. The line representing this limit follows quite accurately the detected sources with minimum mass at all ages. 
Uncertainties on the age and mass values are on average within 0.2 dex. Uncertainties can reach 0.3 dex close to the red supergiant phase (visible as a loop at ages $\sim10-50$ in Fig.~\ref{fig:ccd}).
In order to study only the cluster population of the grand design spiral and avoid the clusters of NGC 5195, we have neglected the clusters with y coordinate bigger than 11600 (in the pixel coordinates of the LEGUS mosaic) both from Fig.~\ref{fig:agemass} and from the rest of the analysis. This cut is similar to removing the UVIS pointing centred on NGC 5195. 

The different classes are not distributed in the same regions of the plot, with class 3 sources having on average smaller masses and younger ages. Previous studies \citep{grasha2015,ryon2017,legus2} have shown that our morphological cluster classification is with good approximation also a dynamical classification. 
Compact associations (i.e. class 3) in NGC 628 are on average younger and less massive than class 1 and 2 clusters \citep{grasha2015}. In addition, the age distributions of class 3 systems suggest that they are more easily and quickly disrupted (see \citealp{legus2} for details), probably because they are not bound. \citet{grasha2015,grasha2017}, focusing on the clustering function of clusters in the LEGUS galaxies, have also shown that class 3 clusters behave differently than class 1 and class 2 clusters. These results contribute to the idea that the morphological classification chosen has also some dynamical implication: class 3 sources seem to be short-lived systems (see Fig.~\ref{fig:agemass}b), possibly already unbound at the time of formation. 

For the remainder of this work we will only analyse class 1 and 2 objects, which we consider to be the candidate stellar clusters, i.e. the  gravitationally bound stellar systems that form the cluster population of M51. In total we have 2834 systems classified as class 1 and 2 out of the 10925 sources with 3 sigma detection in at least 4 LEGUS bands. 

\begin{figure}
 \centering
 \includegraphics[width=\columnwidth]{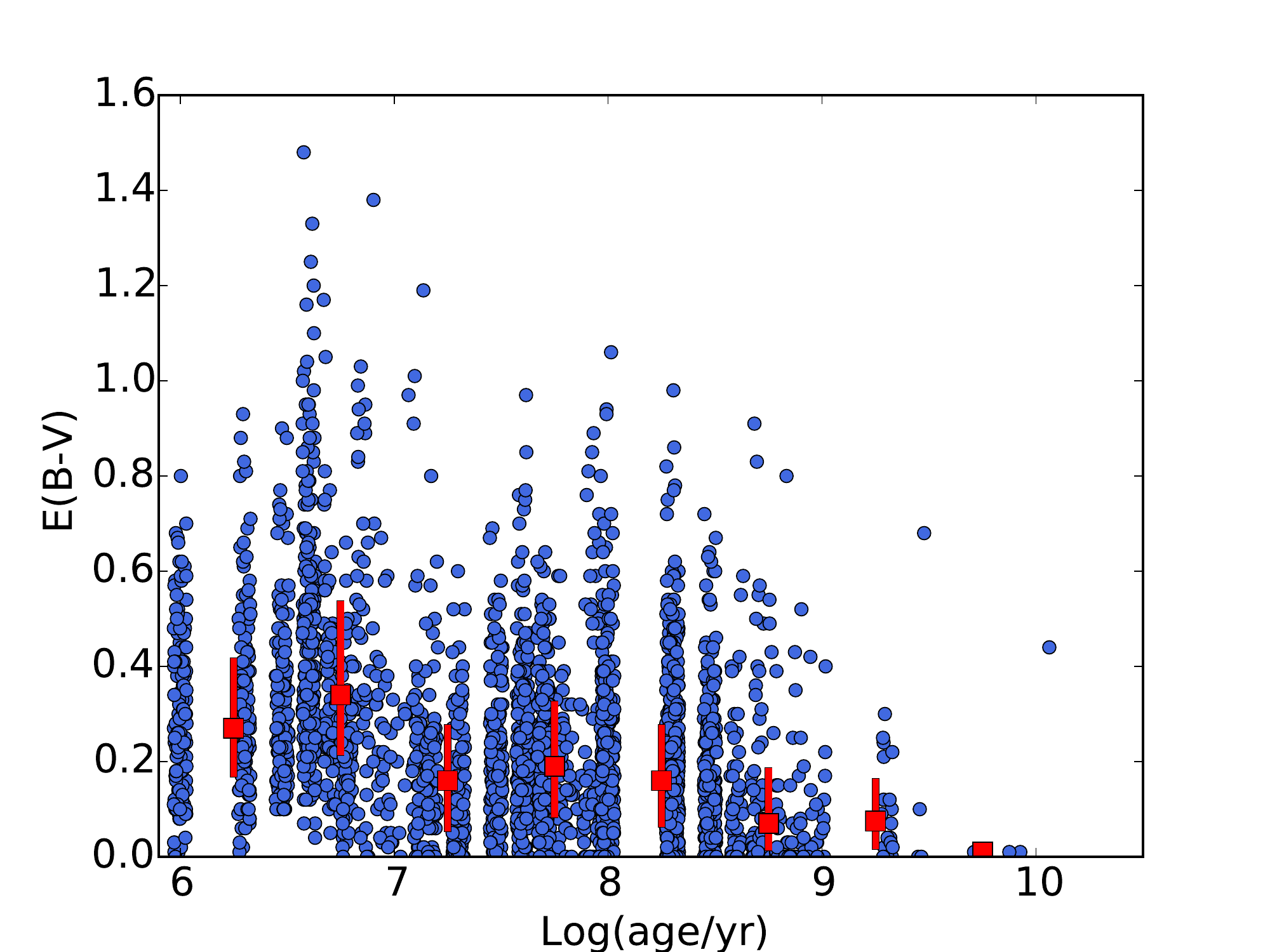}
\caption{Age-extinction diagram for the class 1 and 2 sources in the catalogue (blue circles). The red squares are the median values of E(B-V) in age bins of 0.5 dex. The red lines marks the range between the first and the third quartiles in the same age bins. }
\label{fig:ebv}
\end{figure}

In Fig~\ref{fig:ebv} we show how the recovered extinction changes as a function of cluster age. We see that on average the internal extinction of YSCs changes from $E(B-V)\sim$0.4 mag at very young ages to 0.2 mag for clusters older than 10 Myr, and is even lower ($E(B-V)\sim$0.1 mag) for clusters older than 100 Myr. We observe also a large scatter at each age bin, suggesting that the extinction may not only be related to the evolutionary phase of the clusters but also to the region where the cluster is located within the galaxy. 

\subsection{Comparison with previous catalogues} 
\label{sec:comparison}
Numerous studies of the cluster population in M51 are available in the literature. We compare our catalogue with recently published ones, testing both our cluster selection and agreement in age and mass estimates. These comparisons will be used when we compare the results of our analyses to values reported in the literature.
Among the works that studied the entire galaxy, \citet{scheepmaker2007}, \citet{hwang_lee2008} and \citet{chandar16} used the same $BVI$ data as our work. However, we decided to focus our comparison only on two catalogues for which we have access to estimates of ages and masses.
The first catalogue is the one compiled by \citet{chandar16} (hereafter CH16), using the same $BVI$ observations, plus F658N (\ha) filter observations from the same program (GO-10452, PI: S. Beckwith) and WFPC2/F336W filter ($U$ band) observations (from GO-10501, PI: R.Chandar, GO-5652, PI: R.Kirshner and GO-7375, PI: N.Scoville). The cluster candidates catalogue used in their analysis includes 3812 sources in total (of which 2771 lies in the same area covered by our UVIS observations) and has been made publicly available \citep{chandar16cat}. 
The second catalogue is taken from \citet{bastian2005} (hereafter BA05), covers only the central part of the galaxy and is mainly based on HST observations with the WFPC2 camera. It contains 1150 clusters, 1130 of which are in an area in common with our UVIS pointings.
These two catalogues are very different both in terms of coverage and instruments used. For this reason we compare the catalogue produced with LEGUS with each of them separately.

\subsubsection{Cluster Selection}
\begin{figure*}
 \centering
\subfigure{\includegraphics[width=\textwidth]{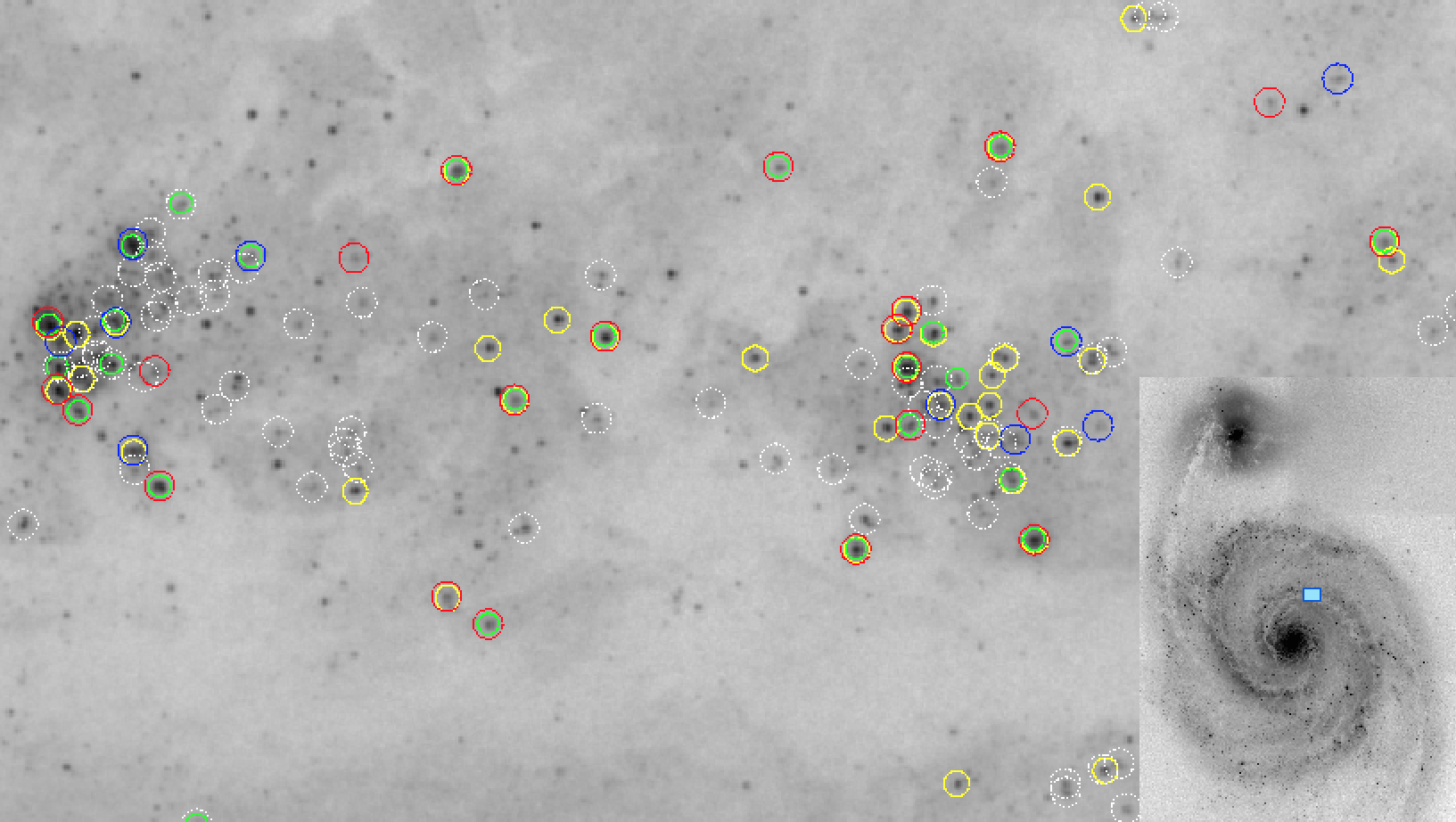}}
\subfigure{\includegraphics[width=\textwidth]{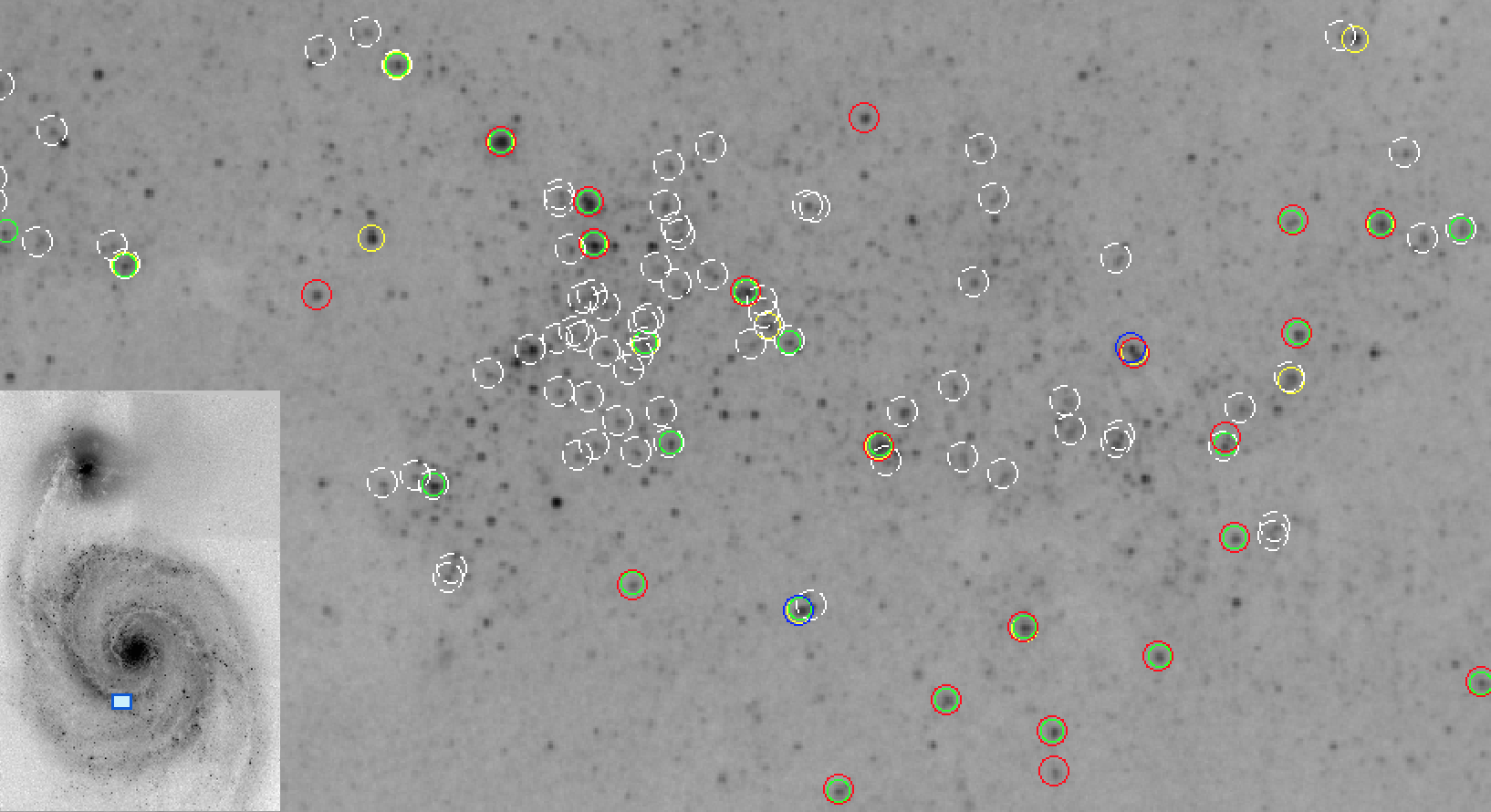}}
 \caption{Comparison between different cluster catalogues in a spiral arm in the galaxy covered by human visual classification (top) and by machine learning classification (bottom) in LEGUS. Green circles are the clusters candidates in \citet{chandar16} catalogue. Yellow circles are the cluster candidates in \citet{bastian2005} catalogue. Red circles are cluster candidates of class 1 and 2 in LEGUS, blue circles are class 3 sources in LEGUS and white dashed circles are sources which have been assigned class 4 in LEGUS. In the bottom right corner an inset shows the position of the zoomed region inside M51.}
\label{fig:comparison}
\end{figure*}

We first compare the fraction of clusters candidates in common between the catalogs. When doing so, we include class 3 sources in the LEGUS sample, as this class of sources is considered in BA05 and CH16 catalogues. Tab.~\ref{tab:comparison} collects the results of the comparison. Among the 2771 candidates detected in the same area of the galaxy by CH16 and LEGUS, 1619 (60\%) systems are in common. We have repeated the comparison in the regions covered by human visual classification in LEGUS, finding a better agreement ($\sim75\%$). We take this last value as reference for the common fraction of candidates and justify the drop observed when considering the entire catalogue as given by the ML misplacing class 3 objects into class 4 (as discussed in Section~\ref{sec:classification}).
Fig.~\ref{fig:comparison} shows a blow-up of the galaxy with the cluster positions of both catalogues. We selected two different regions, one where the sources have been classified via visual classification and one where only ML is available. We notice that some of the CH16 candidates which do not appear in LEGUS catalogue of classes 1, 2 and 3, have been assigned a class 4. This is true for both regions. Those sources were extracted by the LEGUS analysis but were discarded based on their morphological appearance. The differences between the two catalogues are therefore mostly due to source classification and not by the extraction process.

The comparison with the BA05 catalogue indicates a poorer agreement, with less than $40\%$ of sources in common. This discrepancy, observable in Fig.~\ref{fig:comparison}, may be caused by the difference in the data and in the approach used to extract the clusters. BA05 analysis is based on WFPC2 data, whose resolution is a factor of $\sim$2 worse than ACS. In addition they do not apply any CI cut, increasing the contamination from stars.

\begin{table*}
\centering
\caption{Comparison between our cluster catalogue and the catalogues by \citet{chandar16} (CH16) and \citet{bastian2005} (BA05). The number of clusters reported in the columns are: (1) number of cluster candidates detected in the CH16 and B05 catalogues within the same region covered by LEGUS; (2) number of cluster candidates in common between CH16, B05 and LEGUS (class 1,2,3), respectively; (3) number of cluster candidates in the Ch16 and B05 catalogues restricted to the area of the galaxy that has been visually inspected by humans; (4) number of cluster candidates of CH16 and B05 in common with LEGUS class 1, 2 and 3 sources that have been classified by human.
$^1$ These fractions would increase to 95\% if we include cluster candidates classified as 4 within the LEGUS catalogue. $^2$ This fraction would increase to 71\% if class 4 cluster candidates are included. $^3$ The agreement would increase to 67\% if class 4 objects are considered. 
The drop in the common fraction of cluster candidates from column (4) to column (2) is likely due to the ML classification misplacing class 3 sources into class 4 (see Section~\ref{sec:classification}). 
The low fraction of clusters in common with BA05 is most likely due to a different approach for the cluster extraction analysis and low resolution imaging data. BA05 is based on WFPC2 data (a factor of 2.5 worse spatial resolution with respect to WFC3) and does not include a CI cut. This causes a higher contaminations from unresolved sources, which in our analysis were excluded by the CI selection. 
}
\begin{tabular}{lcccc}
\hline
Catalogue			&	\# clusters	&  \# clusters		&  \# clusters		& \# clusters		\\
\				& 	(1) 		& (2)& (3) & (4)\\
\hline
\hline
CH16						& 2711			& 1619 (60\%)$^1$			& 732			& 535 (73\%)$^1$		\\
LEGUS (1,2,3)$_{CH16\ area}$	 	& 3240			& $-$					& 1294			& $-$			\\
\hline
BA05						& 1130			& 388 (35\%)$^2$			& 214			& 83 (39\%)$^3$		\\
LEGUS (1,2,3)$_{BA05\ area}$ 	& 1238			& $-$					& 267			& $-$			\\
\hline
\end{tabular}
\label{tab:comparison}
\end{table*}

\subsubsection{Comparison of Ages and Masses}
\begin{figure*}
 \centering
 \subfigure{\includegraphics[width=0.43\textwidth]{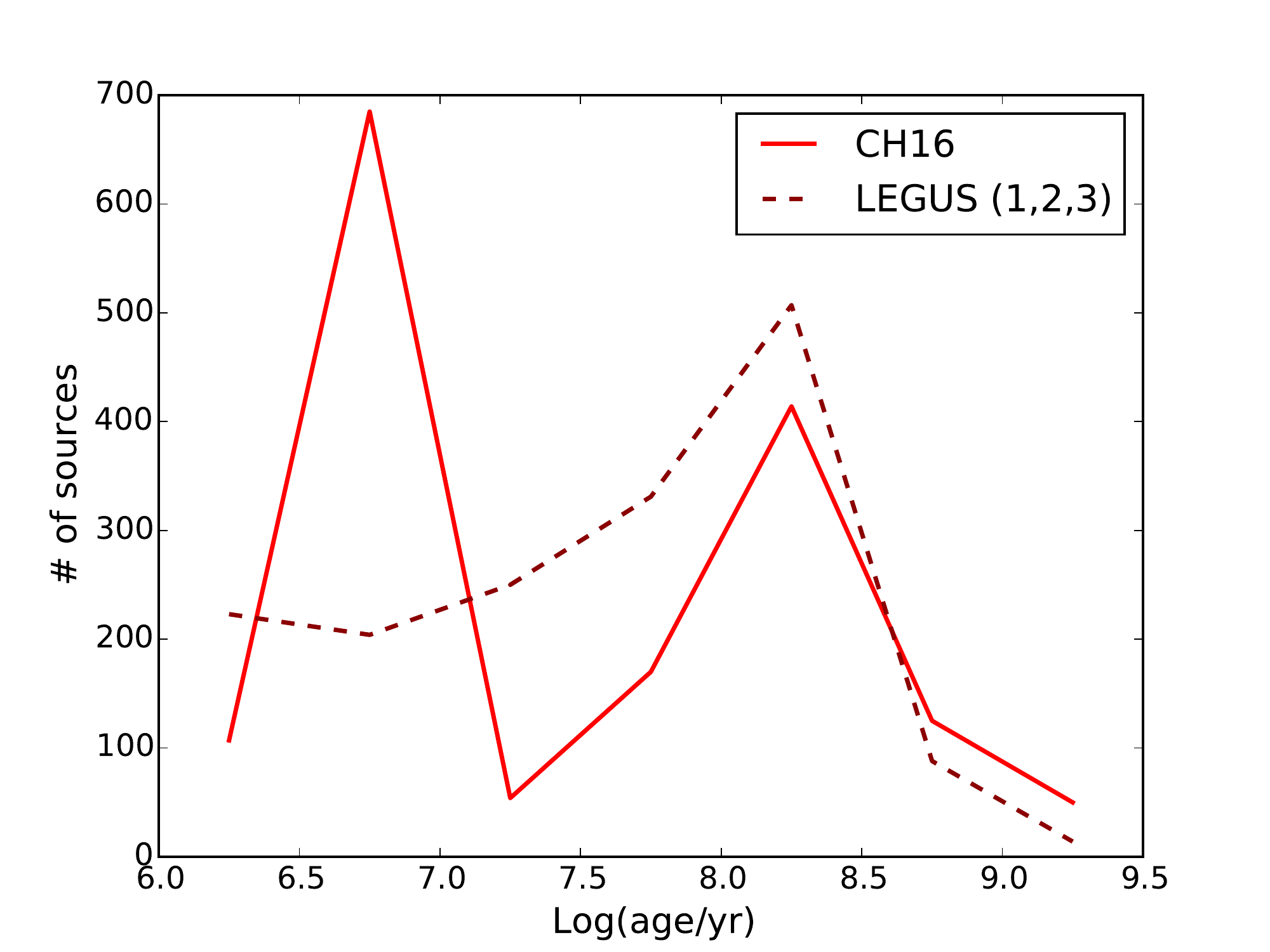}}
 \subfigure{\includegraphics[width=0.43\textwidth]{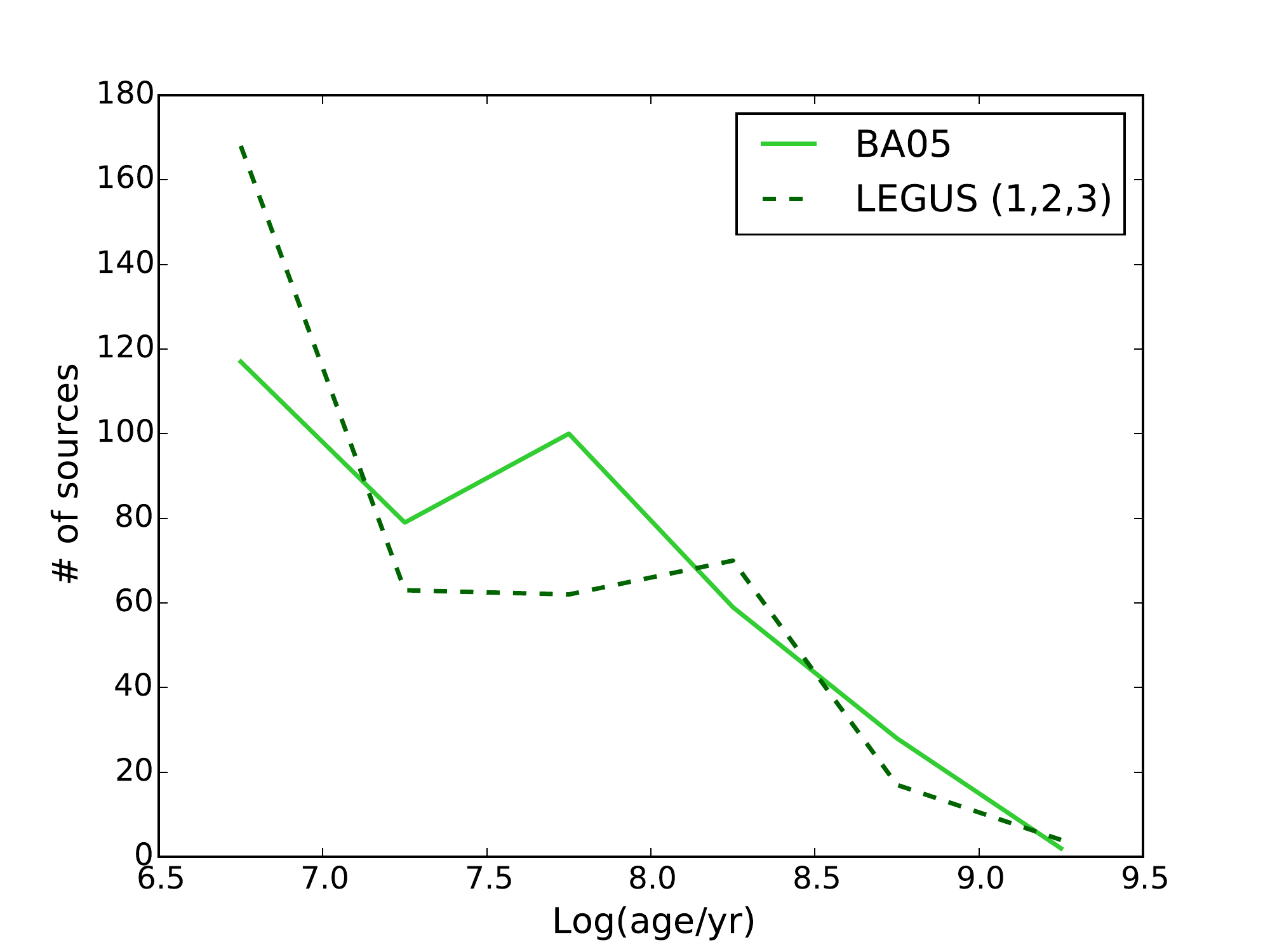}}
 \subfigure{\includegraphics[width=0.43\textwidth]{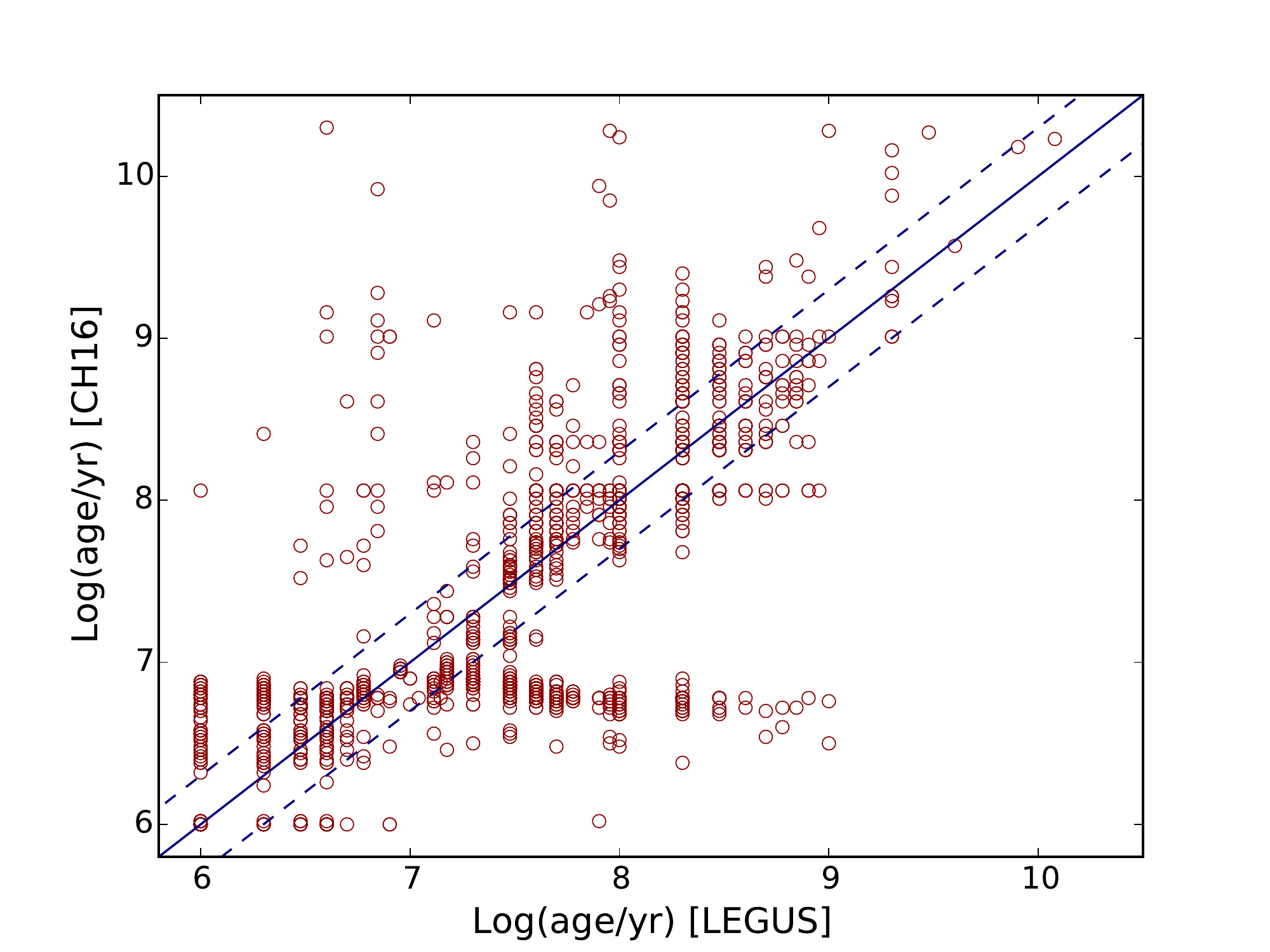}}
  \subfigure{\includegraphics[width=0.43\textwidth]{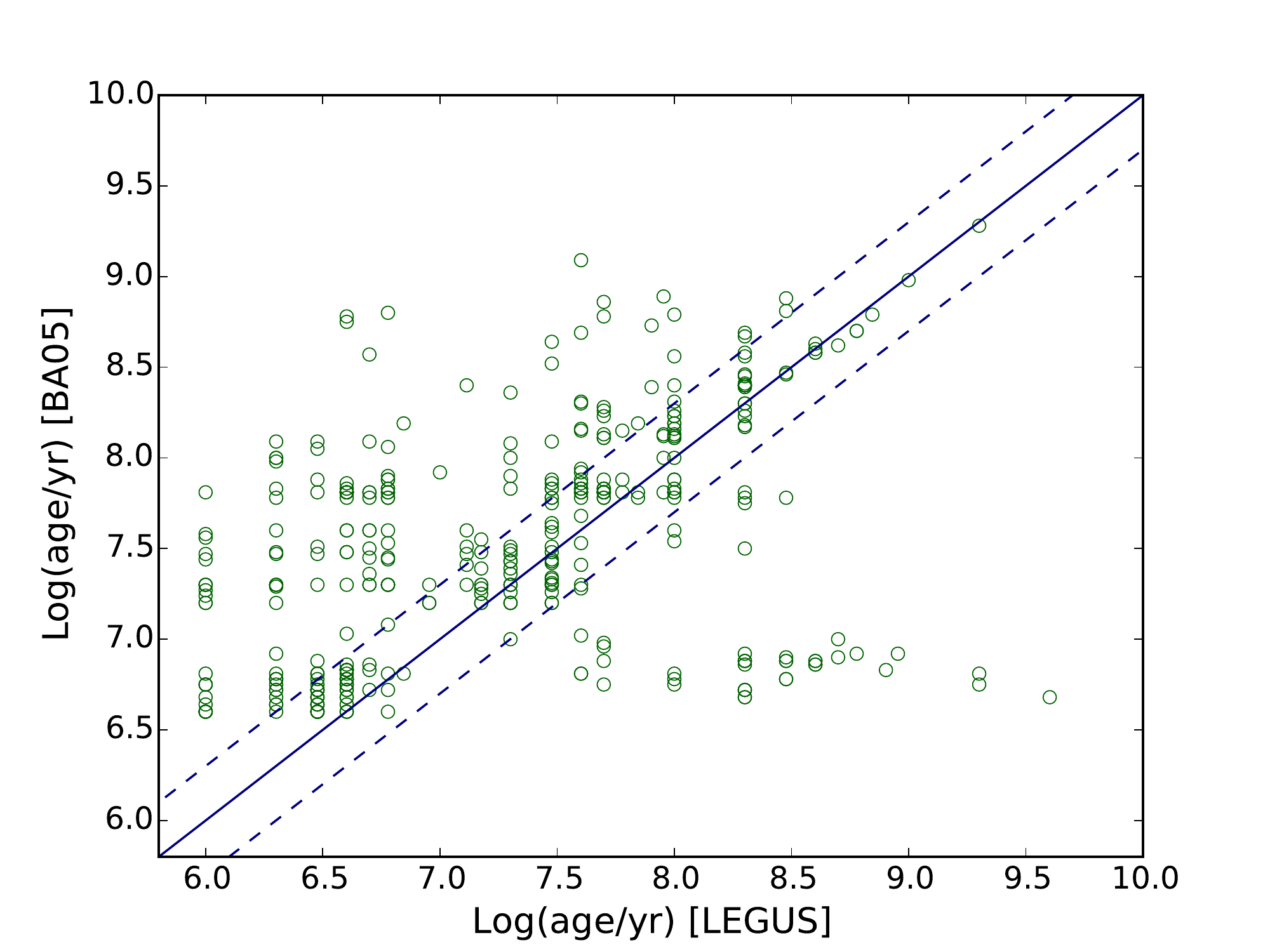}}
 \subfigure{\includegraphics[width=0.43\textwidth]{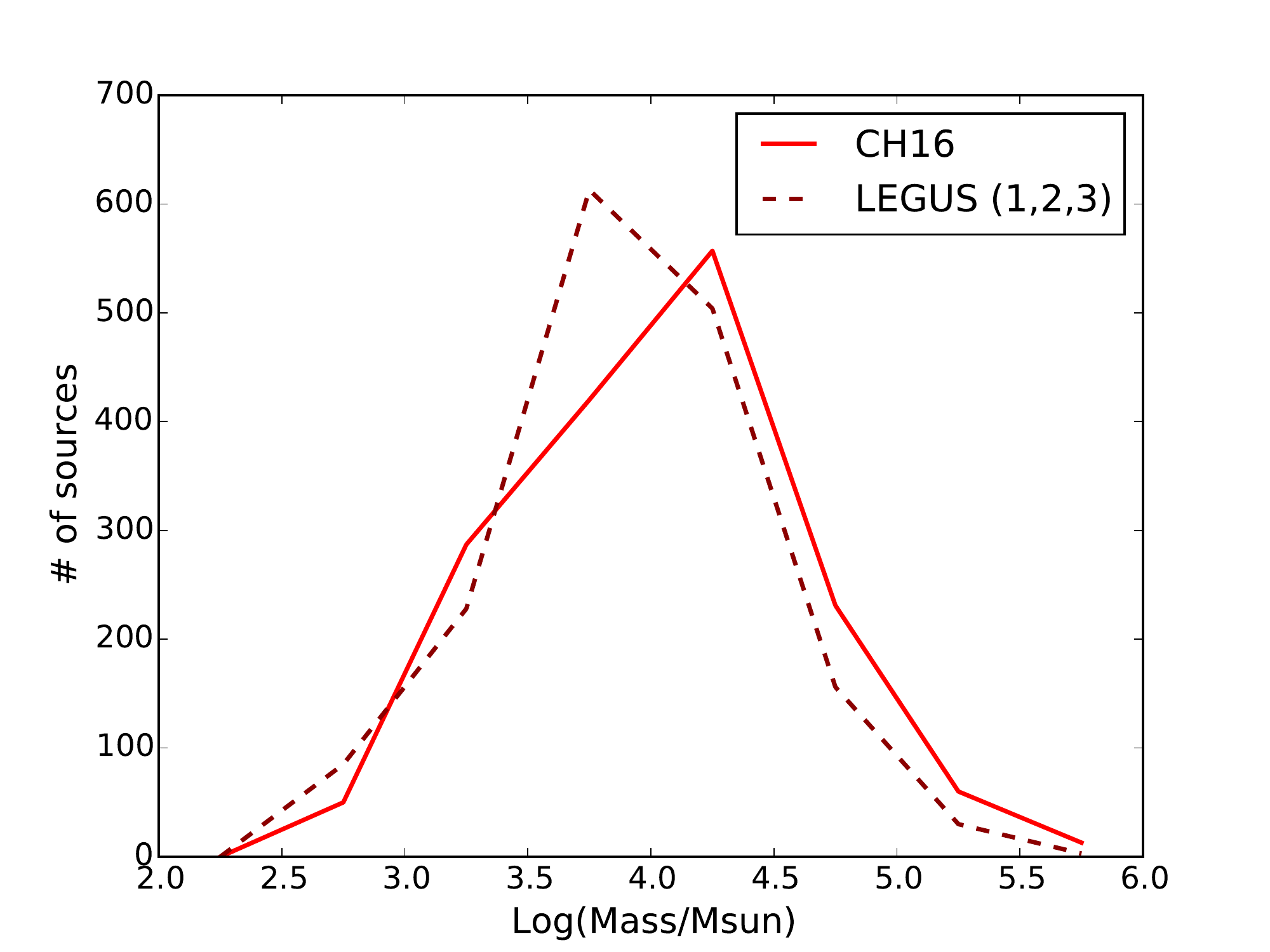}}
 \subfigure{\includegraphics[width=0.43\textwidth]{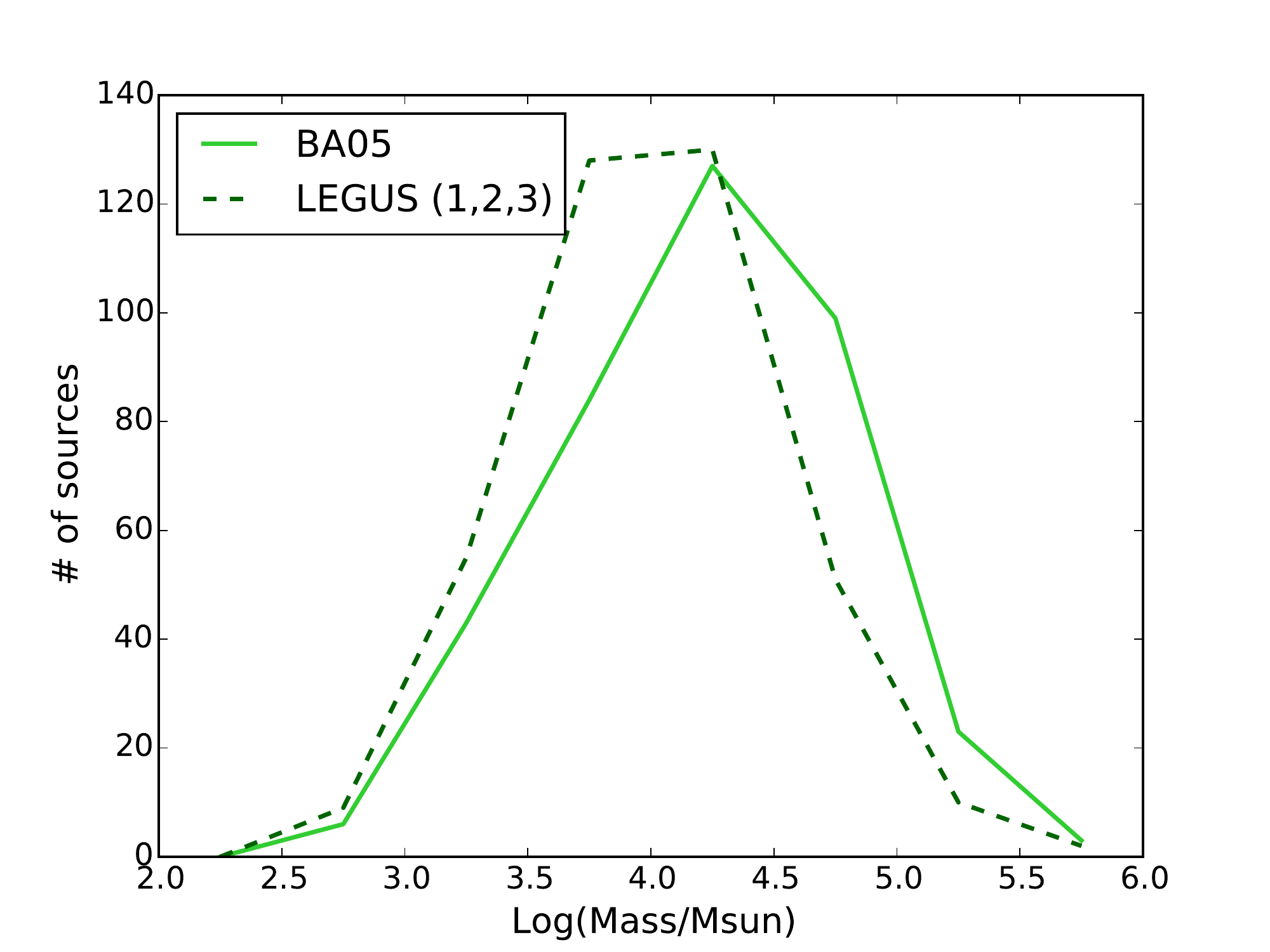}}
 \caption{Comparison of ages and masses retrieved from the broadband SED fitting in the LEGUS analysis and in the works of \citealp{chandar16} (CH16, red, left column) and \citet{bastian2005} (BA05, green, right column). Only the sources in common between LEGUS and CH16 or BA05 are plotted. First row: ages distributions. Second row: 1--to--1 comparison between ages. The 1--to--1 agreement line is shown in solid blue, the 0.3 dex scatter lines in dashed blue. Third row: masses distributions.}
\label{fig:comparison_am}
\end{figure*}
The comparison of age distributions for the sources in common between LEGUS and CH16 is plotted in Fig.~\ref{fig:comparison_am} (top left). The age distribution of CH16 has a strong peak for sources younger than log(age/yr)$=7$ and a subsequent drop in the range $7-7.5$, both of which are not observed in our catalogue. The one--to--one comparison between age estimates in Fig.~\ref{fig:comparison_am} (middle left) shows a large fraction of clusters with young ages in CH16 which have a wide age spread in the LEGUS catalogue. More in general, the differences in the age estimates are mostly caused by the different broadband combinations used in fitting the data, as already noticed in CH16. In addition to the ``standard'' $UBVI$ filter set used for SED fitting of both CH16 and our LEGUS catalogues, we use an extra $UV$ broadband while CH16 use the flux of the narrow band filter centred on the H$\alpha$ emission line, from an aperture of the same size as the broadband ones. Both approaches aim at breaking the age-extinction degeneracy weighting different information. The LEGUS standard approach is to use two datapoints below the Balmer break ($\lambda < 4000$) which give a stronger constraint on the slope of the spectrum, and thus, extinction. The approach used by CH16 is to use the detection of H$\alpha$ emission from gas ionised by massive stars to determine the presence of a very young stellar population in the cluster. From Fig.~\ref{fig:comparison_am} (middle left) we observe that the two methods agree within 0.3 dex in $\sim50$\% of the cases. 
The correlation between the ages derived in the two methods is confirmed by a Spearman's rank correlation coefficient $r_s=0.7$ with a p-value: $10^{-234}$.
We will address in a future work (Chandar et al. in prep.) the systematics and differences in the two methods. In this work we will take into account the differences observed in the age distributions when discussing and comparing our results to those available in the literature.

The mass distributions (Fig.~\ref{fig:comparison_am}, bottom left) show a more similar behaviour, with a broad distribution and a decrease at low masses caused by incompleteness. Note that CH16 retrieve higher mass values at the high-mass end of the distribution. This difference can be important in the study of the mass function shape (Section~\ref{sec:massfunc}).

The comparison of age and mass distributions for the sources in common between LEGUS and BA05 is shown in the right column of Fig.~\ref{fig:comparison_am}.
Since the youngest age assigned by BA05 is log(age/yr)=6.6, for the sake of the comparison in Fig.~\ref{fig:comparison} (top right) we have assigned log(age/yr)=6.6 age to all the sources that in our catalogue are younger. The general trends of the distributions looks similar, but the one--to--one comparison in Fig.~\ref{fig:comparison} (middle right) reveals that the two methods agree within 0.3 dex in $\sim50\%$ of the cases. 
The correlation found with a Spearman's rank test is $r_s=0.5$. A p-value of $10^{-24}$ confirms that this correlation is not random, but the moderate value of $r_s$ is caused by the difference in the age distribution observed in Fig.~\ref{fig:comparison} (middle right). 
BA05 use very different data from our own and allow fits with $BVI$ bands only, with large uncertainties on the recovered properties. For example, the large cloud of systems that sit in the upper left part of the plot has been assigned younger ages in our catalogue. Also in this case we can conclude that the differences in age determinations are mostly caused by the different fitting approach, with our catalogue having more information to break the age-extinction degeneracy. The mass distributions (Fig.~\ref{fig:comparison_am}, bottom right) show the same overall shape, with the BA05 distribution shifted by 0.2 dex to higher values of masses.

In general, for both catalogues, we notice that differences in the derived properties can be also caused by differences in the stellar templates adopted, which are different for all catalogues (CH16 uses \citealp{bruzual_charlot2003} models, while BA05 uses updated GALEV simple stellar population models from \citealp{schulz2002} and \citealp{anders2003}). We will use the differences outlined among these previously  published catalogues and ours when we will discuss the results of our analyses.

\subsection{Cluster Position as a function of age}
\label{sec:position}
\begin{figure*}
 \centering
 \subfigure
   {\includegraphics[width=0.4\textwidth]{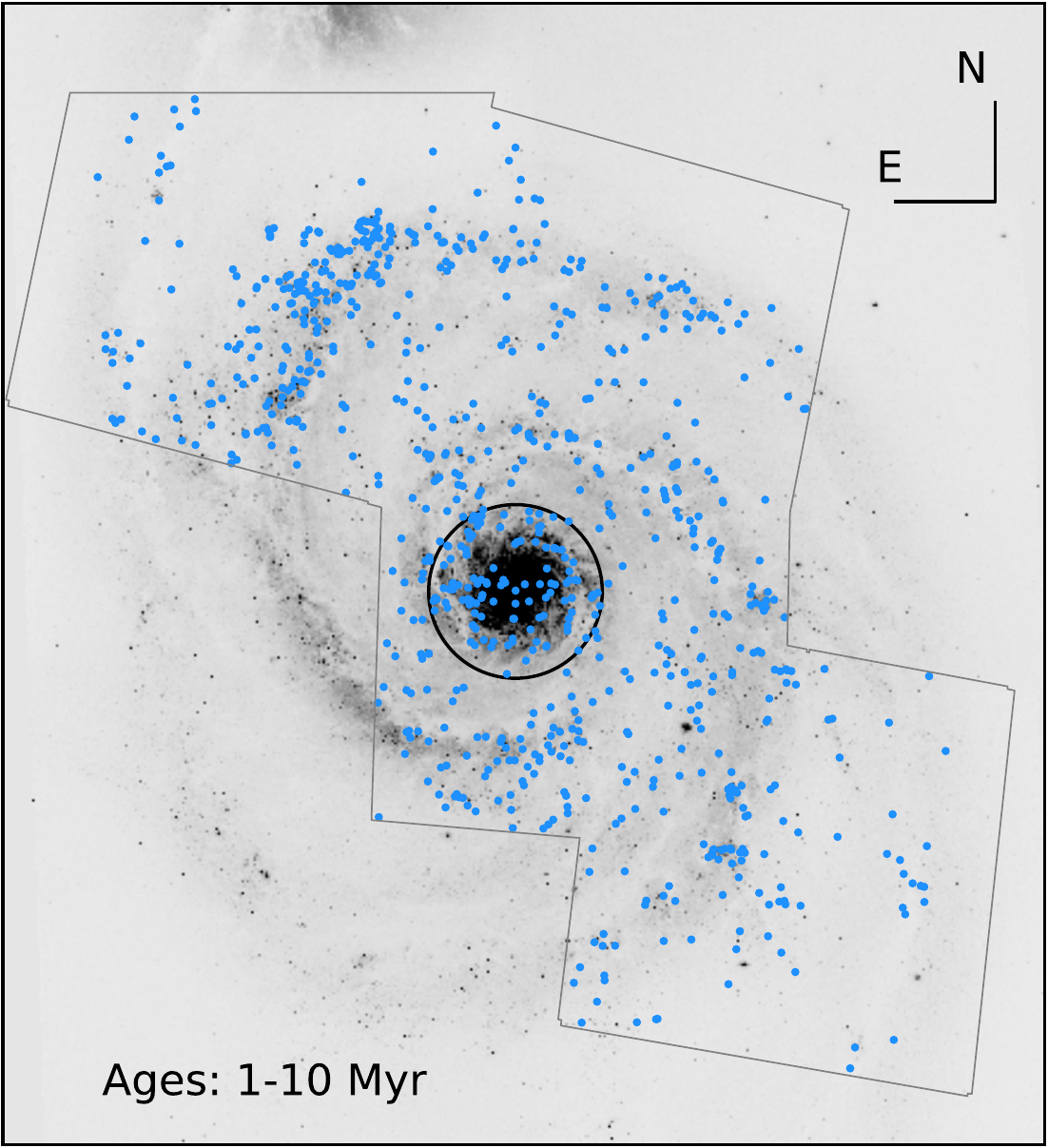}}
 \subfigure
   {\includegraphics[width=0.4\textwidth]{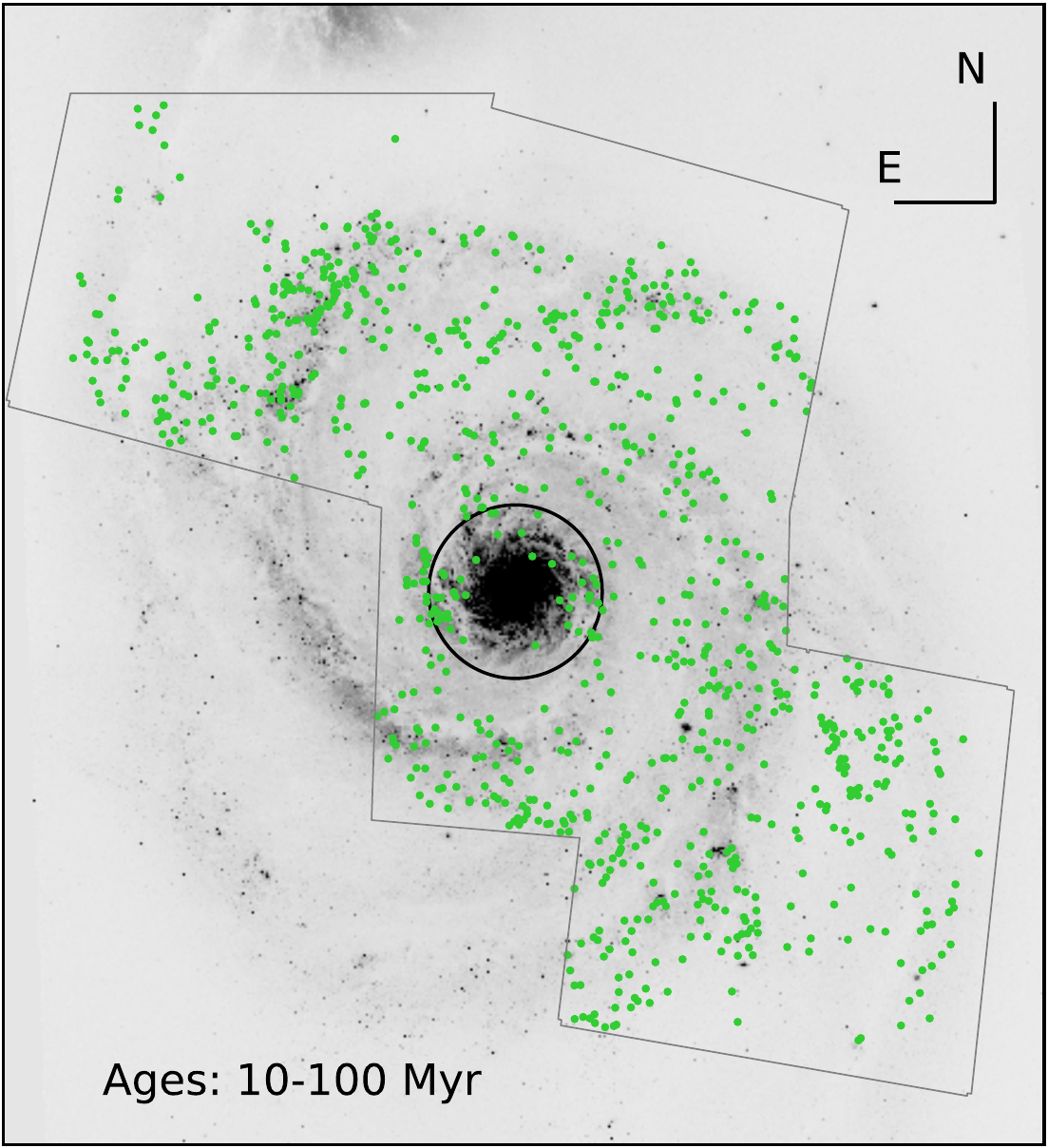}}
 \subfigure
   {\includegraphics[width=0.4\textwidth]{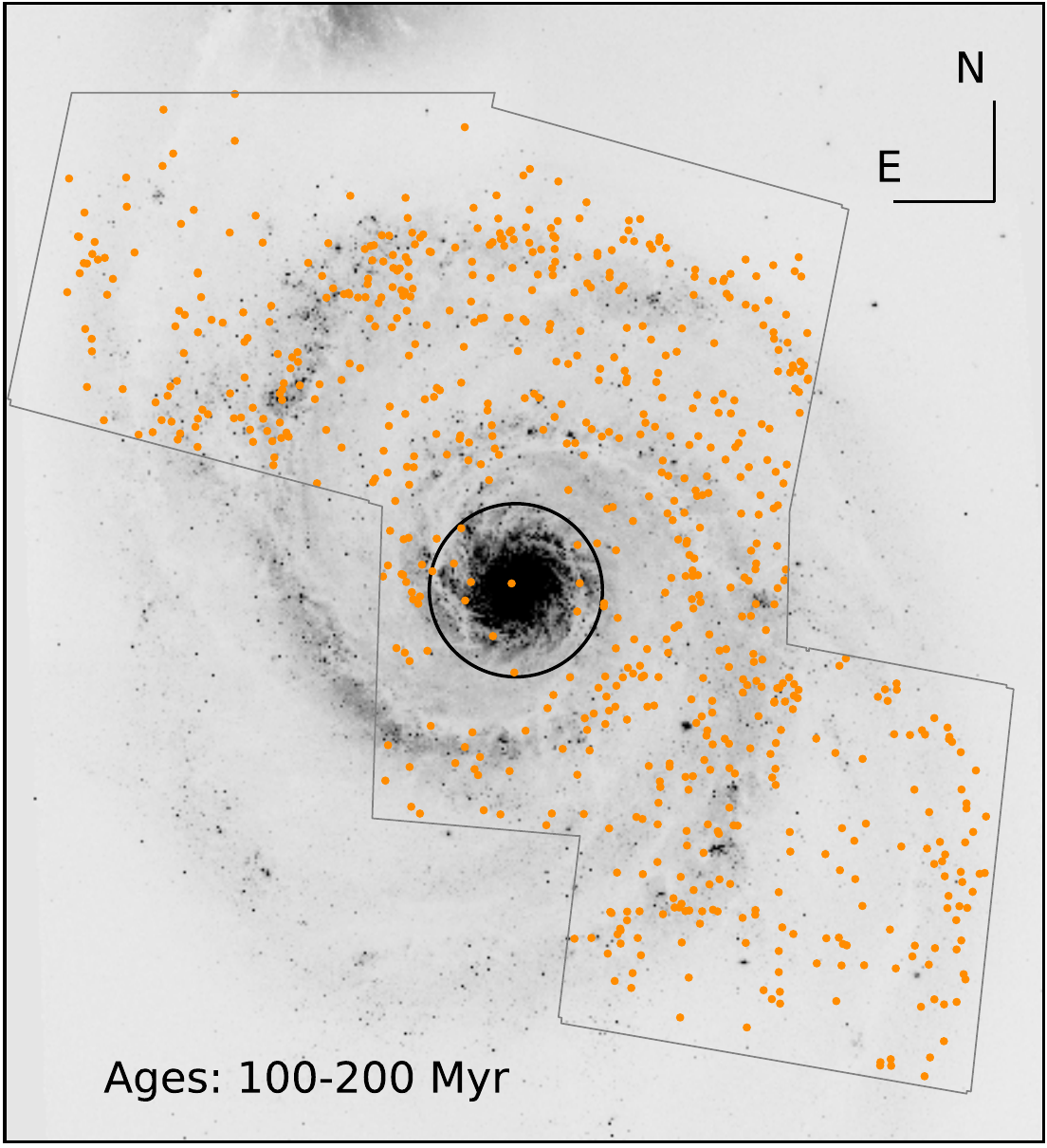}}
 \subfigure
   {\includegraphics[width=0.4\textwidth]{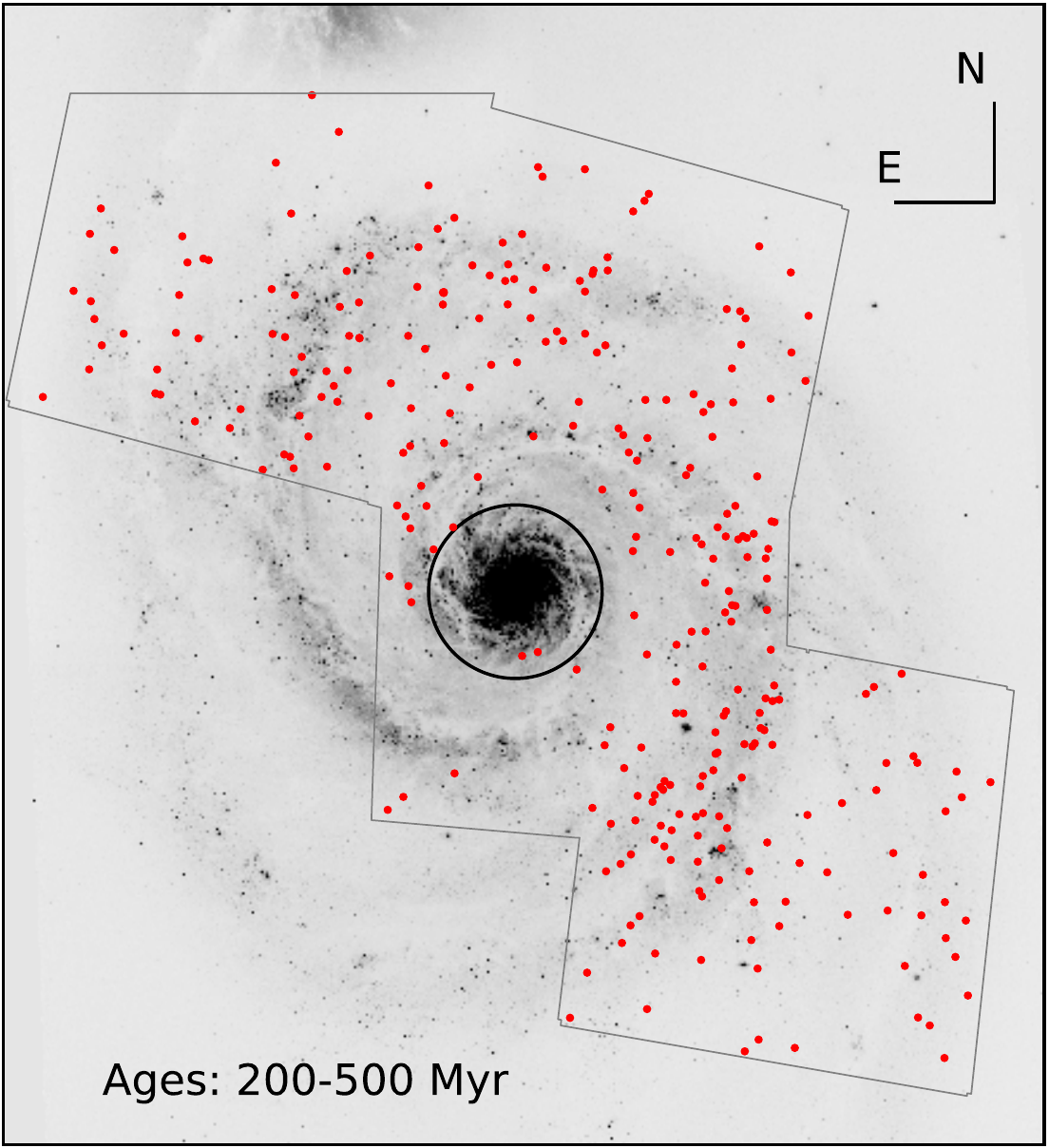}}
 \caption{Position of the clusters divided in age bins in the ranges 1-10 Myr (left), 10-100 Myr (middle-left), 100-200 Myr (middle-right) and 200-500 Myr (right). The central circle of radius 1.3 kpc encloses the region where detection is poorer (see text). The UVIS footprint, restricted to y values below 11600, is overplotted as a grey solid line.}
\label{fig:positions}
\end{figure*}

In order to understand where clusters form and how they move in the dynamically active spiral arm system of M51 we plot the position of the clusters inside the galaxy in Fig.~\ref{fig:positions}. The sample is divided in age bins ($1-10$, $10-100$, $100-200$ and $200-500$ Myr). 
Clusters in our sample are mostly concentrated along the spiral arms. This trend is particularly clear for the very young clusters (age $<10$ Myr) but can also be spotted in the ranges $10-100$ and $100-200$ Myr. In general we observed that young sources are clustered. Moving to older sources the spatial distribution becomes more spread, but it can still be recognized that sources are more concentrated along the spiral arms. In the last age bin, probing clusters older than 200 Myr, the number of available sources is much smaller and is therefore hard to define a distribution, although the sources appear to be evenly spread across the area covered by observations. The strength and age-dependency of the clustering will be further investigated in a future paper (Grasha et al, in prep).

The lack of age gradient as a function of distance from the spiral arm observed in Fig. 8 is in agreement with the detailed study of azimuthal distances of clusters as a function of their ages collected in a forthcoming paper (Shabani et al., in prep.) where the origin of spiral arm and dynamical evolution is investigated.
The observed trend has been predicted by \citet{dobbs2010} which, modeling a spiral structure induced by tidal interactions find that clusters of different ages tend to be found in the same spiral arm without a defined age gradient. 
In a more recent numerical work, \citet{dobbs2017} analyse the evolution of stellar particles in clustered regions, i.e. simulated star clusters within spiral fields. They observe that up to the age range they are able to follow (e.g. 200 Myr) their simulated clusters are mainly distributed along the spiral arms. The trend observed in M51 is thus compatible with that found in \citet{dobbs2017} simulations. Simulations on the evolution of M51 \citep[e.g][]{dobbs2010_m51} suggest that the interaction with the companion galaxy, started $\sim300$ Myr ago, is responsible for creating or strengthening the spiral arms and may have helped keep the old clusters we see now fairly close to the arms.

From Fig.~\ref{fig:positions} we clearly see that our detection is very poor in the centre of the galaxy where the bright background light of the diffuse stellar population is much stronger than in the rest of the galaxy. This effect could explain why we do not detect sources older than 10 Myr (i.e. when cluster light starts to fade), causing a drop in the completeness limit, as already pointed out in Section~\ref{sec:completeness}. For this reason we ignore the clusters within 35'' (1.3 kpc) from the centre of the galaxy from the following analyses.

\subsection{Luminosity function} 
\label{sec:lumfunc}

The luminosity function is intrinsically related to the mass function (luminosity is proportional to mass, with a dependence also on the age) but it is an observed quantity, and therefore, like the colour-colour diagrams, available without any assumption of stellar models and without any SED fitting. 
The luminosity function of YSCs is usually described by a power law function $dN/dL\propto L^{-\alpha}$, with an almost universal index close to $\alpha\approx -2$ as observed in local spiral galaxies (e.g. \citealp{larsen2002,degrijs2003}, see also the reviews by \citealp{whitmore2003} and \citealp{larsen2006b}).

\begin{figure*}
 \centering
 \subfigure[binned]
   {\includegraphics[width=0.48\textwidth]{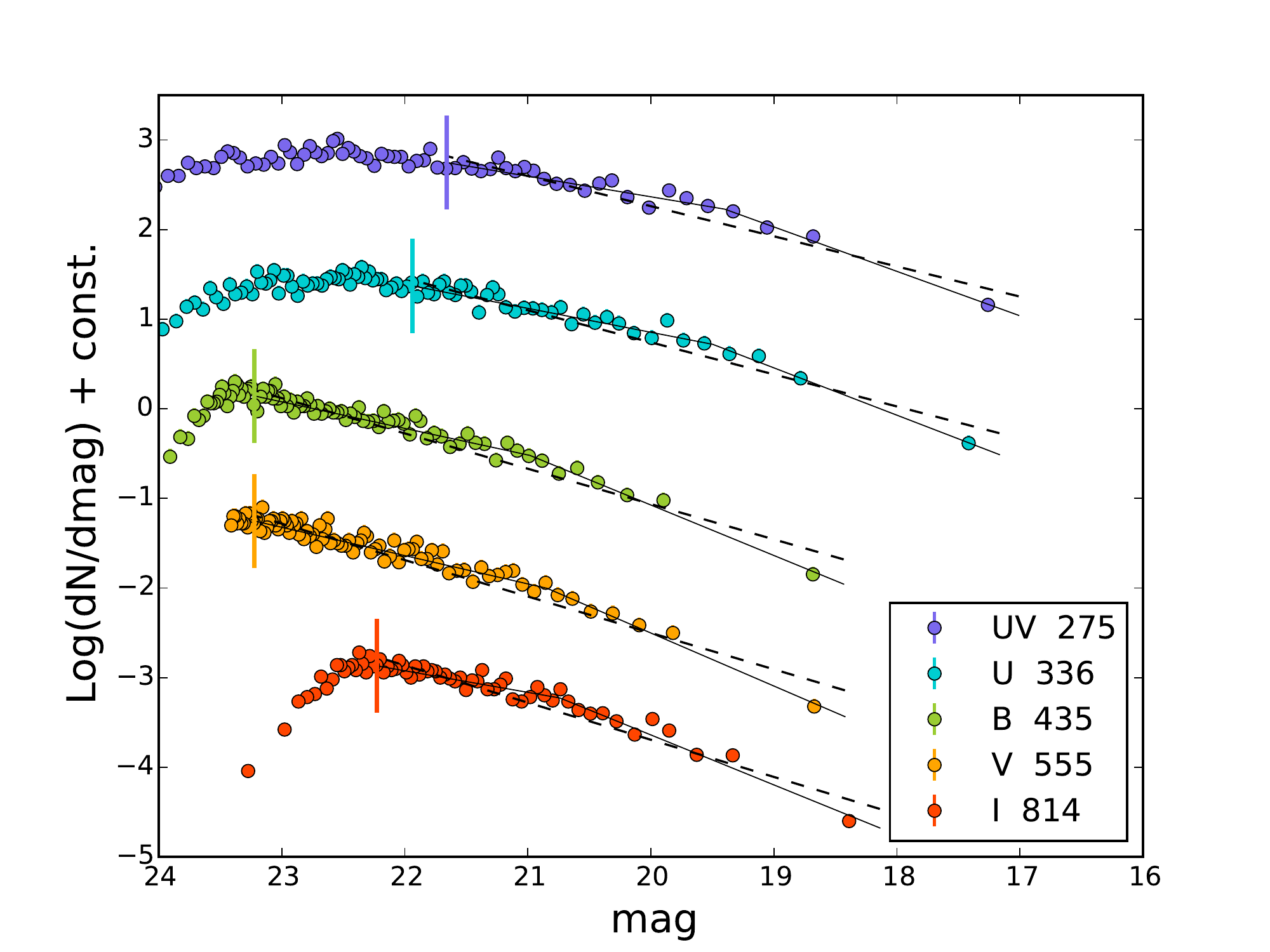}}
 \hspace{0mm}
 \subfigure[cumulative]
   {\includegraphics[width=0.48\textwidth]{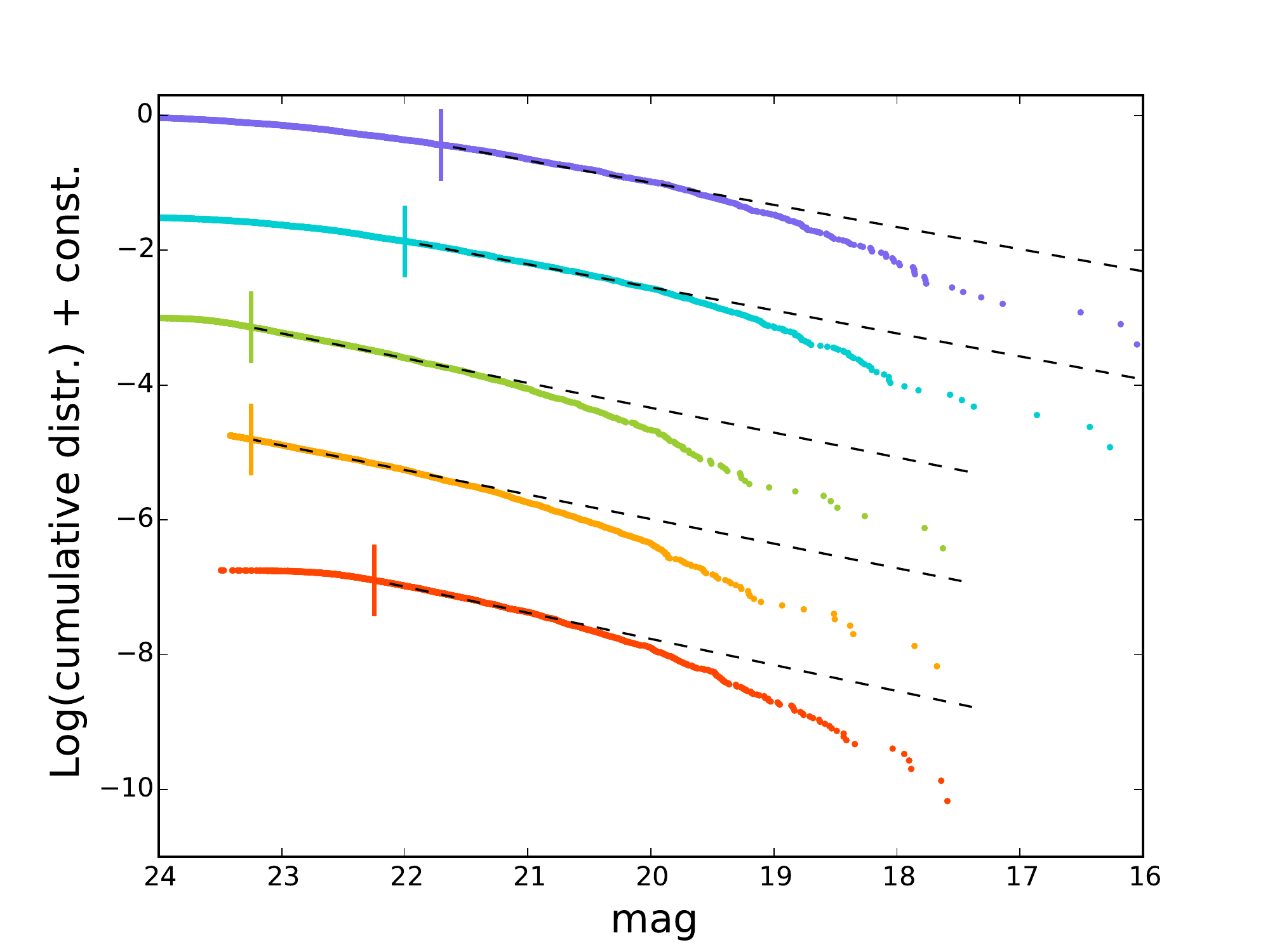}}
 \caption{Binned (a) and cumulative (b) luminosity functions. Fit results are reported in Tab. \ref{tab:lumfit}. The curves are for different filters, from UV band at the top to I band at the bottom. Errorbars in (a) are of the same size of the markers. In both panels the vertical lines mark the completeness limit in each filter.}
\label{fig:lumfit}
\end{figure*}

We analyse the cluster luminosity function by building a binned distribution with the same number of objects per bin, as described in \citet{maiz2005} and performing a least-$\chi^2$ fitting.
The errors on the data are statistical errors given by $\sigma_{\textrm{bin}}=\sqrt{\frac{n_{\textrm{bin}}(n_\textrm{tot}-n_\textrm{bin})}{n_\textrm{tot}}}$, where $n_\textrm{bin}$ is the number of sources in each bin and $n_\textrm{tot}$ is the total number of sources. 
The results of the fits are listed in Tab.~\ref{tab:lumfit} and plotted in Fig.~\ref{fig:lumfit}. 
We have fitted the data up to the completeness limits described in Section~\ref{sec:completeness}.
The function is fitted with both a single and a double power law (PL). The single PL fit gives slopes close to a value of $\alpha=-2$,
however, for all filters, the double power law results in a better fit, as the $\chi^2_{red}$ in this second case is always lower. 
\begin{table*}
\centering
\caption{Results of the fit of the binned luminosity functions.}
\begin{tabular}{|c|c|c|c|c|c|c|c|c}
\hline
\multicolumn{1}{|c|}{Filter} 		& \multicolumn{1}{c|}{$\textrm{Mag}_{\textrm{cut}}$} 	& \multicolumn{2}{l|}{Single PL fit} 	& \multicolumn{4}{l|}{Double PL fit} & \multicolumn{1}{l|}{Cumulative fit}\\
\multicolumn{1}{|c|}{} 		& \multicolumn{1}{c|}{} 		& \multicolumn{1}{c|}{$\alpha$} & \multicolumn{1}{c|}{$\chi^2_{red.}$}		& \multicolumn{1}{c|}{$\alpha_1$}	& \multicolumn{1}{c|}{$\textrm{Mag}_{\textrm{break}}$} 	& \multicolumn{1}{c|}{$\alpha_2$} & \multicolumn{1}{c|}{$\chi^2_{red.}$} & \multicolumn{1}{c|}{$\alpha$}	\\
\hline
\hline
F275W  &	21.71	& $1.84\ _{\pm 0.04}$	& 2.17 &	$1.57\ _{\pm 0.06}$ &	$19.39\ _{\pm  0.25}$ & 	$2.24\ _{\pm 0.10}$  &   0.85 & $2.08\ _{\pm 0.01}$ \\
F336W  &	22.00	& $1.89\ _{\pm 0.04}$	& 1.80 &	$1.67\ _{\pm 0.05}$ &	$19.49\ _{\pm  0.26}$ & 	$2.32\ _{\pm 0.12}$  &   0.92 & $2.10\ _{\pm 0.01}$ \\
F435W  &	23.25	& $1.99\ _{\pm 0.03}$	& 1.72 &	$1.74\ _{\pm 0.04}$ &	$20.97\ _{\pm  0.17}$ & 	$2.41\ _{\pm 0.09}$  &   0.80 & $2.17\ _{\pm 0.01}$ \\
F555W  &	23.25	& $2.02\ _{\pm 0.03}$	& 1.70 &	$1.79\ _{\pm 0.03}$ &	$20.83\ _{\pm  0.17}$ & 	$2.48\ _{\pm 0.10}$  &   0.86 & $2.18\ _{\pm 0.01}$ \\
F814W  &	22.25	& $2.04\ _{\pm 0.04}$	& 2.31 &	$1.60\ _{\pm 0.07}$ &	$20.73\ _{\pm  0.13}$ & 	$2.40\ _{\pm 0.08}$  &   1.03 & $2.28\ _{\pm 0.01}$ \\
\hline
\end{tabular}
\label{tab:lumfit}
\end{table*}

Similar results were found by \citet{haas2008} using a cluster catalogue based only on $BVI$ photometry.
They found that the low-luminosity part of the function could be fitted by a shallow power-law, with slopes in the range $\sim1.7-1.9$, while the high-luminosity end was steeper, with slopes $\sim2.3-2.6$. We similarly found that the low luminosity part of the function is shallower ($\alpha\sim1.6-1.8$) than the high luminosity part ($\alpha\sim2.4-2.5$). 
In both analyses a double power law is a better fit of the luminosity function in all filters. As suggested by \citet{gieles2006}, a broken power law luminosity function suggests that also the underlying mass function has a break.
The possibility that the underlying mass function is truncated is further explored with the study of the mass function in Section~\ref{sec:massfunc} and via Monte Carlo simulations in Section~\ref{sec:montecarlo}.

We compared the binning fitting method with the one presented in \citet{bastian2012}, involving the use of cumulative functions. In the case of a single power law behaviour, the two functions are expected to show the same shape. The cumulative function is given by $y_{\textrm{c}}(m)=\left(1 - \frac{k}{n_{\textrm{data}}}\right)$, where $k$ is the index of the object of magnitude m in the sorted array containing the magnitudes and $n_{\textrm{data}}$ is the length of the array. In case of a simple power law, it has a slope $\alpha_c=\alpha_b-1$ which can  be directly compared to the slope $\alpha_b$ of the binned function. Also in this case a least $\chi^2$ fit is performed. In the cumulative distribution no error is associated with the data, therefore the fit is made with a linear function in the logarithmic space, assigning the same uncertainty to all points. The errors on the fitted parameters have been estimated via a bootstrapping technique: 1000 Monte Carlo realisations of the distribution are simulated, where the luminosity of each cluster is changed using uncertainties normally distributed around $\sigma_{mag}=0.35$ mag. Each realisation was then fitted in the same way as the original one. The standard deviation of the 1000 values recovered for each parameter gave the final uncertainty associated with the recovered slopes.
Results and slopes are collected in Fig.~\ref{fig:lumfit}(b) and Tab.~\ref{tab:lumfit}. We converted $\alpha_c$ into $\alpha_b$ in Tab.~\ref{tab:lumfit}, for an easier comparison with the binned function.
The trends observed in the analysis of the binned distributions are traceable in the cumulative function as well. In particular, we observe that the single PL fit is not a good description of the bright end of the cumulative distributions in all the filters (Fig.~\ref{fig:lumfit}b). 
Also with the cumulative function, the brightest sources fall below the expected curve of a single power law distribution, sign of a break in the luminosity function and therefore also in the underlying mass function.
We note that the fit of the cumulative functions results in steeper slopes than the ones recovered with the binned distributions.  This discrepancy caused by the differences in the two techniques is discussed at length in \citet{legus2}.

\subsection{Mass Function}
\label{sec:massfunc}
The results obtained with the analysis of the luminosity function can be further explored with the study of the properties derived from the SED fit, i.e. the mass and the age distributions.
In the following analyses we use a mass-complete sample, by selecting only clusters above $5000\ \textrm{M}_{\odot}$. This value has been chosen in order to avoid low mass sources, affected by inaccuracies in the SED fitting and by stochastical sampling of the stellar IMF (see e.g. the comparison between deterministic and Bayesian fitting of cluster SEDs in Figure 15 of \citealp{slug}).
The age-mass plot of Fig. \ref{fig:agemass} suggests that we are complete in recovering sources more massive than $5000\ \textrm{M}_{\odot}$ only up to 200 Myr. At older ages, even sources more massive than 5000 \msun can fall below our magnitude detection limit. 
Our mass-limited complete sample therefore contains sources with M $>5000\ \textrm{M}_{\odot}$ and  ages $< 200$ Myr.

The cluster mass function is expected to evolve from a cluster initial mass function (CIMF), usually assumed as a power law $dN/dM \propto M^{\beta}$ with a $\beta=-2$ slope. This slope is interpreted as the sign of the formation of clusters from a turbulent hierarchical medium  \citep{elmegreen2010b}. The initial function is then expected to evolve due to cluster evolution and disruption. 

\begin{figure}
\centering
\includegraphics[width=\columnwidth]{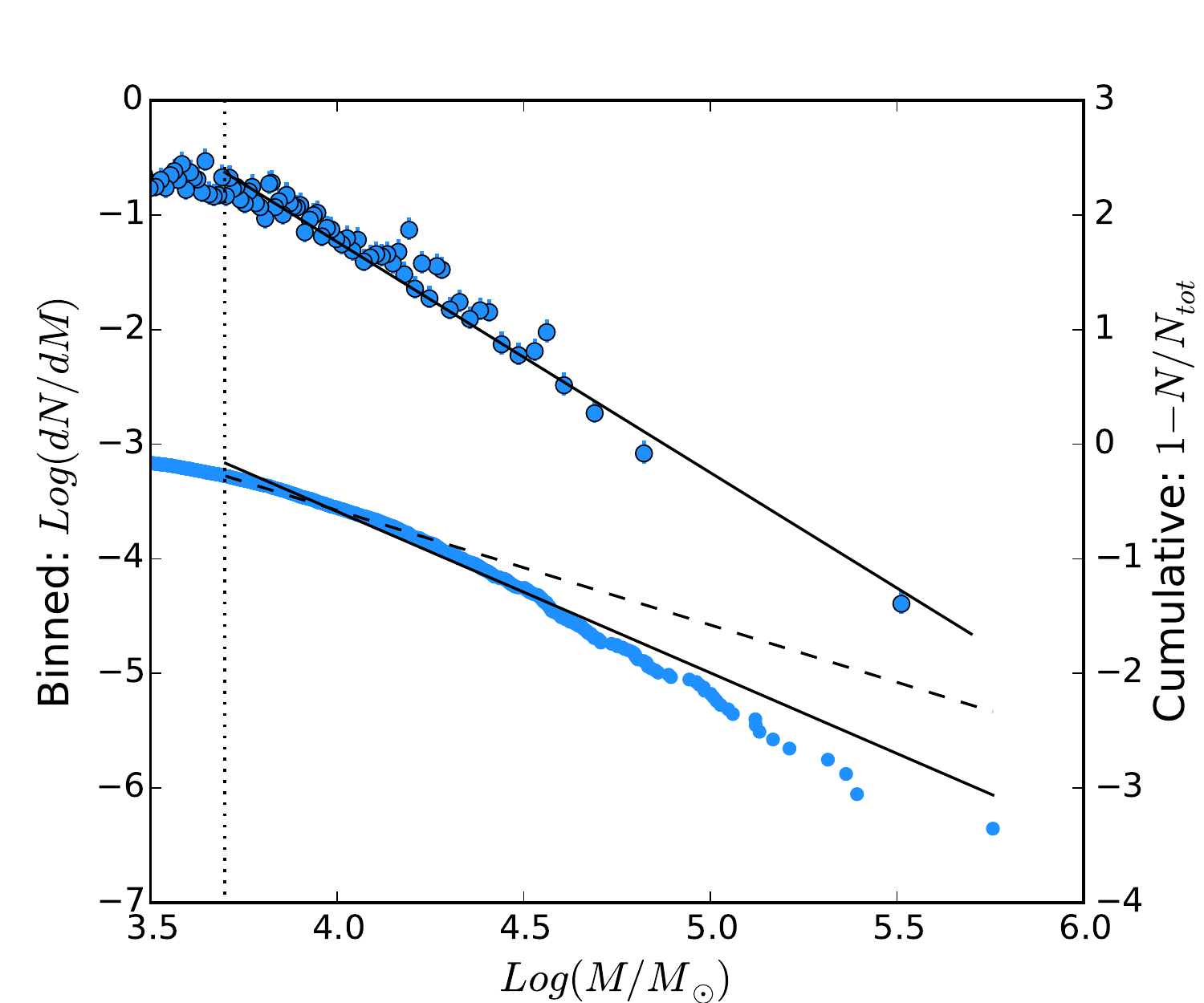}
\caption{Binned (top) and cumulative (bottom) mass function. The solid lines are the best fits with a single power law, the dashed line in the case of the cumulative function shows the slope $-2$ for comparison. The binned function is steeper because the slope of the cumulative function in a simple power law case is $B=\beta+1$, where $\beta$ is the slope of the original function (see Eq.~\ref{eq:cumulative}).}
\label{fig:massfit}
\end{figure}

The cluster mass function of our sample is plotted in Fig.~\ref{fig:massfit}, where bins of equal number of sources were used.
We recover a shape which is well fitted with a single power law of slope $-2.01 \pm0.04$ ($\chi^2_{red.}$  of 1.6), even if a double power law with a  steeper high mass slope fits better the function ($\chi^2_{red.}$ of 1.1, see Tab.~\ref{tab:massfit}). 
\citet{gieles2009} and CH16 found similar slopes, $\beta=-2.09 \pm0.09$ and $\beta=-1.97 \pm0.09$ respectively, considering only clusters in the age range from 10 to 100 Myr. Restricting to the same age range, we find a consistent value of $\beta=-2.03 \pm0.04$ ($\chi^2_{red} = 0.77$, see Fig.~\ref{fig:massfunc_ages}).

As done in the analysis of the cluster luminosity function, we also plot the mass function in a cumulative form (Fig.~\ref{fig:massfit}) and fit it with a pure power law. As already observed for the luminosity functions, the cumulative mass distributions show a steepening at the high-mass end. As observed in \citet{bastian2012} and \citet{legus2}, while the equal number of object binning technique is statistically more robust, it is insensitive to small scales variations, like the dearth of very massive clusters. The cumulative form is therefore more appropriate to study the high-mass end of the mass (and luminosity) function.
\begin{table}
\centering
\caption{Values derived by the fit of the mass function with a least-$\chi^2$ fitting of the binned function. Fits have been performed considering a low mass cut of either 5000 or $10^4$ \msun, as indicated in the second column.}
\begin{tabular}{lccccc}
\hline
Method	& 	M cut 	&  $-\beta_1$			& $M_{break}$	& $-\beta_2$ 	& $\chi^2_{red.}$\\
\		&	[\msun]	& \					& [\msun]		& \			& \			\\
\hline
\hline
Single PL	&	5000		& $2.01_{\pm0.04}$		& $-$			& $-$			& 1.6\\
Double PL	&	5000		& $1.52_{\pm0.12}$		& $1.5\times10^4$	& $2.31_{\pm0.09}$	& 1.1\\
\hline
Single PL	&	$10^4$	& $2.19_{\pm0.07}$		& $-$			& $-$			& 1.6\\
Double PL	&	$10^4$	& $1.71_{\pm0.39}$		& $1.8\times10^4$	& $2.36_{\pm0.13}$	& 1.5\\
\hline
\end{tabular}
\label{tab:massfit}
\end{table}
\begin{table}
\centering
\caption{Values derived by the fit of the cumulative mass function with a maximum-likelihood fit. For a description of the values $N_0$ and $M_0$ see Eq.~\ref{eq:cumulative}. In the last row the result of the fit of the GMC population is reported.}
\begin{tabular}{lccccc}
\hline
Method		& M cut 	& Age 	&$-\beta$				&$N_0$			& $M_0$  \\
\			& [\msun]	& [Myr]	& \					& \				&  [$10^5$ \msun] \\
\hline
\hline
Single PL		& 5000	& 1-200	& $2.30_{\pm0.03}$		& $-$ 			& $-$ \\ 
Truncated 	& 5000	& 1-200	& $2.01_{\pm0.02}$		& $66_{\pm6}$		& $1.00_{\pm0.12}$\\
Truncated		& 5000	& 1-10	& $2.12_{\pm0.22}$		& $10_{\pm7}$		& $0.56_{\pm0.08}$\\
Truncated		& 5000	& 10-100	& $1.97_{\pm0.06}$		& $43_{\pm15}$	& $0.91_{\pm0.16}$\\
Truncated		& 5000	& 100-200	& $2.01_{\pm0.05}$		& $28_{\pm4}$		& $1.15_{\pm0.27}$\\
\hline
Single PL		& $10^4$	& 1-200		& $2.67_{\pm0.03}$		& \ 				& \ \\ 
Truncated		& $10^4$	& 1-200		& $2.34_{\pm0.03}$		& $22_{\pm10}$	& $1.34_{\pm0.24}$\\
\hline
GMC pop 		& $-$	& $-$	& $2.36_{\pm0.16}$     	& $12_{\pm5}$		& $160_{\pm32}$ \\
\hline
\end{tabular}
\label{tab:massfit2}
\end{table}

In order to test the hypothesis of a mass truncation, we have fitted the cumulative distribution with the \texttt{IDL} code \texttt{mspecfit.pro}, implementing the maximum-likelihood fitting technique described in \citet{rosolowsky2007}, commonly used for studying the mass functions of GMCs \citep[e.g.][]{colombo2014a}. The code implements the possibility of having a truncated power law mass function, i.e.
\begin{equation}
\label{eq:cumulative}
N(M'>M)=N_0\left[\left(\frac{M}{M_0}\right)^{\beta+1}-1\right],
\end{equation}
where $M_0$ is the maximum mass in the distribution and $N_0$ is the number of sources more massive than $2^{1/(\beta+1)}M_0$, the point where the distribution shows a significant deviation from a power law (for the formalism, see \citealp{rosolowsky2005}).
A value of $N_0$ bigger than $\sim1$ would indicate that a truncated PL is preferred over a simple one. On the other hand, $N_0<1$ would mean that the truncation mass is not constrained and that a single power law is a good description of the distribution. The resulting parameters of the fit for our sample, considering normally distributed 0.1 dex errors on the masses, are collected in Tab.~\ref{tab:massfit2}. 
The resulting $N_0 = 66 \pm6$ suggests that the fit with a truncated function,with $M_0=10^5$ \msun, is preferred over the simple PL. The best fit for the slope is $\beta=-2.01 \pm0.02$.

In order to test for possible incompleteness at masses close to 5000 \msun, we repeated the analysis of the mass function using a mass cut of $10^4$ \msun.
Results are collected in Tab.~\ref{tab:massfit} and \ref{tab:massfit2}. Different lower limits at the low mass yield steeper than $-2$ power laws but consistent truncation masses. 
The binned function is well fitted with a single power law with $\beta=-2.19 \pm0.07$ ($\chi^2_{red.}=1.55$). The maximum likelihood fit of the cumulative function gives $\beta=-2.34 \pm0.03$, $M_0= (1.34 \pm 0.24) \times10^5$ \msun\ and $N_0=22 \pm 10$, thus a truncation is still statistically significant. 

It has been reported in the literature that the YSC mass function is probably better described by a Schechter function with a $\beta=-2$ slope and a truncation $M_c$ at the high-mass end. In the case of M51, \citet{gieles2006} found that a Schechter function with $M_c = 10^5$ \msun\ would reproduce closely the luminosity function observed. With a very different approach, \citet{gieles2009} derived $M_c = (1.86 \pm0.52) \times 10^5$ \msun\ from the analysis of an evolving mass function. Both results are consistent with our results. 

\citet{chandar2011,chandar16} find that a simple PL  is a good description of the YSC mass function in M51, however they only considered a binned MF. Different mass estimates for high-mass clusters, as noted in Fig.~\ref{fig:comparison_am}, could produce differences in the mass function slopes. Nevertheless, we retrieve the same results of CH16 if a binned function is used. We notice that the bin containing the most massive clusters encompasses the whole range in masses where the truncation mass is found (see Fig.~\ref{fig:massfit}). Thus binning techniques that use equal number of objects are therefore unable to put a constraint on the truncation. 

The truncation mass we recover is smaller but similar to what was found in other spirals, like M 83 ($M_c=(1.60 \pm0.30) \times10^5$ \msun, \citealp{adamo2015}), NGC 1566 ($M_c=2.5 \times10^5$ \msun, \citealp{hollyhead2016}) and NGC 628 ($M_c=(2.03 \pm0.81) \times 10^5$ \msun, \citealp{legus2}). On the other hand, some galaxies still exhibit a truncated mass function but with very different truncation masses. In M31, for example, \citet{johnson2017} found a remarkably small truncation mass of $\sim10^4$ \msun, while the Antennae have a MF that exhibits a PL shape that extends up to masses larger than $10^6$ \msun\ \citep{whitmore2010}. These differences spanning orders of magnitudes suggest that the maximum cluster mass in galaxies must be determined by the internal (gas) properties of the galaxies themselves. \citet{johnson2017} suggested that $M_c$ should scale with the \sigmasfr.
Differences in the recovered truncation mass have also been found within the same galaxy (e.g. \citealp{adamo2015}). We will investigate possible environmental dependencies of the mass function properties of M51 in a follow up work (Paper II).

\subsubsection{Comparison With GMC Masses}
\label{sec:GMC}
We compare our cluster mass function with the mass function of the GMCs in M51 from the catalogue compiled and studied in \citet{colombo2014a}. Clusters are expected to form out of GMCs, via gravitational collapse and fragmentation, and therefore the mass distribution of the latter can in principle leave an imprint on the mass distribution of YMCs.

The mass function of GMCs in M51 steepens continuously going from low to high masses, and cannot be described by a single power law \citep{colombo2014a}, as is instead the case for other galaxies like LMC, M33, M31 and the Milky Way \citep{wong2011,gratier2012,rosolowsky2005}.  We perform a fit of GMC masses with the same code described in the previous section, up to a lower limiting mass of $10^6$ \msun\ (discussed in Section 7.2 and 7.3 on the mass functions in \citealp{colombo2014a}). The resulting best value for the slope and the maximum mass are $\beta=-2.36 \pm0.16$ and $M_0=(1.6 \pm0.3) \times10^7$ \msun. 
The value of $M_0$ implies a truncation mass which is $\sim100$ times bigger in the case of GMCs similar to what has been observed in M83 by \citet{freeman2017}. 
The mass function of GMCs looks steeper than the one of the clusters, within a 3$\sigma$ difference. Analysing the mass function of simulated GMCs and clusters, \citet{dobbs2017} found the opposite trend of a steeper function in the case of clusters. Part of the difference between the simulated and observed trends can be due to the different regions covered within the two surveys: the PAWS survey from which the GMC data are derived covers only the central part of the galaxy, while our clusters also occupy more distant regions from the centre.
The study of the mass function in different regions of M51 in Paper II will enable us to compare the CMF with the GMC one on local scales, testing closely the link between GMC and cluster properties.
 
\subsubsection{Evolution of the Mass Function}
\label{sec:massage}
Cluster disruption affects the mass function and could, in principle, modify its shape: for this reason we study the function in different age bins. 
In order to be able to see how significant the disruption is, we look at the evolution of the CMF normalized by the age range (i.e. $dN/dMdt$).
In case of constant star formation and no disruption the mass functions should overlap. 
Cluster disruption can in principle affect the mass function in different ways according to the disruption model considered. 

Two main empirical disruption scenarios have been proposed in the literature and they differ in the dependence with the cluster mass. A first model, firstly developed to explain the age distribution of clusters in the Antennae galaxies (see \citealp{fall2005} and \citealp{whitmore2007}), proposes that all clusters lose the same fraction of their mass in a given time. This implies that the disruption time of clusters is independent on the cluster mass and we therefore call this model mass-independent disruption (MID). It is characterized by a power-law mass decline and therefore by a disruption rate which depends linearly on the mass \citep{fall2009}, i.e.
\begin{equation}
M(t)\propto t^\lambda, \ \ \ \ \frac{dM}{dt}\propto M
\end{equation}
On the other hand, the mass dependent disruption (MDD) time scenario assumes that the lifetime of a cluster depends on its initial mass, with a relation $t_{dis}\propto M^k$ (with $k=0.65$, i.e. less massive clusters have shorter lifetimes).
Initially suggested by \citet{boutloukos2003} considering only instantaneous disruption to explain the properties of the cluster populations in the SMC, M33 and M51, this model has been updated to account also for gradual mass-loss in \citet{lamers2005}. This model is characterized by a disruption rate which depends sub-linearly on the mass as:
\begin{equation}
\frac{dM}{dt}\propto M^{1-k}
\end{equation}
The two scenarios predict different evolutions for the cluster mass function \citep[e.g.][]{fall2009}. In the MID model, the mass function shape is constant in time, it only shifts to lower masses due to all clusters losing the same fraction of mass. In the MDD model, instead, low-mass clusters have shorter lifetimes and this results in a dearth of clusters at the low-mass end of the function, as the time evolves.

\begin{figure}
\centering
\includegraphics[width=1.\columnwidth]{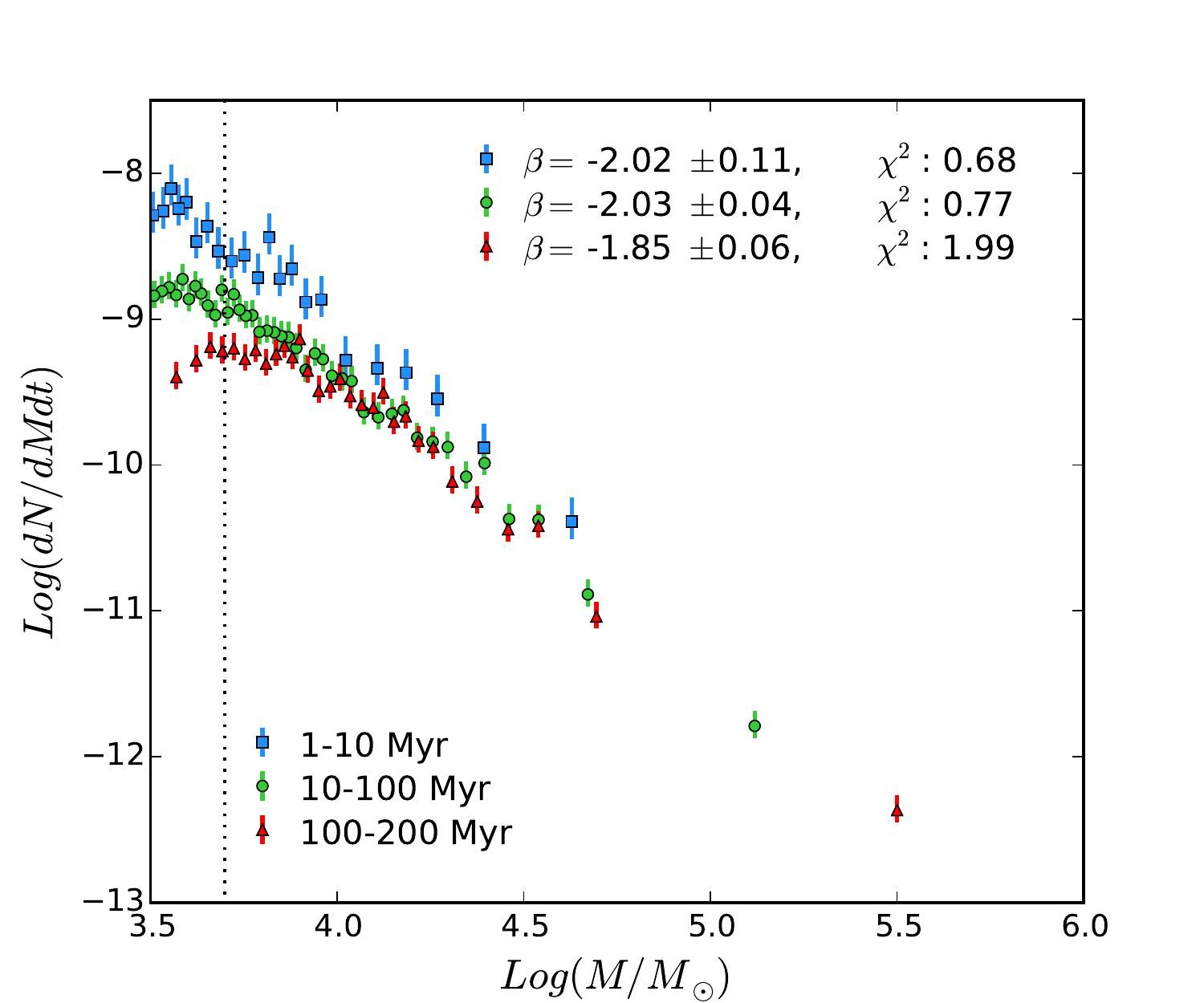}
\caption{Mass function divided in age bins and normalized by the age range in each bin. The black dotted line is the low-mass limit of 5000 \msun. The shift in normalisation between the young function (blue) and the others suggest that cluster disruption is already happening between 10 and 100 Myr. The old function (red) flattens at low masses, but it is difficult to separate the effects of disruption and incompleteness.}
\label{fig:massfunc_ages}
\end{figure}

In Figure~\ref{fig:massfunc_ages} we observe that the normalised CMF at ages below 10 Myr is detached from the CMFs of the other two age bins, suggesting a stronger drop in the number of sources. In the age range $10-100$ Myr, compared to the range $100-200$ Myr, the main difference between the two CMF is seen at low masses as a bend in the CMF of the oldest clusters, i.e. $100 - 200$ Myr. This trend would suggest a shorter disruption time for low mass clusters, however as can be seen in Fig~\ref{fig:agemass}, at these ages also incompleteness could start affecting the data. So the flattening could be the result of both mass dependent disruption time and incompleteness. On the other hand, at high masses the functions seem to follow each other quite well. 

Each function is fitted with a least-$\chi^2$ approach. Single power laws are fitted and the resulting slopes for the age bins $1-10$, $10-100$ and $100-200$ Myr are $\beta=-2.02 \pm0.11$,  $-2.03 \pm0.04$ and $-1.85 \pm0.06$, respectively. We can compare these values with the results of CH16 and \citet{gieles2009} as both of them studied the mass function in age bins. CH16  found a slope  $\beta=-2.06 \pm 0.05$ for sources younger than 10 Myr and $\beta=-1.97 \pm 0.09$ for sources in the range 10-100 Myr. For older sources they consider a bin with ages in the range 100-400 Myr finding a slope of $\beta=-2.19 \pm 0.06$. 
Using age bins of $4-10$ Myr, $10-100$ Myr and $100-600$ Myr, \citet{gieles2009} found slopes of $\beta=-2.08 \pm 0.08$, $-2.09 \pm 0.09$ and $-2.76 \pm 0.28$ respectively, using the cluster catalogue of B05. While up to 100 Myr those values are comparable with what we find, at old ages their results seem to strongly deviate from our own. A first reason for this deviation may be the smaller size of our last age bin, which extends only to 200 Myr and therefore neglects older sources. 
However, a more likely explanation can be found in the definition of the minimum mass considered in each age range: because of our completeness limit, we always consider sources more massive than 5000 \msun, while the cut of the older bin in \citet{gieles2009} is 6$\times10^4$ \msun\ and in CH16 is $\approx 10^4$ \msun. In both cases the low-mass part of the function is not considered in the fit, thus their fit may be more sensitive to the presence of a bend in the form of a truncation. As seen in Tab.~\ref{tab:massfit}, fitting only the high mass part of the CMF results in stepper slopes also in our catalogue, even if a shorter age range is used.

Fitting cumulative instead of binned distributions with the maximum-likelihood fit with the \texttt{mspecfit.pro} code yields slopes and truncation masses collected in Tab.~\ref{tab:massfit2}. Results for age bins $10-100$ Myr and $100-200$ Myr are very similar to the results found for the whole population. In both age ranges the presence of a truncation ($N_0 >>1$) is statistically significant. The CMF in the age range $1-10$ Myr has a fitted $M_0$ which is a factor 2 smaller. The statistical significancy $N_0$ of the latter fit is, within uncertainties, not much larger than 1 ($N_0=10 \pm7$). This result seems driven by size--of--sample effects. Uncertainties in this last case are larger because the sample is small, counting only 140 clusters, compared to the other two age bins hosting more than 500 clusters each. These uncertainties prevent to statistically test the truncation for the mass function in the bin $1-10$ Myr.

\subsection{Age Function}
\label{sec:agefunc}
We can investigate the cluster evolution analysing the age distributions of the clusters. The YSC age function is determined by the star (and cluster) formation history (SFH and CFH) convolved with cluster disruption.

In first approximation, the SFH of spiral galaxies can be assumed constant  for extended periods, unless external perturbations (like interactions, minor, or major mergers) change the condition of the gas in the galaxy. 
YSC disruption is usually inferred by changes in the number of clusters as a function of time, assuming that the SFH has been constant. In the presence of enhancement in SF, the change brought by the increasing SFR can be misinterpreted as disruption. Thus it is of fundamental importance to know the recent  SFH of the host galaxy. 
The easiest assumption of a constant star formation rate allows a very straightforward interpretation of the age function which it is not necessarily true. In the case of M51, we know it is an interacting system and that tidal interactions can enhance the star formation \citep{pettitt2017}. Many simulations of the M51 evolution have suggested a double close passage of the companion galaxy, the older approximately $400-500$ Myr ago and a more recent one $50-100$ Myr ago \citep[see][]{salo2000,dobbs2010_m51}. Whether the enhancement of star formation during the close passages with the interacting galaxy has a visible impact on the age function is difficult to assess, without an accurate star formation history. Our analysis is limited to an age range $<200$ Myr, hence we expect our analysis to be only partially affected.

\begin{figure}
\centering
\includegraphics[width=\columnwidth]{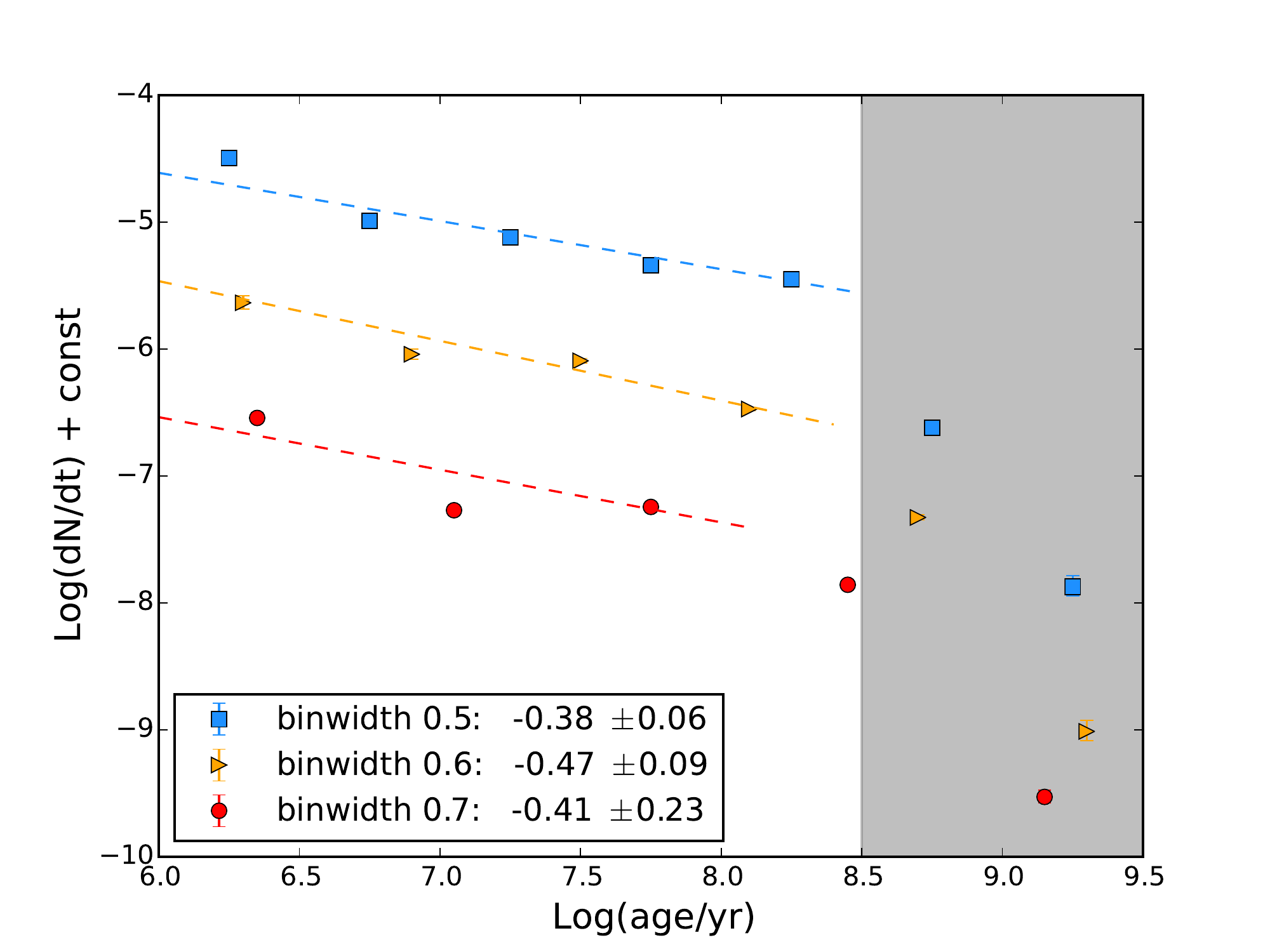}
\caption{Age function of the cluster catalogue, comparing the effect of different bin widths, namely widths of 0.5 dex (blue squares), 0.6 dex (orange triangles) and 0.7 dex (red circles). The grey-shaded area marks the ages at which incompleteness causes a steepening of the slope, preventing the study of the function.}
\label{fig:agefit}
\end{figure}
We build an age function dividing the sample in age bins of 0.5 dex width and taking the number of sources in each bin. This number is normalised by the age range spanned by each bin (Fig.~\ref{fig:agefit}). Points have been fitted with a simple power law $dN/dt \propto t^{-\gamma}$ up to $\textrm{log(age/yr)}=8.5$. After that the incompleteness strongly affects the shape of the function, which starts declining steeply. The resulting slope is $\gamma=0.38 \pm0.06$,  smaller than reported  by CH16 who found $\gamma$ in the range $0.6-0.7$ using a set of mass selections and age intervals. 
Note that, in their cluster selection, CH16 do not remove clusters in the internal part of the galaxy, which happen to have a much steeper slope. However the main difference can probably to be attributed to the different age modeling of the two catalogues. As can be seen from Fig. \ref{fig:comparison} the ages of the CH16 catalogue have a prominent peak between logarithmic ages of 6.5 and 7. The resulting cluster ages are therefore younger on average and result in a steeper decline.

A source of uncertainty in the recovered slope of the age function is related to the binning of the data. The distribution of the ages is discrete, and therefore binning is necessary, but the choice of the bins can affect the recovered slopes. To test possible variations we repeat our analysis changing the bin size. Results are shown in Fig.~\ref{fig:agefit}. Differences on the recovered slopes are within the errors and therefore in this case the age function is statistically not sensitive to the choice of binning.

\begin{figure}
 \centering
 \includegraphics[width=\columnwidth]{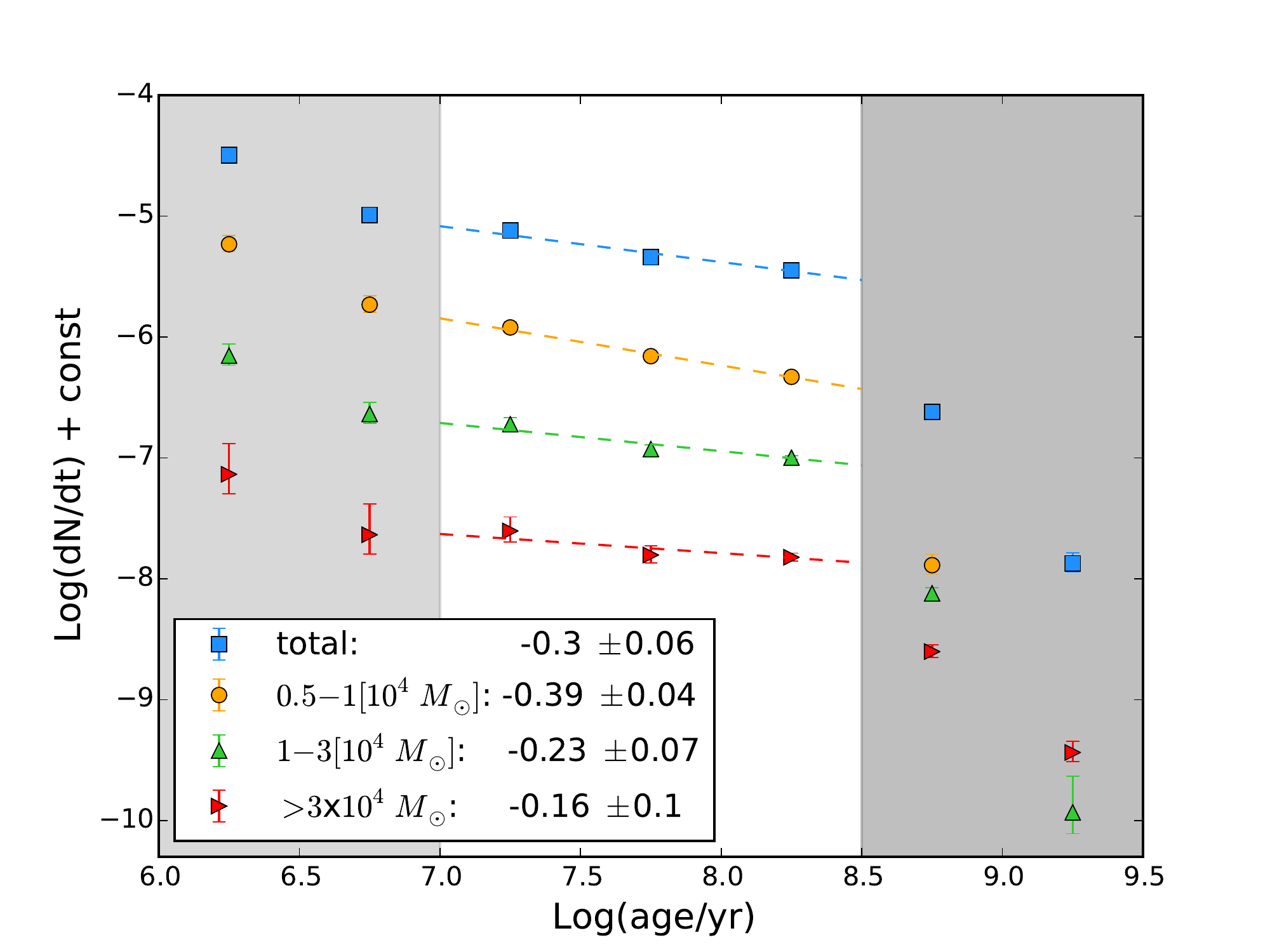}
\caption{Age function divided in mass bins. Dashed lines represent the best fitting curve for the bins in the range Log(age/yr)$=7-8.5$. The grey-shaded areas mark the part of the functions excluded from the analysis due to incompleteness (old ages) and possible contamination by unbound sources (young ages).}
\label{fig:agefit_mass}
\end{figure}

Caution must be taken when considering the age function at young ages: in the literature it has been proposed that an ``infant mortality'' (introduced by \citealp{lada2003}), caused by the expulsion of leftover gas from star formation, could in principle cause a rapid decline in the number of clusters surviving after $\sim10$ Myr. However numerical simulations show that gas expulsion does not have strong impact of the dynamical status of the stars within a gravitationally bound cluster (see \citealp{longmore2014} review). As already discussed in \citet{legus2}, at young ages it could be easier to include in the sample sources that are unbound at the origin. We are not considering the sizes of clusters, or their internal dynamics, therefore we are unable to assess the boundness of clusters. However, given a typical size of a few parsecs for the cluster radius \citep{ryon2015,ryon2017}, we know that clusters older than 10 Myr have ages larger than their crossing time, and we can consider them bound systems. 
The inclusion of unbound sources would cause a strong decline in the age function because they contaminate our sample at ages younger than 10 Myr. If we exclude the youngest age bins below 10 Myr, we find that the fit results in a shallower slope, $\gamma=0.30 \pm0.06$ (compare the resulting slopes in Fig.~\ref{fig:agefit} and Fig~\ref{fig:agefit_mass}). In the hypothesis of constant SFR over the last 200 Myr, we conclude that disruption is not very significant in the age range $10-200$ Myr. 

In order to test if the disruption time is mass dependent, we divided the catalogue in three subsamples of increasing mass, in the ranges: $5000-10^4$ \msun, $10^4-3\times10^4$ \msun\ and masses $>3\times10^4$ \msun. The slopes of the resulting age functions (Fig.~\ref{fig:agefit_mass}) present small differences, with more massive sources having flatter slopes. Incompleteness at ages $\approx200$ Myr may start affecting the less massive sources ($\sim$5000 \msun) which seem to have significantly more disruption. The differences between the two more massive bins are within 1 $\sigma$, thus very similar.
CH16 found slopes compatible within 1 $\sigma$ in different mass bins, i.e. $\gamma=-0.71 \pm0.03$, $-0.64 \pm0.20$ and $-0.62 \pm0.07$ for mass ranges log(M/\msun)$=3.8-4.0$ , $4.0-4.5$ and $4.5-5.0$ respectively. These slopes are systematically steeper than what we find, with differences close to $\sim3 \sigma$ from our values.

The difficulty in retrieving the correct model for mass disruption can be also due to the simultaneous action of different processes dispersing the clusters mass.
\citet{elmegreen2010} propose a model in which clusters are put into an hierarchical environment in both space and time and show that, under reasonable assumptions, many different processes of mass disruption (or the combination of them) can reproduce an age function with a power law decline, as generally observed.

\begin{figure}
 \centering
\subfigure{ \includegraphics[width=0.95\columnwidth]{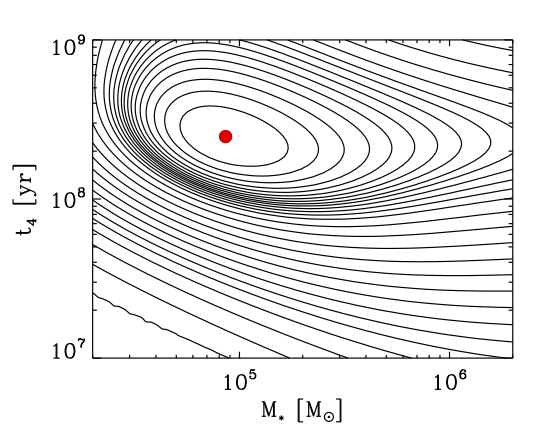}}
\subfigure{ \includegraphics[width=1.\columnwidth]{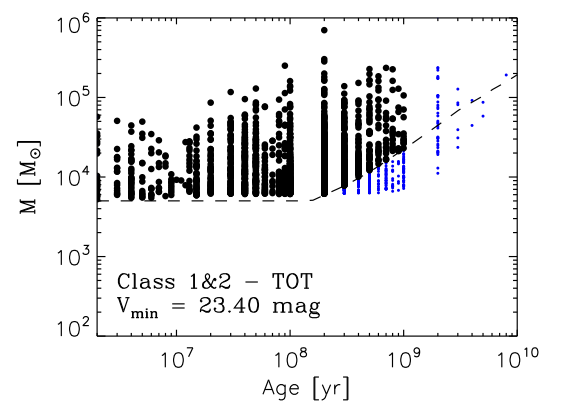}}
\caption{Probability distributions for the Schechter mass $M_*$ and for the typical disruption time of a $10^4$ \msun\ cluster, $t_4$, in the maximum-likelihood test (top). The red dot indicates the best fit. The plots on the bottom collect the age-mass distributions, with a dashed line indicating the limit (both due to the imposed mass cut and the limiting magnitude) above which we have selected the sources for the analysis. Black points are the selected sources, while blue points are the total sample. 
}
\label{fig:agefit_maxlk}
\end{figure}

Under the assumption of a MDD time, we can derive a typical value for cluster disruption. We consider $t_4$, i.e. the time necessary to disrupt a cluster of $10^4$ \msun, as it is an indicative physical value. We use a maximum-likelihood fitting technique, introduced by \citet{gieles2009}, where we assume an ICMF described by a $-2$ power law with a possible exponential truncation at $M_*$ (which is left as a free parameter) and a disruption process which is mass dependent in time, with a timescale-mass relation given by $t_{dis}\propto M^{0.65}$. In this analysis we considered sources with ages up to $10^9$ yr, limiting the sample to $M>5000$ \msun\ and $V_{mag}<23.4$ mag (see Fig.~\ref{fig:agefit_maxlk}). The mass cut, as previously pointed out, allows us the study of a mass-complete sample (up to 200 Myr). The cut in magnitude instead allows us to consider sources older than $200$ Myr, accounting for the fading of old sources below the magnitude completeness limit.

The results of this maximum-likelihood analysis are $M_*=(8.6\ \pm\ 0.5)\times10^4$ \msun\ and $t_4=230\ \pm\ 20$ Myr (probability distributions are given in Fig.~\ref{fig:agefit_maxlk}). 
The Schechter truncation mass is compatible to what we found in the previous section within the uncertainties. The disruption timescale is of the same order of $t_4\simeq130$ Myr found in the analysis of \citet{gieles2009}, slightly above the interval $100\lesssim t_4\lesssim200$ Myr retrieved for different assumptions for the cluster formation history in \citet{gieles2005}. However, the analysis of the age function suggests that some disruption, possibly also with mass independent time, may have been effective from the beginning, as we see a mild decrease in $dN/dt$ already at young ages.

\section{Simulated Montecarlo Populations}
\label{sec:montecarlo}
In order to better understand the results of the previous sections, we perform simulations, producing Monte Carlo populations of synthetic clusters, with properties similar to what we observe for the cluster population of M51. We use different initial sets of parameters for the mass distribution and the disruption and compare the synthetic populations with the observed one. This is particularly useful for the luminosity function, which, while easy to obtain, is the result of many generations of YSCs formed and evolved within the galaxy.
Analytical and semi-analytical models have tried to derive an expected LF shape from basic assumptions on the CIMF and on the age distribution and have been applied to the studies of cluster populations (e.g. \citealp{haas2008} in M51, \citealp{fall2006} in the Antennae, \citealp{hollyhead2016} in NGC 1566). We will also refer to those studies in order to compare expectations, simulations and observations.

\subsection{Simulated Luminosity Functions}
We simulate populations of clusters with masses larger than 200 \msun. The number of clusters per simulation is set such that we have the same number of clusters with $M>5000$ \msun\ as the observed population, i.e. $\sim1200$. The star (and cluster) formation history is assumed constant for 200 Myr, which is also the maximum age we assign to the simulated clusters. After simulating age and mass for each of the synthetic clusters we assign a magnitude to them in the $B$, $V$ and $I$ bands, using the same models adopted for the SED fitting and described in Section~\ref{sec:sed}. We can then build the luminosity functions, both in the binned and in the cumulative way (see Sec~\ref{sec:lumfunc} for ref.), and fit them with a power law. The models used for the initial mass function and for the disruption are:
\begin{description}
\item[\textbf{PL-2:}] pure $-2$ power law, no disruption.
\item[\textbf{PL-2\_MDD:}] pure $-2$ power law, mass dependent disruption time model with $k=0.65$ and $t_4=230$ Myr (as the results of the maximum-likelihood fit of Sec~\ref{sec:agefunc} suggest, see the same Section also for the formalism).
\item[\textbf{SCH:}] Schechter function\footnote{The Schechter mass function, when not specified, is assumed to have a slope $-2$ in the power law part, i.e. $dN\propto M^{-2}\ e^{-M/M_*} dM$} with $M_*=10^5$ \msun, no disruption.
\item[\textbf{SCH\_MDD:}] Schechter function with $M_*=10^5$ \msun, mass dependent disruption time model with $k=0.65$ and $t_4=230$ Myr.
\end{description}
We have not included MID in the models because it will only  change the normalisation of the LF, not the shape. For this reason the two models without disruption (PL-2 and SCH) can be used to also study the expected LF in the case of mass independent disruption time model.

\begin{figure*}
\centering
\includegraphics[width=0.9\textwidth]{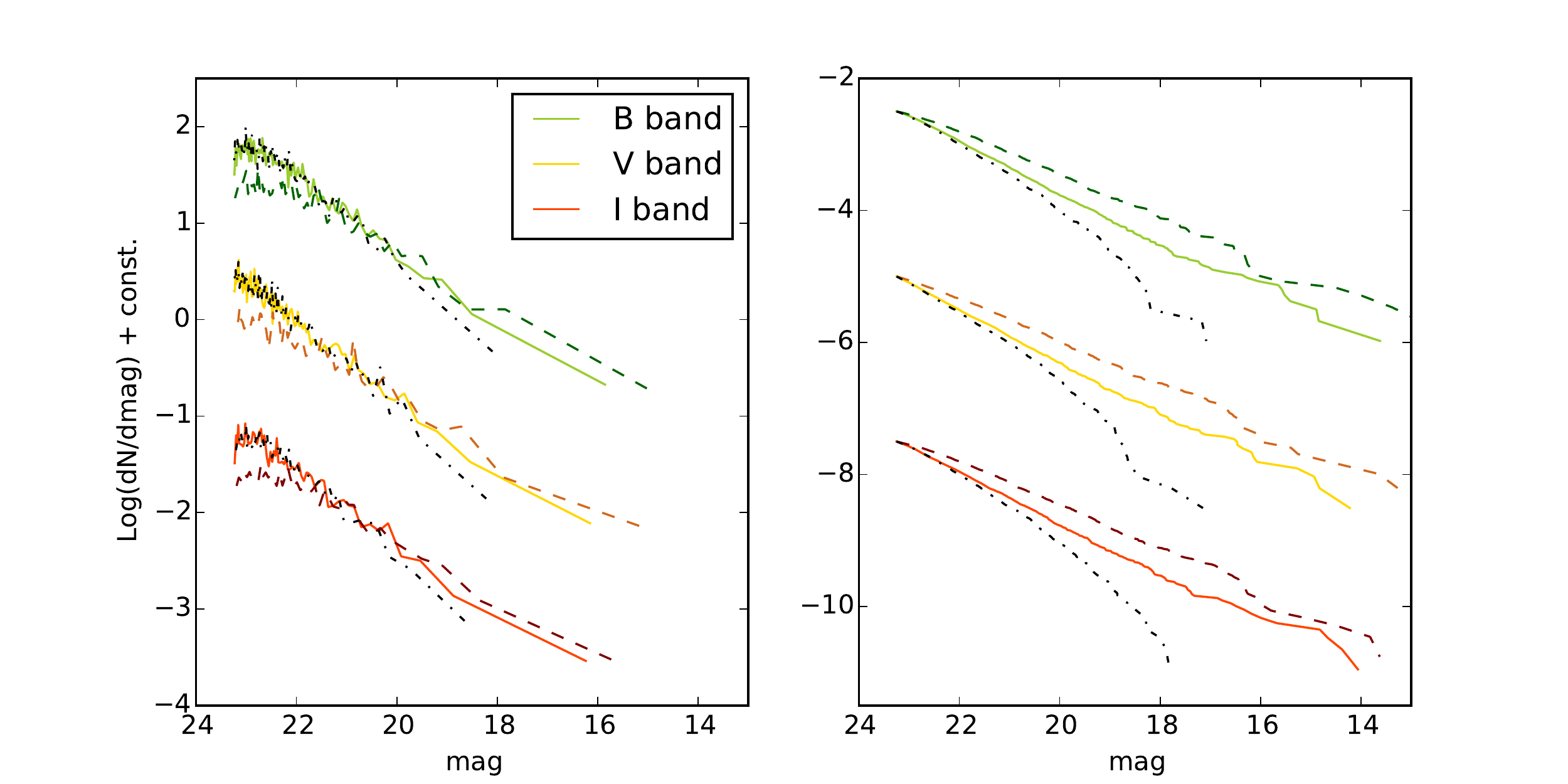}
\caption{Luminosity functions of the simulated populations in the binned (left) and cumulative (right) form. $BVI$ bands are plotted in both cases. The models plotted are $-2$ power law (PL-2, solid line), $-2$ power law including MDD (PL-2\_MDD, dashed) and Schechter (SCH, dashed-dotted) functions. Even if the underlying mass functions are very different, the luminosity functions are graphically quite similar in the binned form. When plotted cumulatively, instead, the differences are clear. In this second case the SCH model has the most similar LF shape to the one observed in the data.}
\label{fig:lumfit_simulated}
\end{figure*}

The results of the analysis are collected in Tab~\ref{tab:mc_lum} and plotted in Fig~\ref{fig:lumfit_simulated}.  
As expected, a $-2$ power law mass function has a luminosity function with the same shape, when no disruption is considered. 
Slopes close to $-2$ are retrieved in all filters with both a binned and a cumulative function fit. The values for the cumulative function are slightly steeper, but both methods gives comparable results.
\begin{table}
\centering
\caption{Results of the fit of the luminosity functions derived from the simulated cluster populations with a pure power law. The models are described in the text. The fit of the observed luminosity function (with only sources younger than 200 Myr) is given for comparison. The errors, not reported in the table, are on the order of $\pm0.03$ mag. The magnitude limits used are the same of the real data, listed in Tab.~\ref{tab:lumfit}.}
\begin{tabular}{llllllll}
\hline
\multicolumn{1}{l}{MF}	&  \multicolumn{3}{l}{binned function} 	& \multicolumn{1}{c}{\ }	& \multicolumn{3}{c}{cumulative function} \\
\					& B 		& V 		& I				& \ 					& B 		& V 		& I 				\\
\hline
\hline
OBS.				& 1.94	& 1.97	& 1.99			& \					& 2.14	& 2.14	& 2.25			\\
\hline
PL-2					& 1.90	& 2.00	& 1.96			& \					& 2.02	& 2.03	& 2.06			\\
PL-2\_MDD			& 1.65	& 1.70	& 1.76			& \					& 1.78	& 1.79	& 1.83			\\
SCH					& 1.94	& 2.06	& 2.13			& \					& 2.19	& 2.20	& 2.38			\\
SCH\_MDD			& 1.81	& 1.90	& 2.02			& \					& 2.08	& 2.09	& 2.29			\\
\hline
\end{tabular}
\label{tab:mc_lum}
\end{table}

Considering disruption, MID would not change the shape of the mass or luminosity function, as discussed in Section~\ref{sec:massage}. On the other hand a MDD would remove more quickly low-mass sources, modifying the luminosity function. The recovered slopes in this case are shallower, indicating that many low-luminosity sources have been removed (or have fallen below the completeness limit). 
The effect of MDD is producing shallower LF in all filters. 
We also observe that the slope is steeper in redder filters, as was also observed in the real data. This trend is mainly due to the difference in the the magnitude range fitted. The $I$ band, having a brighter completeness limit, is fitted only up to a magnitude of 22.25 mag, which is less affected by disruption than less bright magnitudes. If all filters will be fitted up to the same limiting magnitude the trend would not appear.  

When considering a Schechter mass function, the corresponding LF also has a steeper end. This is reflected in the recovered slopes, which in many cases have values more negative than $-2$. Including MDD still produces the effect of bending the low-luminosity end of the LF, but in this case, with slopes shallower than $-2$ at low luminosity and steeper than $-2$ at high luminosities, the resulting slope with a single power law can still be more negative than $-2$, as observed in the real data.

We can conclude that in order to produce luminosity functions with slopes steeper than $-2$ we need the underlying mass function to be truncated, or at least steeper than $-2$. MDD can affect the luminosity function, but only producing shallower functions.

\subsection{Age-Luminosity relation}
Another characteristic of the luminosity function that can be tested is the relative contribution of sources of different ages to the total luminosity in each magnitude bin. It has been proposed in the literature that, for a truncated mass function, the median age of clusters vary as function of luminosity, with the brightest clusters being on average younger than the faintest \citep[see e.g. ][]{adamobastian2015}. This expectation has been confirmed with semi-analytical models \citep[e.g.][]{larsen2009,gieles2010} and compared successfully with observations \citep{larsen2009,bastian2012b}. 

\begin{figure*}
\centering
\includegraphics[width=1\textwidth]{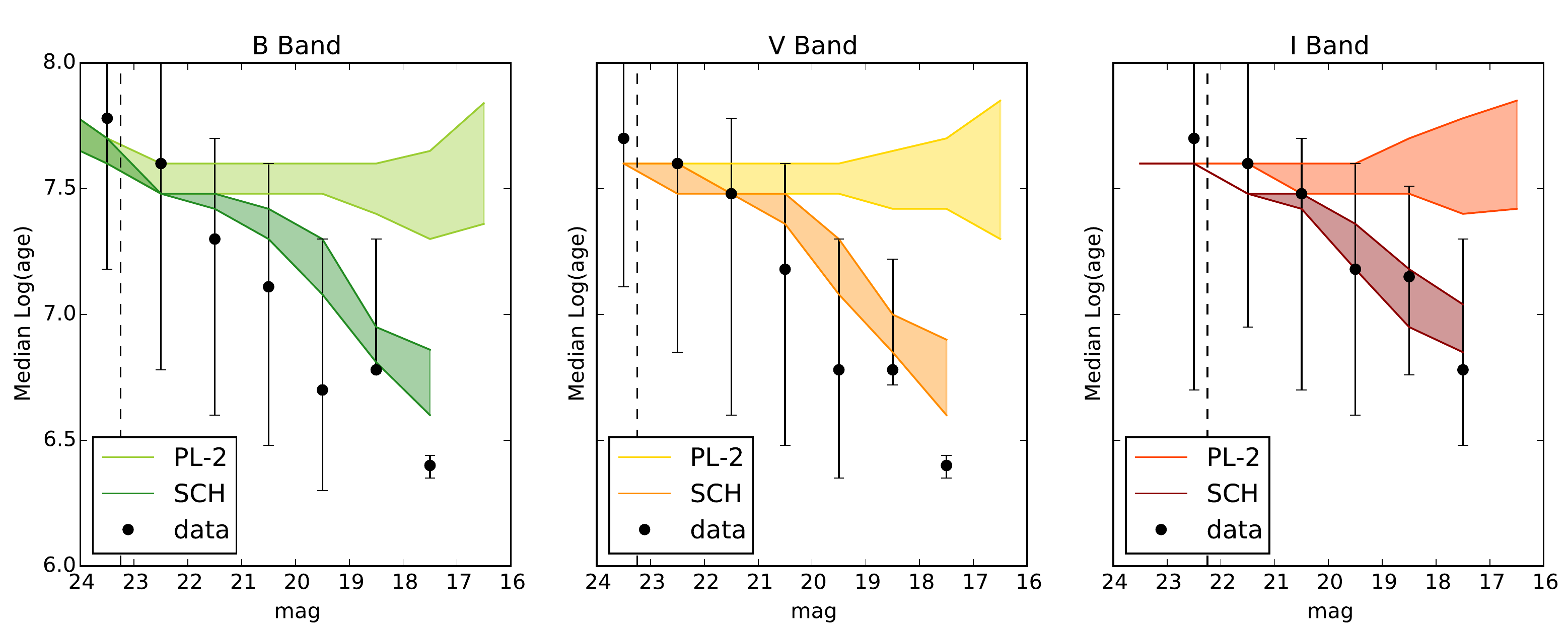}
\caption{Median ages of the clusters in magnitude bins of 1 mag. The observed data (black points, with bars extending from the 1st to the 3rd quartiles) are plotted over the expectations from the models. The shaded areas encompass the central 50\% distributions of medians from the simulations. The light areas (with an almost uniform behavior) are from the PL-2 model, the (declining) darker areas are from the SCH model. The data show a trend for younger ages associated in brighter clusters.  This is in agreement with expectations from a Schechter mass function with a mass truncation at $10^5$ \msun.}
\label{fig:lumfit_ages}
\end{figure*}

To verify that a similar trend is visible also with our catalogue we use the simulated cluster population, considering the PL-2 and SCH runs. 
For both cases we divide the sources in magnitude bins of 1 mag and we take the median age of the sources inside each bin. We repeated this process 100 times and in Fig.~\ref{fig:lumfit_ages} we plot the area covered by the distribution of the central 50\% median ages per magnitude bin. In the case of PL-2 the median ages are more or less constant at all magnitude, even if towards the bright end the spread between the percentiles increases, due to the lower number of sources there. On the other hand, for SCH, a trend with younger ages towards brighter bins is clear. 
We also include the median and 25th and 75th percent intervals obtained from the observed luminosities of the clusters. They show the same decreasing trend as the SCH model, bringing additional support to it. Similar conclusions are reached studying the age-luminosity relation for $\sim$50 of the brightest clusters of M51 with spectroscopically-derived ages in a forthcoming paper (Cabrera-Ziri et al., in prep).

We have not considered disruption in this simple comparison. Anyway we do not expect the disruption to change drastically the results: MID is unable to produce the observed trend, and MDD could in principle only produce an opposite trend, with brighter sources being on average older \citep{larsen2009}. We must conclude that the trend we see between ages and luminosities is another sign of an underlying truncated mass function.

\subsection{Simulated Mass Function}
\label{sec:mcmass}
\begin{figure*}
\centering
\subfigure{\includegraphics[width=0.325\textwidth]{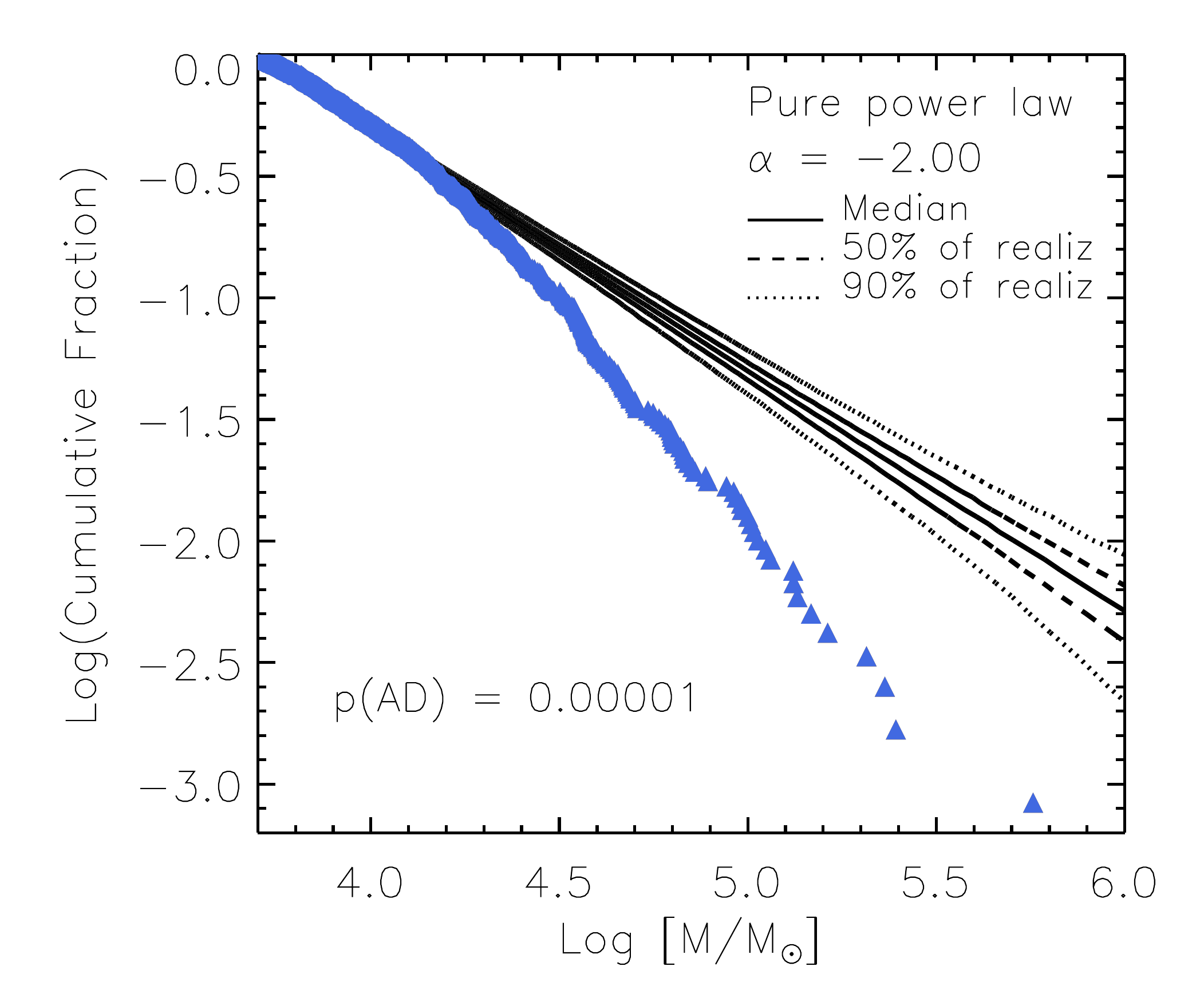}}
\subfigure{\includegraphics[width=0.325\textwidth]{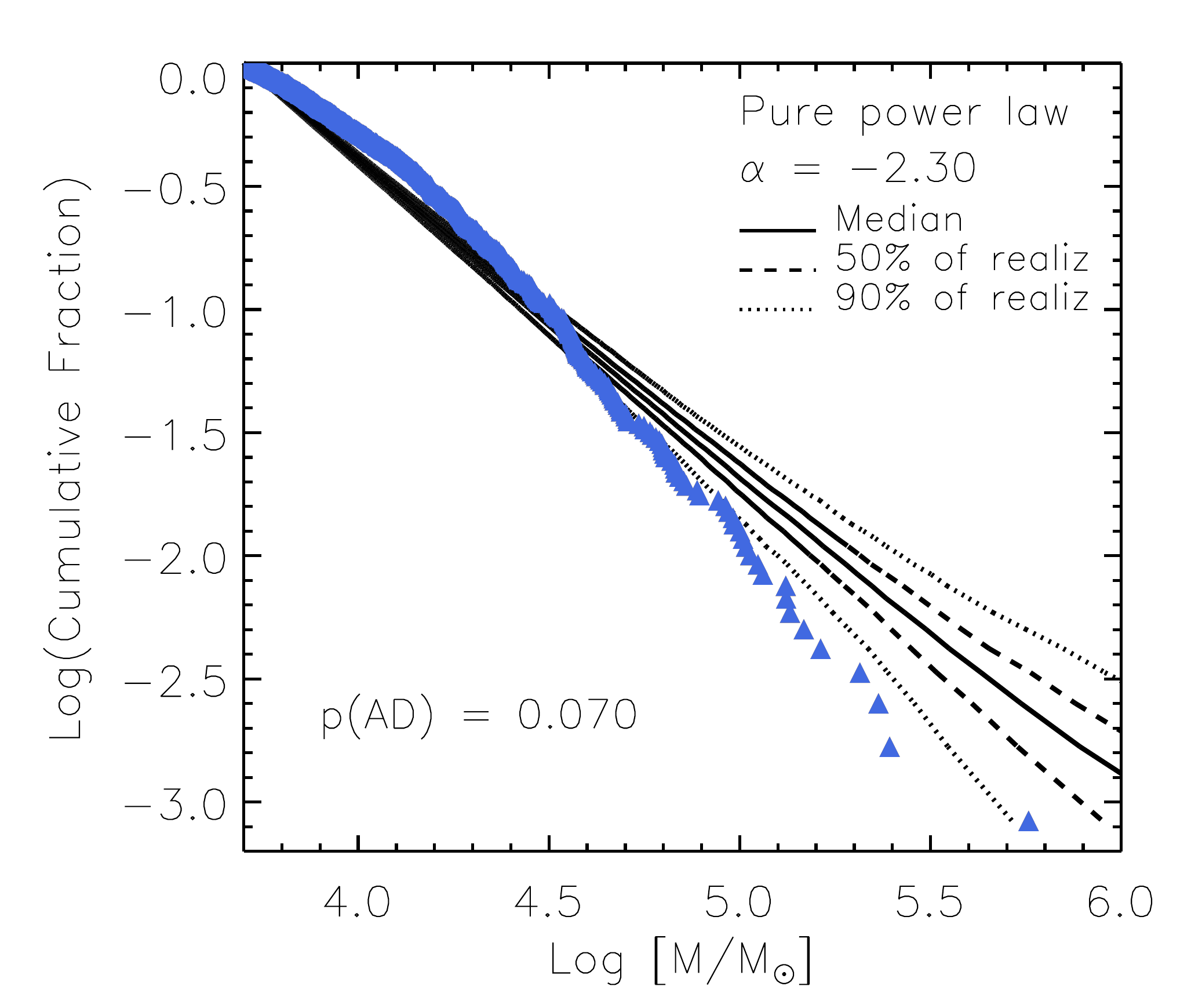}}
\subfigure{\includegraphics[width=0.325\textwidth]{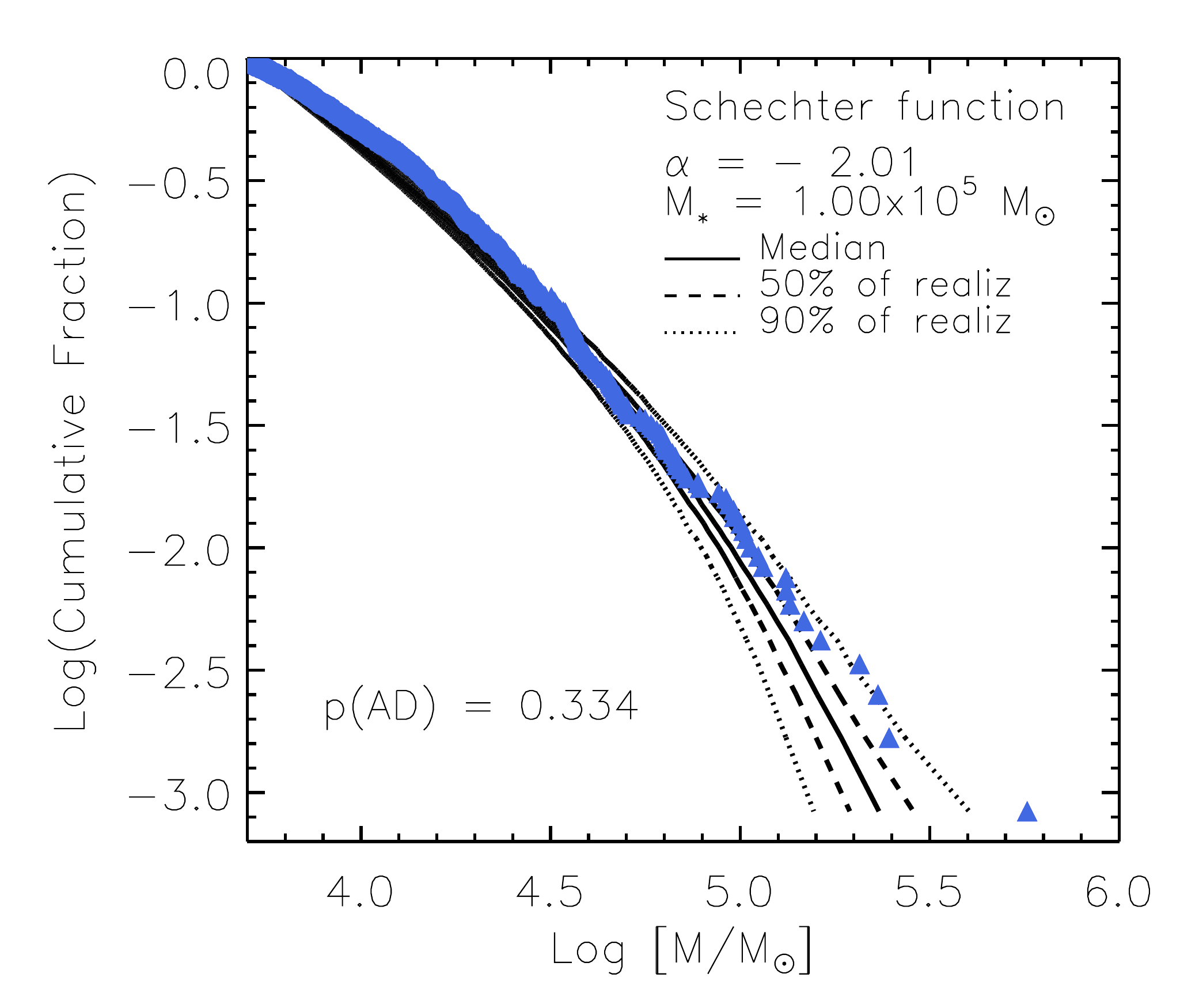}}
\caption{Monte Carlo simulations (black lines) compared to the observed mass function (blue triangles). 1000 Monte Carlo populations were simulated. The median mass distributions (solid lines) and the limits within 50\% (dashed) and 90\% (dotted) of the simulations are plotted. The resulting probability of the AD test is also reported (see text for description). The models for the simulated clusters are pure power laws with different slopes (left and center) and Schechter mass functions (right).}
\label{fig:massfit_simulated}
\end{figure*}

We compare the simulated mass functions with the observed one. For each model we set the number of simulated clusters in order to be the same of the observed ones. In doing so we are able to test for the effect of random sampling from the mass function, which could produce a ``truncation-like'' effect (see \citealp{dasilva2012}). We repeat the simulations 1000 times and compare the observed mass function with the median, the 50\% and the 90\% limits of the simulated functions in Fig.~\ref{fig:massfit_simulated}. 
We plot the mass functions in the cumulative form as we have seen that in this way the differences are graphically easier to spot. The models of the mass function considered for the simulations are a simple $-2$ power law, the single power law best fit of the cumulative function and a Schechter function with truncation mass $1.0\times10^5$ \msun\ (best fit found in Tab.~\ref{tab:massfit2}).
When comparing the mass functions behaviours in Fig.~\ref{fig:massfit_simulated} we immediately notice that a simple $-2$ power law (left panel) overestimates the number of cluster at high masses. A steeper power law (middle panel) follows better the observed data on average, but it underestimates the low mass clusters and overestimates the high mass ones. 
The Schechter case (right panel) instead follows quite nicely the observed mass function at all masses.

We test the null hypothesis that the observed masses are described by our models. We are mainly interested in the upper part of the mass function, which is the only part that possibly deviates from a simple power law description. We run the Anderson-Darling (AD) test comparing the observed masses larger than $1.0\times10^4$ \msun\ with the ones produced in the Monte Carlo simulations. The AD test returns the probability that the null hypothesis (the two tested samples are drawn from the same distribution) is true and a typical value for rejecting the null hypothesis is $p\sim10^{-4}$. The resulting probabilities of our test are collected in the plots of Fig~\ref{fig:massfit_simulated}. 
They confirm that $-2$ power law is a poor description for the massive part of the function (p $\approx10^{-5}$) while the other cases perform better (p $=0.070$ for the steeper power law and p $=0.334$ for the Schechter function). Both for the steeper power law and for the Schechter function the null hypothesis is not rejected. The test gives a better agreement with the Schechter function.

We can conclude that the analysis of the mass function suggests that there is a mechanism that inhibits the formation of clusters at very high masses. As pointed out in Section~\ref{sec:GMC}, the same is seen for GMCs, and this mass cut could therefore come from the progenitor structures.

\section{Cluster Formation Efficiency}
\label{sec:cfe}
\begin{figure}
\begin{center}
\includegraphics[width=\columnwidth]{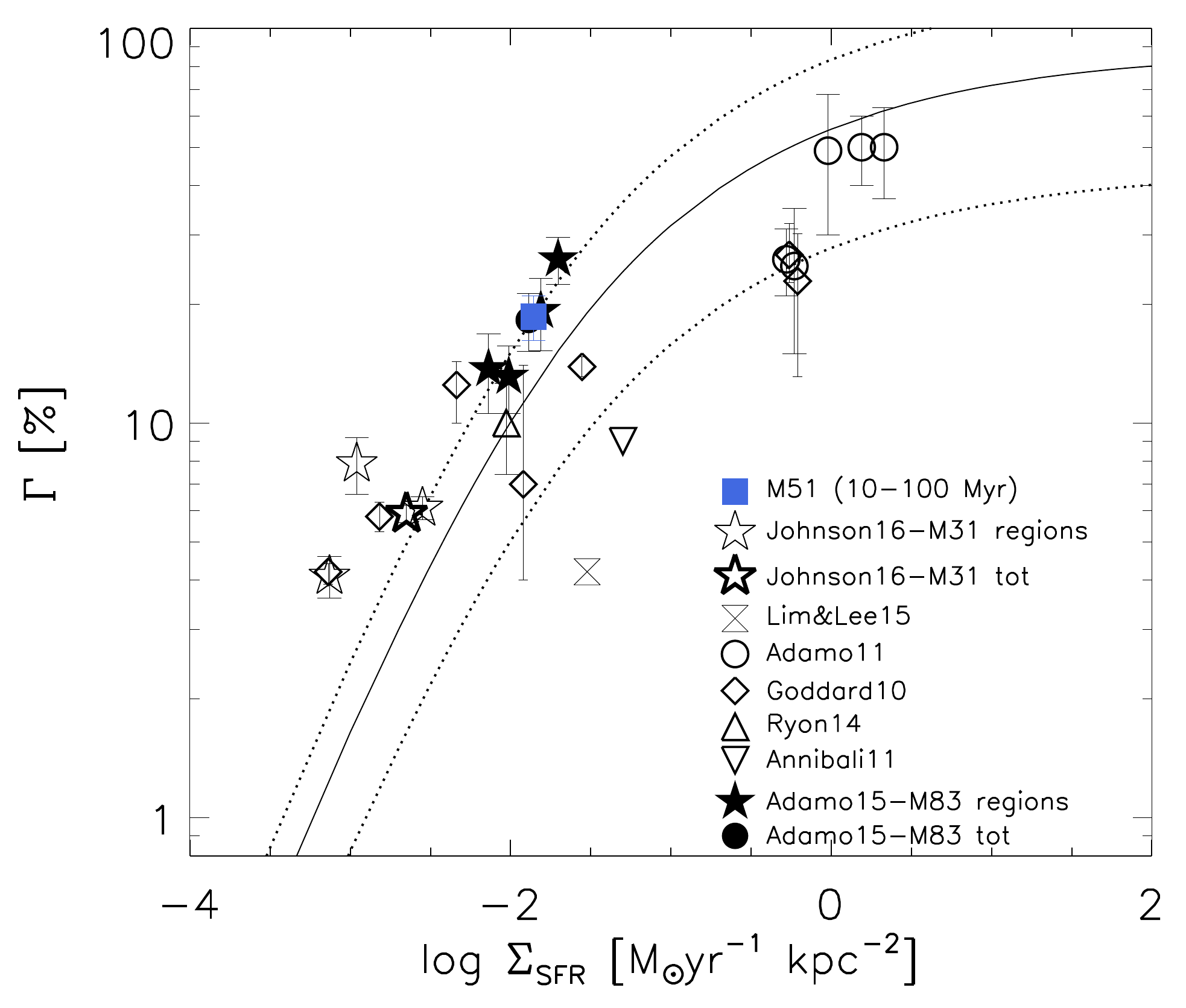}
\caption{Cluster Formation Efficiency $\Gamma$ in function of the average $\Sigma_{\textrm{SFR}}$ derived in the age range 10-100 Myr (blue square). Values for other galaxies \citep[taken from][]{goddard2010,adamo2011,annibali2011,ryon2014,adamo2015,lim2015,johnson2016} are shown for comparison. The black solid line is the $\Sigma_{\textrm{SFR}}-\Gamma$ model presented in \citet{kruijssen2012} with a 3$\sigma$ uncertainty enclosed by the dotted lines.}
\label{fig:gamma}
\end{center}
\end{figure}

The Cluster Formation Efficiency, CFE (also called $\Gamma$), is the fraction of star formation happening within bound clusters \citep[see][]{bastian2012}. Previous studies have found that $\Gamma$ varies positively with the SFR density, \sigmasfr. Galaxies with higher \sigmasfr\ also have on average a larger $\Gamma$ (see \citealp{goddard2010,adamo2011,ryon2014} and Fig. \ref{fig:gamma}). Variations of CFE values have been observed also inside single galaxies, e.g. in the study of M83 \citep{silvavilla2013,adamo2015}, suggesting that the main dependence of the efficiency is on the \sigmasfr.

In order to calculate $\Gamma$, we compared the SFR with the cluster formation rate (CFR). We calculated the SFR from FUV, correcting for dust using 24$\mu$m emission according to the recipe from \citet{hao2011} and assuming a $0.1-100$ \msun\ Kroupa IMF. The resulting value is collected in Tab.~\ref{tab:gamma}.

The CFR has been calculated in the age ranges, $1-10$, $1-100$, $1-200$ and $10-100$ Myr.
The sum of the masses of clusters in our catalogue with $M>5000$ \msun\ in the selected age range, divided by the age spanned, gives the CFR of the mass limited sample of sources. In order to correct for the missing mass of clusters less massive than 5000 \msun, we have assumed a model for the ICMF and derived the contribution of the low-mass clusters down to 100 \msun.
The assumed ICMF is a Schechter function with a truncation mass of $1.0\times10^5$ \msun, as resulting from the fit in Section~\ref{sec:massfunc}.

Values for measured masses, CFRs and $\Gamma$ are collected in Tab.~\ref{tab:gamma}. 
Errors on $\Gamma$ have been estimated considering uncertainties of 0.1 dex on age and mass values, a Poisson error 
on the number of clusters used for calculating the total mass, the errors on the mass function given in Tab.~\ref{tab:massfit2} and assuming an uncertainty of $10\%$ on the SFR value. 
$\Gamma_{1-10}$ has been derived using two different tracers for the SFR. In one case \ha+24$\mu$m has been used following the recipe by \citet{kennicutt2009}. This recipe assumes that SFR has been constant for $\sim100$ Myr. In the second case \ha\ has been used without correction for obscured SF. This second case traces only the SF younger than 10 Myr but a 50\% uncertainty is associated to the SFR value.
\begin{table}
\centering
\caption{CFR and CFE ($\Gamma$) values for M51. $^a$ SFR derived from FUV+24$\mu$m, associated uncertainty of $\sim10\%$. $^b$ SFR derived from \ha+24$\mu$m, associated uncertainty of $\sim10\%$. $^c$ SFR derived from \ha\ only, not corrected for obscured SF, associated uncertainty of $\sim50\%$.}
\begin{tabular}{ccccc}
\hline
Age range		& CFR		& SFR		& $\langle\Sigma_{\textrm{SFR}}\rangle$ 				& $\Gamma$	\\ 
\ [Myr]		& [\msun/yr]	& [\msun/yr]	& $\left[\textrm{M}_{\odot}/\textrm{yr kpc}^{-2}\right]$		& [\%]	\\
\hline
\hline
$10-100$		& 0.305		& 1.636$^a$			& 0.0139		& 18.6$\ _{\pm2.4}$		\\
\hline
$1-10$		& 0.465		& 1.437$^b$			& 0.0132		& 32.4$\ _{\pm12.1}$	\\
$1-10$		& 0.465		& 0.734$^c$			& 0.0062		& 63.3$\ _{\pm39.0}$	\\
$1-100$		& 0.321		& 1.636$^a$			& 0.0139		& 19.6$\ _{\pm2.5}$		\\
$1-200$		& 0.264		& 1.636$^a$			& 0.0139		& 16.2$\ _{\pm1.9}$		\\
\hline
\end{tabular}
\label{tab:gamma}
\end{table}

$\Gamma_{10-100}=(18.6 \pm2.4)\%$ is probably the value that is less affected by systematics. The absence of clusters younger than 10 Myr effectively avoid the possible inclusion of unbound sources in the sample, while the restriction to ages younger than 100 Myr lowers the effects of the cluster disruption on the $\Gamma$ derivation \citep[see][]{kruijssen2016}.
However, Tab.~\ref{tab:gamma} shows that all $\Gamma$ are consistent within 2$\sigma$ with a 20\% value. $\Gamma_{1-10}$ and $\Gamma_{1-100}$ are bigger due to the possible contamination by unbound sources already discussed in Section~\ref{sec:agefunc}. $\Gamma_{1-200}$ is slightly lower, but we know from the age function analysis (Section~\ref{sec:agefunc}) that disruption is affecting the cluster population.

$\Gamma_{10-100}$ is compared with CFEs from other galaxies in Fig.~\ref{fig:gamma}. Our value for M51 is similar to the $\Gamma$ values found for other local galaxies, in particular for similar spiral galaxies like M83 \citep{adamo2015}. More in general, it fits well into the \sigmasfr$-\Gamma$ relation modeled by \citet{kruijssen2012}, which predicts the amount of star formation happening in the cluster to increase with increasing surface density of star formation. 
In a recent work on M 31, \citet{johnson2016} suggested a comparison of the $\Gamma$ values with the surface density of molecular gas. Their results show that the cluster formation efficiency scales with $\Sigma_{\textrm{gas}}$, with the mid-plane pressure of the galactic disk and with the fraction of molecular over atomic gas. These finding suggests that environments with higher pressure can form denser gas clouds which in turn result in a higher fraction of star formation happening in a clustered form. The environmental analysis of Paper II, based on regions with different \sigmasfr\ and $\Sigma_{\textrm{gas}}$ inside M51, will help testing the cluster formation efficiency scenario.

\section{Conclusions} 
\label{sec:conclusions}
Using LEGUS \citep{legus1} broadband observations of M51 we have built a new catalogue of young stellar clusters. The new WFC3 coverage allows very accurate photometry in the $UV$ and $U$ bands, necessary for deriving precise ages of the young sources. 
The cluster catalogue is automatically extracted using the steps described in \citet{legus2}. A critical parameter of the extraction is the minimum concentration index considered (1.35 in our analysis). The SEDs of the extracted sources are fitted with \textit{Yggdrasil} SSP models \citep{yggdrasil} in a deterministic approach. 

Sources with detection in at least 4 filters are initially visually classified by humans. This subsample has been used as a training set for a ML algorithm that has classified the entire catalogue.
We focus our analyses on 2834 sources which are compact and uniform in color, neglecting multi-peaked sources.

Clusters are spatially associated with the spiral arms of the galaxy up to ages of 200 Myr, as recently shown in a simulation of spiral galaxies by \citet{dobbs2017}.
The luminosity, mass and age functions of the cluster sample are analyzed and compared to the results of simulated Monte Carlo cluster populations, in order to be able to better interpret the observed features. 

We list hereafter the main results of this work.
\begin{itemize}
\item A double power law provides the best fit for the luminosity function, suggesting a truncation at bright magnitudes which is better observed when the function is plotted in a cumulative way. A trend of steeper slopes with redder filter is observed, as already pointed out by \citet{haas2008}, and, in each filter, brighter sources have on average younger ages. 
\item The mass function has been directly studied with a maximum--likelihood fit, supporting the hypothesis of a truncated function: the cluster population of M51 is consistent with a Schechter function with slope $-2$ and a mass truncation at $M_{*} = 10^5$ \msun. 
The analysis is repeated considering only high mass clusters (M$>10^4$ \msun). A steeper slope is retrieved but the fit still gives preference to a truncated function with $M_{*}\sim10^5$. 
\item The age function indicates the presence of a moderate disruption over the range $10-200$ Myr.
The study of disruption in the first 10 Myr is precluded by the contamination of possibly unbound sources. Under the assumption of a mass-dependent disruption in time, a typical timescale for the disruption of $10^4$ \msun\ clusters is derived, $t_4=230$ Myr.  
\item Simulated Monte Carlo populations are used to test the luminosity and mass function analyses. The trends observed in the luminosity function are recovered when a cluster population with an underlying Schechter mass function is analysed. A cluster population simulated with a pure power law mass function fails instead to produce the observed trends. 
We notice that a cumulative function is the preferred way to study the bright end of the luminosity function and to investigate a possible deviation from a simple power law. The inclusion of cluster disruption in the simulations have also an impact on the luminosity function, but only at the faint end.
Monte Carlo population are also used for a direct study of the mass function. A careful study of the high mass end of the function in the comparison with the simulated populations rejects the hypothesis of the function following a simple $-2$ power law, with a probability of $\simeq0.001\%$ given by an Anderson-Darling statistics. 
\item We derive the fraction of star formation happening in clusters, which for M51 is $\simeq20\%$. This value is in line with the model of \citet{kruijssen2012} linking the cluster formation efficiency with the SFR density, as well as consistent with observations for $\gamma$ within local star-forming galaxies. 
\end{itemize}

In a forthcoming paper (Paper II) clusters will be analyzed as a function of environments inside M51, enabling us to link cluster properties with the interstellar medium and GMC properties. 



\section*{Acknowledgements}
Based on observations made with the NASA/ESA Hubble Space Telescope, obtained  at the Space Telescope Science Institute, 
which is operated by the Association of Universities for Research in Astronomy, Inc., under NASA contract NAS 5--26555. 
These observations are associated with program \# 13364.
Support for program \# 13364 was provided by NASA through a grant from the Space Telescope Science Institute.
We acknowledge the help and suggestions provided by M. Gieles and N. Bastian.
CLD acknowledges funding from the European Research Council for the FP7 ERC starting grant project LOCALSTAR. 
DAG kindly acknowledges financial support by the German Research Foundation (DFG) through program GO 1659/3-2.
MF acknowledges support by the Science and Technology Facilities Council 
[grant number  ST/P000541/1].





\bibliographystyle{mnras.bst}
\bibliography{biblio_p1}





\bsp	
\label{lastpage}
\end{document}